\documentclass[12pt,preprint]{aastex}

\slugcomment{To appear in the Astrophysical Journal}

\lefthead{Skinner et al.}
\righthead{NGC~2024}

\begin{document}

\title{A Deep {\it Chandra} X-ray Observation of the Embedded  
                            Young Cluster in NGC 2024}

\author{Stephen Skinner}
\affil{CASA, Univ. of Colorado, Boulder, CO 80309-0389 }

\author{Marc Gagn\'{e} and Emily Belzer}
\affil{Dept. of Geology \& Astronomy, West Chester Univ., West Chester, PA 19383-2130}


%
\newcommand{\ltsimeq}{\raisebox{-0.6ex}{$\,\stackrel{\raisebox{-.2ex}%
{$\textstyle<$}}{\sim}\,$}}
\begin{abstract}
We present results of a sensitive 76 ksec {\it Chandra} observation of
the young stellar cluster in NGC 2024, lying at a distance of $\sim$415
pc in the Orion B giant molecular cloud. Previous infrared observations
have shown that this remarkable cluster contains several hundred embedded 
young stars, most of which are still surrounded by circumstellar disks.
Thus, it presents a rare opportunity to study X-ray activity in a large
sample of optically invisible protostars and  classical T Tauri stars (cTTS)
undergoing accretion. {\it Chandra} detected 283 X-ray sources of which
248 were identified with  counterparts at other wavelengths, mostly in
the near-infrared. Astrometric registration of  {\it Chandra} images against 
the Two Micron All Sky Survey (2MASS) resulted in positional 
offsets of $\approx$0.25$''$ near field center,
yielding high confidence  indentifications of infrared counterparts.
The {\it Chandra} detections are characterized by hard heavily-absorbed
spectra and spectacular variability. Spectral analysis of more than 100 of the
brightest X-ray sources yields a mean extinction 
 $\langle$A$_V$$\rangle$ $\sim$ 10.5 mag and typical plasma energies
$\langle$kT$\rangle$ $\sim$ 3 keV.  The range of variability includes
rapid impulsive flares and persistent low-level  fluctuations
indicative of strong magnetic activity, as well as slow rises and falls
in count rate whose origin is more obscure. Some slowly-evolving outbursts
reached sustained temperatures of kT $\sim$6 - 10 keV.  {\it Chandra}
detected all but one of a subsample of 27 cTTS identified from previous
near and mid-IR photometry, and their X-ray and bolometric luminosities
are correlated. We also report the X-ray detection of IRS 2b, which
is thought to be a massive embedded late O or early B star that may
be the ionizing source of NGC 2024. Seven millimeter-bright cores
(FIR 1-7) in NGC 2024 that may be protostellar were not detected,
with the possible exception of faint emission near the unusual
core FIR-4.
\end{abstract}



\newpage

\section{Introduction}

The conspicuous HII region NGC 2024 lies about 15$'$ east of the bright
O9.5Ib supergiant $\zeta$ Orionis in the Orion B molecular cloud (= Lynds 1630).
Its estimated distance is 415 pc (Anthony-Twarog 1982). 
Near-infrared (IR) observations have uncovered a remarkably dense 
cluster of young stars in the molecular cloud associated with
NGC 2024 (Grasdalen 1974; Barnes et al. 1989 (B89); Comer\'{o}n, Rieke, \&
Rieke 1996; Lada et al. 1991; 
Haisch, Lada \& Lada 2000 (HLL); Haisch et al. 2001 (HLP);
Meyer 1996 (M96)).
The K-band survey of Lada et al. showed that the
cluster covers an area of $\approx$180 arc-min$^{2}$ and contains
more than 300 IR sources. The more sensitive survey of Comer\'{o}n
et al. detected $\approx$150 sources in the central 30 arc-min$^{2}$
region of the cluster down to a completeness limit of K $\approx$ 17,
of which about two-thirds showed IR excesses characteristic of
circumstellar disks. Using sensitive L-band observations, 
HLL found a somewhat higher infrared excess fraction $\geq$86\%. 
A recent near-IR adaptive optics survey detected 73 stars,
of which 3 are binaries and one is a triple system (Beck, Simon,
\& Close 2003 (B03)). Despite the high concentration of IR sources,
there are no optically visible stars toward the cluster center.
The cluster is undoubtedly young with age estimates ranging from 
$\sim$0.3 My (Meyer 1996) up to a few My (Comer\'{o}n et al. 1996).

Millimeter and submillimeter observations provide evidence that star-formation is
still underway in NGC 2024. Seven compact condensations of dust and gas denoted as 
FIR 1-7  were detected at 350 - 1300 $\mu$m by Mezger et al. (1988 (M88),
1992 (M92)), who argued that the condensations are isothermal protostars.
These compact cores have been  subsequently studied by numerous authors
as summarized by Lai et al. (2002). Their evolutionary status is still
a subject of debate, but the discovery of a faint near-infrared source 
at the FIR 4 position (Moore \& Chandler 1989; Moore \& Yamashita 1995) 
and a unipolar redshifted outflow (Chandler \& Carlstron 1996) suggests 
active star formation  in the vicinity of FIR 4.
Polarization observations and
Zeeman observations of  absorption lines toward the cloud with radio interferometers
have detected a magnetic field and  there is some evidence
that the field morphology around FIR 5 has been influenced by gravitational collapse 
(Crutcher et al. 1999; Lai et al. 2002).

At X-ray wavelengths, {\it ROSAT} observations of NGC 2024
were analyzed by Freyberg \& Schmitt (1995). The most sensitive image
obtained was a 42 ks HRI exposure in which 52 X-ray
sources were detected. A shorter 22 ks PSPC image showed that
most X-ray sources have  large absorption 
N$_{\rm H}$ $\sim$ 10$^{22}$ cm$^{-2}$ and typical unabsorbed 
X-ray luminosities were L$_{\rm x}$ $\sim$ 10$^{30}$ ergs s$^{-1}$.
The {\it ROSAT} detections were well-correlated with
known K-band sources and the X-ray properties suggested
that most were embedded young stars. 
The {\it ROSAT} observations clearly revealed a large
population of X-ray emitting sources  in 
NGC 2024. But, because of the heavy X-ray absorption
below $\sim$1 keV and  rather limited PSPC spectral
resolution and bandpass ($\approx$0.2 - 2.5 keV), spectral
parameters for individual sources were not easily 
determined from {\it ROSAT} data.

The  NGC 2024 region is an ideal target for
X-ray studies of star-formation for several reasons.
It contains a diverse range of objects including 
class I protostars
\footnote{We adopt the the IR classification scheme
used in Haisch et al. (2001) based on the IR spectral
index $\alpha$ = $d$~log($\lambda$F$_{\lambda}$)/$d$~log$\lambda$
evaluated in the 2.2 - 10 $\mu$m range. Class I sources
have  $\alpha$ $>$ 0.3, flat spectrum sources have
$-$0.3 $\leq$ $\alpha$ $<$ 0.3, class II sources
have $-$1.6 $\leq$ $\alpha$ $<$ $-$0.3, and class III
sources have $\alpha$ $<$ $-$1.6.}, classical T Tauri stars (cTTS),
weak-lined T Tauri stars (wTTS), the luminous infrared source
IRS 2b which may be a massive young OB star, and seven millimeter
condensations that may be protostellar. NGC 2024
lies at  galactic coordinates  ($b$,$l$) $\sim$
($-$16$^{\circ}$, 207$^{\circ}$) so source confusion from 
the galactic plane or galactic center is not an issue.
In addition, the high absorption
toward the cluster center reduces background 
contamination. 

We present here the results of a sensitive 76 ks {\it Chandra}
observation of NGC 2024 which provides higher angular
and spectral resolution, broader energy coverage, and longer time 
monitoring than has previously been obtained. 
Our primary goals were: (i) to obtain a sensitive 
high angular resolution X-ray census of 
NGC 2024, (ii) to compare the X-ray population with 
existing IR catalogs and identify potential cloud
members that might have escaped IR detection, 
(iii) to quantify the X-ray spectral and timing 
properties of the X-ray population,
(iv) to determine if the total X-ray luminosity 
contributes significantly to the ionization of the
visible HII region, and (v) to search for X-ray
emission from the compact millimeter cores FIR 1-7.
In the following
we discuss the spatial, temporal, and spectral properties
of the X-ray sources in NGC 2024 and compare the X-ray and
infrared populations.

\section{{\it Chandra} Observation and Data Reduction}

\subsection{Observation}

{\it Chandra} observed NGC 2024 from 8 August 2001
at 06:37 UT until 04:29 UT on 9 August (ObsId 1878), spanning 
21.9 hours. 
Exposures were obtained in faint data mode
with a 3.2 s frame time using the ACIS-I imaging
array as the primary detector. ACIS-I consists of
four front-illuminated 1024 $\times$ 1024 pixel CCDs 
with a pixel size of $\sim$0.492$''$ and a combined 
field-of-view (FoV) of $\approx$16.9$'$ $\times$ 
16.9$'$. The S2 and S3 CCDs in the ACIS-S array were also
enabled, but our discussion here will focus  on the 
ACIS-I data. The exposure livetime varied slightly between CCDs
but was in the range 76,655 - 76,658 s.
The {\it Chandra} aimpoint was at
nominal RA = 5$^h$ 41$^m$ 46.3$^s$, DEC = $-$1$^{\circ}$ 55$'$ 11.7$''$ 
(J2000), which is $\approx$32$''$ from the strong millimeter 
source FIR-5 (M88). More detailed information
on {\it Chandra}  and its instrumentation can be found 
in the {\it Chandra Proposer's Observatory Guide (POG)}
\footnote{http://cxc.harvard.edu/udocs/docs/docs.html}.

\subsection{CIAO Data Reduction}

We followed standard data reduction procedures for ACIS 
data using 
CIAO software vers. 2.2.1\footnote{Further information on 
{\it Chandra Interactive
Analysis of Observations (CIAO)} software can be found at
http:$//$asc.harvard.edu/ciao/}. 
Our data reduction
was based on the Level 1  files generated during standard  
processing by the {\it Chandra} X-ray Center (CXC) using
vers. 2.6 of the {\it Chandra} calibration database.
The CIAO data reduction  included
(i) application of the most recent aspect and focal
length corrections, (ii) event selection 
based on good {\it ASCA} grades 0,2,3,4,6 and a clean status 
bit (status = 0) and removal of afterglow events, and
(iii) energy filtering. For  image analysis and source
identification we used events in the 0.5 - 7 keV range
to reduce particle background.
Pixel randomization, which is introduced
by CXC as part of the standard data processing to remove possible
spatial aliasing effects, was retained. The algorithm to partially correct
for the effects of charge transfer inefficiency (CTI) was not
yet implemented in CIAO vers. 2.2.1 and we thus did not apply
CTI corrections. 

\subsection{Spatial Resolution and Limiting Sensitivity}

The on-axis spatial resolution of ACIS-I is limited by the 
physical pixel size of $\sim$0.492$''$. For on-axis point sources,
approximately 90\% of the encircled energy lies within 4 pixels
(1.97$''$) of the center pixel at 1.49 keV ({\it POG}). The limiting 
sensitivity of ACIS depends on many factors including 
the exposure time, background level, location of the
source on the CCD-array, and intrinsic source properties.
To obtain a rough sensitivity estimate for this observation,
we consider an on-axis point
source and adopt a 6 count detection threshold (or 
7.83 $\times$ 10$^{-2}$ c/ks for 
a livetime of 76.656 ks). We assume that the source has
spectral properties typical of those sources detected in
NGC 2024 (Sec. 4.3), namely an optically thin thermal plasma 
with  kT $\approx$ 2.8 keV,  absorption column density
N$_{\rm H}$ = 2.3 $\times$ 10$^{22}$ cm$^{-2}$, and abundances 
$\sim$0.3 solar. For this typical source at an assumed 
distance of 415 pc, the unabsorbed
luminosity detection limit from the Portable Interactive
Multi-Mission Simulator ({\it PIMMS}) software 
is log L$_{\rm x}$ $\approx$ 28.8 ergs s$^{-1}$ (0.5 - 7 keV).
The detection limit is higher for off-axis sources, or for
sources with heavier absorption or lower temperatures.
Conversely, the detection limit is lower for  sources
having less absorption or higher temperatures.

\section{Data Analysis}

The X-ray data analysis consisted of (i) {\it Chandra} image 
generation and registration against IR positions from the 2MASS data base,
(ii) identification of X-ray 
sources and their probable IR counterparts using existing
catalogs, (iii) extraction and analysis
of event lists for each source to determine source-specific
quantities  such as counts, mean photon energy, absorbed flux, and
the Kolmogorov-Smirnov (KS) variability statistic, (iv) light curve 
extraction for all sources, and 
(v)  extraction and fitting of spectra for brighter sources
to obtain the absorption column density, characteristic
plasma temperature, and unabsorbed luminosity. 
The analysis procedure is described in more detail 
below.

\subsection{Image Analysis and Source Identification}

We used the level 2 event file created by the CIAO processing
described above to generate  full-resolution 2800 $\times$ 2800 
pixel images (0.492$''$ pixels) as well as binned images.
Light curves from on-chip background regions were inspected for large
background fluctuations that might have resulted from solar flares
and none were found. To insure accurate identification of infrared
counterparts, we registered {\em Chandra} positions against IR
positions in the  the 2MASS 
database\footnote{
http:$//$www.ipac.caltech.edu/2mass/}.
To do this, we identified approximately 100 2MASS sources with obvious
counterparts in the initial {\em Chandra} images. We then used
the {\scshape astrom} program in the Starlink software package
to derive a 4-coefficient plate scale solution using
2MASS J2000 positions and corresponding {\it Chandra} physical 
pixel positions. After applying the positional correction to
the {\it Chandra} data, we found typical offsets between
X-ray and 2MASS positions near the center of the ACIS-I
detector to be $\sqrt{\Delta{\rm RA}^2 + \Delta{\rm DEC}^2}$
$\approx$ 0.25$''$. These small offsets permitted IR counterparts
to be identified  with very high confidence.

After applying the astrometric correction, we used the
CIAO wavelet source detection program {\scshape wavdetect} to search 
for X-ray sources.
Several {\scshape wavdetect} passes were made using scale factors
of 2,4, and 8 and  false-alarm probabilities of 10$^{-5}$ and
10$^{-6}$. The source lists from each {\scshape wavdetect}
pass were compared and the {\it Chandra} images were visually
inspected for missed or spurious detections. 

We identified 283 X-ray
sources  of which 248 were associated with
known IR, optical, or radio counterparts based on searches of the
SIMBAD data base, the 2MASS all-sky release Point Source Catalog,
source lists from previous IR studies (HLL; HLPTL; B89; B03; M96),
and a list of radio detections obtained in a recent 3.6 cm VLA
survey conducted by Rodriguez et al. (2003). Table 1 lists
the X-ray detections and their counterparts along with other
source information. Table 1 is also supplied in electronic
format.

The positions and PSF-corrected 95\% encircled energy regions
around each source were used to extract source events. For
the brightest sources a significant number of photons can
fall outside the 95\% encircled energy region, so larger
99\% encircled energy regions were used for event extraction in
bright sources.
For on-axis sources the 95\% and 99\% encircled energy regions
are nearly circular and have radii R$_{\rm 95EE}$ $\approx$
2$''$ and R$_{\rm 99EE}$ $\approx$ 8$''$ (Feigelson et al. 2002).
In the crowded cluster core region, it was necessary in some cases 
to use smaller extraction regions to avoid overlap from nearby
sources. 

A background count rate of 0.17 counts s$^{-1}$ per ACIS-I CCD 
in the 0.5 - 7 keV band (Table 6.6 of {\it POG}) was used
along with the size of the source extraction
region and exposure time to estimate the number of background counts for  
each source. The net source counts were then obtained by 
subtracting the number of background counts from the  
total number of counts in the source extraction region.
The net counts for each source are given in Table 1 along
with other source information. The number of source counts
that might have fallen outside the extraction region 
is within the net count uncertainty quoted in Table 1.
Sources with fewer than 6 net counts were not included
in Table 1.

Figure 1  shows a smoothed broad-band  {\em Chandra} image
of a 4$'$ $\times$ 4$'$ region in the central part of the 
cluster where the X-ray source density is highest.
Figure 2 is an unsmoothed image of the region in the 
vicinity of the infrared source IRS 2b, the possible 
ionizing source of NGC 2024 (Bik et al. 2003).
Figure 3 is a zoomed smoothed broad-band image of
the region near the millimeter sources FIR 1-7, which were 
not detected by {\it Chandra} with the possible exception of 
weak emission near FIR-4. This weak emission (4 counts) is 
shown in Figure 4. The astrometric-corrected positions
of all 283 {\em Chandra} detections are plotted in Figure 5,
overlaid with the positions of all 912 catalogued IR sources
in the ACIS-I FoV. The X-ray source density is clearly
higher near the center of the FoV, within the boundaries
of the known IR cluster.

\subsubsection{Unidentified {\it Chandra} Sources}

Figure 6 shows the  positions of the 35 X-ray sources
which lack counterparts at other wavelengths. About
one-third of the unidentified sources have positions 
projected within the  boundaries of the IR cluster and could be 
cluster members. However, the majority of the
unidentified sources are probably extragalactic.

The contribution of X-ray emitting foreground 
stars to the unidentified sample is expected to
be small because of the absence of optical counterparts
(see also Barger et al. 2002 for a breakdown of 
{\em Chandra} Deep Field North sources by object type).
The faint hard emission of the
unidentified X-ray sources also suggests a predominantly
extragalactic origin. None of the 35 unidentified
sources has more than 100 counts, while 43\% of
the identified sources have more than 100 counts.
All but four of the  unidentified sources have
mean photon energies above the average computed
for all X-ray detections (Sec. 4.2).

Using the 2 - 8 keV X-ray number counts from 
{\it Chandra} Deep Field  observations  
(Cowie et al. 2002), we estimate $\sim$16 extragalactic
sources in the ACIS-I FoV above our detection limit.
This estimate should be treated with  caution
since (i) a reduction in the number of  background
sources detected toward NGC 2024 will occur as a result 
of absorption  below a
few keV from intervening molecular cloud material, which
is non-uniform due to the north-south molecular
ridge that transects NGC 2024, and (ii) the
accuracy of the log N - log S distribution 
for extragalactic background sources from {\it Chandra} 
Deep Field observations at the lower Galactic latitude
observed here ($b$ $\approx$ $-$16$^{\circ}$) is not yet
known, as already pointed out in a discussion of
{\it Chandra} Orion Nebula observations 
(Feigelson et al. 2002).

In Figure 6, we have circled six unidentified {\it Chandra}
sources that could be extragalactic based on their 
exceptionally faint ($\leq$25 counts), 
hard ($\langle$E$\rangle$ $>$ 4 keV), non-variable emission. 
Excluding these extragalactic candidates, there are
seven remaining unidentified sources within the approximate
IR cluster boundaries that may be young stars
and for which follow-up observations at other
wavelengths would be useful. These are sources
20, 21, 157, 168, 180,  197, and 223. The latter
source may have a 2MASS  IR counterpart (Table 1).

\subsection{Timing Analysis}

To obtain a quantitative measurement of the variability for each
source, we have computed the non-parameteric Kolmogorov-Smirnov (KS) 
statistic from the event list of each source (Table 1).
The KS statistic is given by
KS = $\sqrt{n}$ sup$|f_{i}(t) - f_{0}(t)|$ 
where $n$ is the number of events, $f_{i}(t)$ is the normalized
observed cumulative distribution and $f_{0}(t)$ is the
normalized model cumulative distribution, assuming a constant flux.
A discussion of the KS method and
a comparison with the more familiar $\chi^2$ test is given by Rohatgi 
(1976) and additional information can be found in Babu \& Feigelson (1996)
and Press et al. (1992).

If $z$ is the value of the KS statistic then the null hypothesis
probability P($z$), or probability of constant flux, is given by
a convergent infinite series (Press et al. 1992): \\

\begin{equation}
{\rm P}(z) = 2 \sum_{j=1}^{\infty} (-1)^{j-1}e^{-2j^{2}z^{2}} 
\end{equation}

Truncating the series at six terms gives a few representative
values:   P(0.5) = 0.96, P(1.0) = 0.27,
P(1.5) = 0.02,  P(1.7) = 0.006,
and P (2.0) = 6.7 $\times$ 10$^{-4}$.
In the discussion below we adopt  $z$ = KS = 1.7 as the
dividing line between variable and non-variable sources,
where KS = 1.7 implies a variability probability
P$_{var}$ = 0.994. That is, any source in Table 1
with KS $>$ 1.7 is considered to be variable.
The KS values in Table 1 can be
converted to the corresponding null hypothesis probability using 
the above equation. We note that the KS statistic
is  less reliable for diagnosing
variability in weak sources ($<$25 counts)
as discussed further below (Sec. 4.1).

The KS statistic identifies those sources that were 
variable but does not provide information on variability
profiles or timescales. To obtain this information,
light curves were generated 
for all sources. Figure 7 shows
light curves for ten sources in the ACIS-I FoV that 
demonstrate the wide range of
variability detected.  Figure 8  illustrates the
continuous variability that was detected in source 94 throughout the 
observation. In Figure 9, we show the hard variable  emission
in source  207 with mean photon energies in excess of
4 keV and very high plasma temperatures.
The interesting variability in these sources 
is discussed in more detail below (Sec. 4.5).

\subsection{Spectral Analysis}

We have used two different  approaches to 
determine source spectral properties. First, the event
list of each source was analyzed to determine the 
mean photon energy $\langle$E$\rangle$ and the
absorbed flux F$_{\rm x,abs}$ (0.5 - 7 keV) values listed 
in Table 1. Second, spectra were extracted for the 
brightest sources and fitted with thermal plasma models
to determine the absorption column density (N$_{\rm H}$),
plasma temperature (kT), and unabsorbed luminosity 
L$_{\rm x}$(0.5 - 7 keV) values given in Table 1. 

These two methods are complementary. The advantage of
event list analysis is that it provides information
on spectral hardness (or $\langle$E$\rangle$) and
F$_{\rm x,abs}$ for all detections, including faint
sources. Spectral fitting provides additional 
information on N$_{\rm H}$, kT, and L$_{\rm x}$
but is only practical for brighter sources
($\geq$90 counts).

The advantage of using both techniques is that
the column density N$_{\rm H}$ from spectral fits
of brighter sources can be incorporated into the 
event list analysis. In that case, the event
list analysis provides information on kT and
L$_{\rm x}$, as well as $\langle$E$\rangle$ and
F$_{\rm x,abs}$. If the event list of a specific 
source is first
partitioned by time, then the time evolution of
these four quantities can be plotted in 
``light curve'' format, as shown in Figures 8
and 9.

\noindent {\em Event List Analysis:}
Our event list analysis followed the procedures described in
Gagn\'{e}, Daniel, \& Skinner (2003), which we  summarize
here.  The mean energy $\langle$E$\rangle$ is obtained 
from the event list, which contains the arrival time and 
energy  of each photon.
To compute  the  unabsorbed flux F$_{\rm x,abs}$, 
we generated an auxiliary
response file (ARF) for each source, which contains 
information on effective area as a function of photon energy,
taking into account the source position on the detector.
The absorbed flux is then given by

\begin{equation}
   F_{\rm x,abs} = \frac{1}{t} \sum_{i=1}^{n} \frac{E_i}{a_i(E)} 
   - F_{\rm bkg}A,
\end{equation}
where $E_i$ is the energy of each photon, $a_i(E)$ is the corresponding
effective area, $t$ is the livetime of the observation,
A is the area of the source extraction region, and
$F_{\rm bkg} = 3.84 \times 10^{-18}$~ergs~cm$^{-2}$~s$^{-1}$~pixel$^{-1}$
is the 0.5-7.0~keV background flux as determined from 
the CXC software tool {\scshape make\_acisbg}.

For brighter sources with $\geq$90 counts, additional information on the 
column density N$_{\rm H}$ is available from spectral fits (see below)
and the event list analysis can be taken a step further to obtain
plasma temperature T and L$_{\rm x}$. This is accomplished by generating 
a grid of simulated spectra
using the 1T {\scshape vapec} model in XSPEC vers. 11, with 
column densities in the range log N$_{\rm H}$ = 20.0 - 24.0 cm$^{-2}$
and temperatures in the range log T = 6.0 - 9.0 (K) in increments of
0.1 dex. The grid is parameterized as a function of the 
mean photon energy $\langle$E$\rangle$, which is known from the
event list of each source.
Simulations show that $\langle$E$\rangle$ increases smoothly with
N$_{\rm H}$ and T. The grid is then entered with the values of
N$_{\rm H}$ and $\langle$E$\rangle$ to determine T. Once T is
determined it can be used along with N$_{\rm H}$ to compute
the conversion factor from absorbed to unabsorbed flux
C(T,N$_{\rm H}$) =  F$_{\rm x}$/F$_{\rm x,abs}$. The unabsorbed luminosity is then
L$_{\rm x}$ = 4$\pi$d$^{2}$F$_{\rm x}$ = 4$\pi$d$^{2}$F$_{\rm x,abs}$C(T,N$_{\rm H}$).

\noindent {\em Spectral Fits:}
Spectra for sources with $\geq$90 counts were extracted
along with source-specific response matrix files (RMF)
and ARF files  using the 
CIAO tool {\scshape psextract}. Background spectra were extracted
from regions in the immediate vicinity of the source.
Spectra of weaker sources with $\sim$90 - 300 counts
were rebinned to 10 counts per bin prior to spectral
fitting, while brighter sources ($>$300 counts) were rebinned to
15 counts per bin. We have restricted our spectral 
analysis to sources with $\geq$90 counts since our
test fits show that the uncertainties in spectral parameters
derived for weaker sources are often too large to permit
definitive analysis.

All sources with $\geq$90 counts were fitted with a single-temperature 
absorbed optically thin plasma model (1T {\scshape vapec})  
using {\scshape xspec} vers. 11. A representative 1T {\scshape vapec} fit 
of variable source  85 is shown in Figure 10, along with
its pre-outburst spectrum.  Brighter sources ($>$300
counts) were also fitted with a two-temperature
2T {\scshape vapec} model and in some cases the differential
emission measure (DEM) model {\scshape c6pvmkl} was applied.
During spectral fitting with {\scshape vapec} models, the free parameters were
the equivalent neutral hydrogen column density 
(N$_{\rm H}$),   X-ray temperature(s) (kT), and normalization
factors used to reproduce the observed count rate.

Abundances were held fixed at solar values relative
to Anders \& Grevesse (1989), except for iron which
was fixed at a lower value Fe = 0.3 $\times$ solar.
The choice of a subsolar iron abundance was based on
variable abundance fits of the brightest sources which 
invariably converged
to a low Fe abundance near Fe $\approx$ 0.3 $\times$ solar.
For example, we cite the relatively strong non-variable source 
75 (3092 counts). Fits
of this source with a 1T {\scshape vapec} model converged to 
Fe = 0.22 [0.04 - 0.40] solar and 2T  {\scshape vapec} models
gave a very slight improvement with  Fe = 0.35 [0.13 - 0.58]
solar, where brackets enclose 90\% confidence intervals.
A comparison of 1T {\scshape vapec} fits of different sources 
indicates that solar abundance models tend to overestimate
N$_{\rm H}$ and unabsorbed  L$_{\rm x}$ relative to the improved fits
obtained with subsolar Fe.

The unabsorbed X-ray flux in the 
0.5 - 7 keV range was measured from 
best-fit models after setting N$_{\rm H}$ = 0 and then converted
to an equivalent unabsorbed luminosity L$_{\rm x}$ using a 
distance of 415 pc. Best-fit values of N$_{\rm H}$ and kT (with
90\% confidence limits from the {\scshape xspec} {\it steppar} routine) 
and unabsorbed L$_{\rm x}$ for the brightest sources are given in Table 1.

We emphasize that  kT values determined from the single-temperature
1T {\scshape vapec} models  are  only used to define a characteristic
time-averaged temperature that allows a large number of  sources
to be compared. This should not be construed to mean that
the X-ray emitting plasma is  isothermal. It is
clear from our analysis  that large
temperature fluctuations did occur in some sources 
(Table 1 notes; see also Figs. 8 and 9). However, in sources
with no statistically significant variability we only found 
definite improvement in spectral fits using a two-temperature
model (2T {\scshape vapec}) in a few cases (Table 1 notes). 
This may be attributed to the strong absorption below $\sim$1 keV
that is present in most spectra, effectively masking any
soft emission that could be present. The spectrum of source  85
in Figure 10 clearly illustrates this low-energy absorption.

\subsection{Implications of Photon Pileup}

Pileup occurs when two or more photons are detected within
the same pixel during one 3.2 s readout period. Pileup is
a consideration for high count-rate sources at small off-axis
angles where the point-spread function (PSF) is narrowest, even
for rapidly flaring  sources where the count rate may increase to high levels
only for short periods of time. In general, the effects of pileup are to
make the spectrum appear harder than it actually is and to
artificially reduce the count rate. However, there are other
implications as discussed in the {\it Chandra POG}.

We have identified those bright sources in NGC 2024 where
significant pileup ($>$10\%) could have occurred. Fortunately,
pileup is not an issue for this observation except for a 
few sources, and then only at moderate levels. Three strongly
variable sources located within $\approx$1$'$ of the {\it Chandra}
aimpoint had peak count rates that exceeded $\approx$0.09 c s$^{-1}$
 and are suspected to have piled at the
$\approx$10\% - 30\% level for short periods. These 
sources are nos. 174, 192, and 210. Another source which
had a relatively high but steady count rate was no. 120,
whose mean count rate was 0.10 c s$^{-1}$ with 
short-term fluctuations up to $\approx$0.15 c s$^{-1}$. However,
it is offset $\approx$2$'$ from the aimpoint so its pileup
should not have exceeded $\approx$10\% - 15\% due to 
off-axis broadening of the PSF.  Because of 
moderate pileup in these four sources, their mean photon
energies and flare temperatures given in Table 1 may be 
slightly overestimated, and total counts may be slightly
underestimated.

\section{X-ray  Properties}

We summarize here the global X-ray properties of the
{\it Chandra} sources in NGC 2024 based on temporal and
spectral analysis.

\subsection{Time Variability}
A total of 72 out of 283 sources (25\%) showed variability
as inferred from KS $>$ 1.7, and further information on 
sources with the strongest variability is given in the notes
to  Table 1.  
However, the variability fraction appears to be much larger among
brighter sources.   Only 5\% of those
sources with less than 25 counts showed variability, 
whereas 64\% of sources with more than 
400 counts were variable. This trend is obvious
in Figure 11,  which shows a monotonic increase in 
variability fraction with source counts. On physical
grounds, there is little reason to suspect that 
the variability fraction should increase monotonically
with the number of source counts. Instead, this 
trend  can be attributed to the increased difficulty in detecting
variability in weaker sources. It is thus likely that
the true variability fraction of the 283 detections
is greater than 25\% due to undetected variability in faint
sources, and this value should be interpreted as a lower limit.

The range of variability shown in the light curves (Fig. 7)
is truly amazing, consisting of rapid onset flares
(e.g. sources 27, 81 and 174), variations with slow rises
and decays (175, 192, and 207), and sources with slowly
declining or increasing count rates throughout the observation
(94, 123, 269). Although
rapid-onset flares are suggestive of solar-like behavior attributable
to magnetic reconnection events,  the origin of the slower variability 
is more obscure. In some cases the slow variability may be   
dynamically-induced, as for example by  rotation of an X-ray bright
active region on the star across the  line-of-sight. 
Longer observations spanning  days or weeks would be
needed to search for  periodic variability.

\subsection{Spectral Hardness (Mean Photon Energy) }

Figure 12 shows the distribution of mean photon
energy $\langle$E$\rangle$ for all 283 detections. 
The average value  is 
$\langle$ $\langle$E$\rangle$ $\rangle$ =
2.7 keV. The distribution in Figure 12 is 
noticeably truncated at low energies below
$\langle$E$\rangle$ $\sim$ 1.4 keV. This is a
result of the heavy absorption toward the cloud
which effectively absorbs soft photons and skews
the  distribution.

Figure 13 is a plot of $\langle$E$\rangle$ against
the KS variability statistic. We have partitioned
the diagram into four quadrants using KS = 1.7
to separate variable and non-variable sources and
$\langle$E$\rangle$ = 2.7 keV to distinguish 
between sources with above average or below average
mean photon energies.

Several interesting features are apparent in
Figure 13. First, all four quadrants are well-populated
and there is no obvious relationship between 
variability (KS) and $\langle$E$\rangle$ for the
sample as a whole. 
It is apparent that variability is not confined
to hard sources since many sources with 
KS $<$ 1.7 have $\langle$E$\rangle$ $>$ 2.7 keV.
The Kendall's $\tau$ test in the 
{\scshape asurv} statistical software package gives a probability 
of correlation between KS and $\langle$E$\rangle$ of
$p$ = 0.74 and the Cox test gives $p$ = 0.93. Thus,
statistical analysis does not provide conclusive
evidence for a correlation.
Second, it is obvious from Figure 13 that most of the
35 unidentified {\it Chandra} sources lie in the 
lower right part of the diagram and have hard non-variable
emission. These sources are also faint (Sec. 3.1.1)
and the hard emission suggests that most are extragalactic 
background sources but some could be
 deeply embedded sources in the cloud.
Third, there is a large spread in $\langle$E$\rangle$
among the {\it Chandra} sources, ranging from 
$\langle$E$\rangle$ = 0.92 keV (source 181, which
has the IR counterpart HLL-116) to the faint hard 
non-variable source 197 with $\langle$E$\rangle$ = 4.77
keV and no known counterpart.

\subsection{Spectral Temperature and Absorption  Estimates}

Since very few of the {\it Chandra} detections have
optical counterparts, reliable information on the 
visual extinction A$_V$  is generally not available.
We have not made extensive use of methods for estimating A$_V$
based on JHK photometry since these methods 
make {\it a priori} assumptions about dereddened
colors of the star in order to extract the color 
excess and A$_V$. The errors which can be incurred
in A$_V$ are substantial, as discussed in more 
detail by Gagn\'{e} et al. (2003).

Figure 14 is a plot of N$_{\rm H}$ and kT derived from
spectral fits of more than 100 of the brightest X-ray sources.
It is apparent from Figure 14 that those sources with
the highest absorption and highest temperatures are
all variable. However, the converse is not true.
Variability was also detected in  sources that
are situated in the lower left part of the diagram,
where kT and N$_{\rm H}$ are relatively low. Thus, X-ray variability 
alone does not provide reliable information on 
either time-averaged temperature or absorption.

The median absorption determined from 117 sources is log N$_{\rm H}$ = 
22.18 cm$^{-2}$ and the mean is log $\langle$N$_{\rm H}$$\rangle$ =
22.37 cm$^{-2}$. The average  
temperature derived from 1T {\scshape vapec} fits of sources with $\geq$90
counts is $\langle$kT$\rangle$ = 2.8 keV. In computing this
average we have attempted to exclude time intervals during
large rapid flares or slowly varying outbursts, although this was 
not possible in some cases where non-flare counts were low.

The mean absorption column density log $\langle$N$_{\rm H}$$\rangle$ =
22.37 cm$^{-2}$ corresponds to a  mean visual extinction
$\langle$A$_V$$\rangle$ = 10.5 mag using the conversion
A$_V$ = 4.5 $\times$ 10$^{-22}$~N$_{\rm H}$ mag (Gorenstein 1975).
This X-ray derived value is in excellent agreement with 
the value A$_V$ $\approx$ 10.5 mag determined for cluster
members using IR data (HLL). We note a rather large spread
in N$_{\rm H}$ ranging from no detectable absorption in 
source 282 (= 2MASS 05422123-0159104) to a maximum of 
log N$_{\rm H}$ = 23.18 cm$^{-2}$ (A$_V$ $\approx$ 68 mag)
in the variable source 160.

Estimates of A$_V$  based on IR data have
been derived for 27 class II sources (cTTS)  
in NGC 2024 and the class III source (wTTS) 
HLP-24, as listed in Table 2 of HLP.
We derived independent estimates of N$_{\rm H}$ for 21 of 
the class II sources from {\it Chandra} spectral fits (Table 1).
Using the results of  Gorenstein (1975) to convert from
N$_{\rm H}$ to A$_V$, we  generally find good agreement between 
the X-ray and IR determinations of A$_V$ subject to rather
large uncertainties in the X-ray estimates for fainter
sources. But significant differences were found for two sources.
For source 31 (187 counts) whose  counterpart is
the IR source HLP-64, we obtain A$_V$ = 3.1 [1.8 - 4.5; 90\% conf.] 
mag while the IR data give A$_V$ = 7.1 mag.
For source 69 (115 counts) the  counterpart is
HLP-14 which lies in a crowded IR field and the X-ray data 
yield A$_V$ = 5.8 [3.6 - 8.6] mag
but the IR data  give a smaller value A$_V$ = 0.8 mag.

\subsection{X-ray Luminosities}

Since reliable estimates of the visual extinction are not
available for most of the embedded sources in NGC 2024,
the intrinsic (unabsorbed) X-ray luminosities of many of
the fainter {\it Chandra} detections are not yet known.
We thus restrict our discussion of X-ray luminosities to
the subsample of class II and flat-spectrum sources for which A$_V$ estimates
are available, and to an estimate of the total cluster 
X-ray luminosity.

\subsubsection{X-ray Luminosities of Class II Sources (cTTS)}

NGC 2024 contains 27 class II sources identified from previous
near and mid-IR observations (Table 2 of HLP). 
Their spectral energy distributions peak in the near IR
and decline more slowly toward the mid-IR than do
normal stellar photospheres, indicative of circumstellar
disks.

{\it Chandra} detected 26 of the 27 class II sources
listed in HLP.
The only class II source not clearly detected was 
HLP-100 (= 2MASS 05413545-0152288). However, we do
see a weak X-ray source (5 counts) at an offset of
0.75$''$ from HLP-100, and we classify it as
a possible detection.

Figure 15 shows the cumulative X-ray luminosity 
distribution function for the 27 class II sources.
This distribution function was obtained using the 
Kaplan-Meier (KM) estimator in the {\scshape asurv} 
software package. We have estimated unabsorbed L$_{\rm x}$
for the faintest class II detections using the PIMMS
simulator and the absorbed fluxes in Table 1. The PIMMS
simulations assumed a 1T Raymond-Smith optically thin 
plasma model with a typical temperature for sources in
NGC 2024 and an absorption  N$_{\rm H}$ computed from
the  A$_{V}$ values given in HLP and the
Gorenstein (1975) transformation.
The possible detection of HLP-100 was treated as an
upper limit (5 counts) with log L$_{\rm x}$(0.5-7 keV) $\leq$
28.63 ergs s$^{-1}$.

A hard tail is clearly evident in the 
luminosity distribution function shown in Figure 15. This is due to
the anomalously high L$_{\rm x}$ values of four class II
sources with  L$_{\rm x}$ $\geq$ 30.7 ergs s$^{-1}$ which
exhibited rapid-onset flares or other prominent variability 
(source nos. 41, 94, 133, 174).

Estimates of the bolometric luminosity (L$_{bol}$) were
given for the 27 class II sources by HLP, ranging from
L$_{bol}$ = 61 L$_{\odot}$ for HLP-74 (source 174)
to L$_{bol}$ = 1 L$_{\odot}$ for both HLP-98 (source 220)
and HLP-100 (not detected).
We used their L$_{bol}$ values along with the unabsorbed
L$_{\rm x}$(0.5 - 7 keV) measurements from {\it Chandra} to check for 
a correlation and to obtain a 
linear regression fit using the Buckley-James
algorithm in {\scshape asurv}. We excluded the 
four strongly variable sources noted above 
from the correlation test. The Kendall's $\tau$ and
Cox tests both give  a correlation probability
$p$ =  0.98 between L$_{bol}$
and unabsorbed L$_{\rm x}$(0.5 - 7 keV). 
The regression analysis
gives  log L$_{\rm x}$ (ergs s$^{-1}$) = 29.86 $+$ 0.54 ($\pm$0.19)
log [L$_{bol}$/L$_{\odot}$], with a standard 
deviation of 0.36 in log  L$_{\rm x}$. For a typical class II
source with L$_{bol}$ $\sim$ 1 L$_{\odot}$ this relation
yields log [L$_{\rm x}$/L$_{bol}$] $\sim$ $-$3.7. 

We consider
the results of the above regression fit to be preliminary
since  it is  based only on the IR-bright (K $\leq$ 10.5)
class II sources in NGC 2024 identified by HLP
and  lacks  information on faint sources whose IR classification
is not yet known.
Even so, the regression fit is
consistent at the $\approx$1$\sigma$ level 
with results from other young clusters having more
complete optical/IR data such as IC 348 (Preibisch 
\& Zinnecker 2002), and is nearly identical to the 
average  $\langle$log[L$_{\rm x}$/L$_{bol}$]$\rangle$ =
$-$3.9 $\pm$ 0.7  found for a much larger sample of young
stars in the Orion Nebula Cluster (Feigelson et al. 
2002). Thus, based on this preliminary analysis, we find
no reason to believe that the L$_{\rm x}$/L$_{bol}$ ratios
of class II sources in NGC 2024 are substantially different 
from those in other young clusters.

Previous studies of star-forming regions have reported 
correlations  L$_{\rm x}$ $\propto$ L$_{\rm J}$ $\propto$  L$_{bol}$,
where L$_{\rm J}$ is the J-band luminosity (Casanova et al. 1995).
Since the soft X-ray absorption in the energy range kT $\approx$ 
1.1 - 1.4 keV is comparable to that at
J-band, it is possible to compare the observed (absorbed)
soft-band X-ray  and  J-band luminosities.
To compare with results for other star-forming regions,
we have plotted  the absorbed luminosity L$_{\rm x,abs}$ (1.1 - 1.4 keV) 
versus observed J magnitude from HLP for the known class
II sources in NGC 2024 (Fig. 16). Neither parameter has been
corrected for extinction. 
We have excluded the four highly-variable class II sources
noted above. There is a clear trend for brighter J-band
sources to have larger L$_{\rm x,abs}$
with the notable exception of one outlier, namely the 
bright IR source HLP-2 ({\em Chandra} no. 77).
Excluding the outlier and the four strongly variable sources, 
the Cox and Kendall's $\tau$
tests  give a probability of correlation $p$ = 0.99
between observed  J magnitude and L$_{\rm x,abs}$.
Using the Buckley-James algorithm and again excluding the 
outlier, a regression fit
gives log L$_{\rm x,abs}$[1.1 - 1.4 keV] (ergs s$^{-1}$) = 
$-$0.37 ($\pm$0.11)(J$-$12) $+$
28.27  with a standard deviation of 0.56 in  
log L$_{x,abs}$.
Considering the uncertainties, the above regression  slope 
$-$0.37 ($\pm$0.11) agrees  quite well with the results 
for other star-forming regions  (Casanova et al. 1995;
Getman et al. 2002). 

It is noteworthy that several studies have shown obvious
exceptions to the L$_{\rm x,abs}$ $\propto$ L$_{\rm J}$
relation. In NGC 2024, the IR-bright source
HLP-2 (J = 7.93, K = 6.56) is more than  an order of magnitude 
less luminous in L$_{\rm x,abs}$ than predicted by the
above regression fit. We have undertaken an analysis similar to
the above using K magnitudes  and observed X-ray fluxes in the harder
3.2 - 4.2 keV band, for which the absorptions are comparable.
Again, HLP-2 shows up as an outlier with a luminosity 
L$_{\rm x,abs}$(3.2 - 4.2 keV)  that is at least an order of 
magnitude below that expected for its  L$_{\rm K}$.
Exceptions to the L$_{\rm x,abs}$ $\propto$ L$_{\rm J}$ relation
were also seen in
NGC 1333 (Fig. 11 of Getman et al. 2002) and 
the $\rho$ Ophiuchi core (Fig. 4 of Casanova et al. 1995).
The exceptions that are bright at J-band but weak in X-rays
such as HLP-2 and the objects SR1 and SR3 in $\rho$ Ophiuchi
appear to have high bolometric luminosities and are thus more
massive stars or perhaps unresolved multiple systems. 
 
\subsubsection{X-ray Luminosities of Flat-Spectrum Sources}

Six flat-spectrum sources were identified in Table 1 of HLP,
of which three were detected by {\em Chandra}. These were
HLP-1  (source 187), HLP-12 (source 71), and HLP-38
(source 125). HLP-1 is the brightest K-band source in
the HLP catalog (K $\approx$ 5.5 - 5.8).  
The IR-bright flat-spectrum source HLP-6
was not detected but it fell in a gap between CCDs.
Although the sample is small, the 50\% detection rate of
these six flat-spectrum sources is lower than 
the 96\% detection rate for known class IIs.
The X-ray luminosities of the three  flat-spectrum
detections were similar and in the range
log L$_{\rm x}$ = 30.14 - 30.37 ergs s$^{-1}$ (Table 1).
Because of the small sample size, statistical analysis
of the flat-spectrum sources was not undertaken.

\subsubsection{Ionizing Source and Total Cluster X-ray Luminosity}

The mean absorbed X-ray flux of all 283 detections from Table 1 is 
$\langle$F$_{\rm x,abs}$ (0.5 - 7 keV)$\rangle$ = 4.6 $\times$ 10$^{-14}$ 
ergs cm$^{-2}$ s$^{-1}$. Assuming a 1T optically thin plasma model with
source properties typical of NGC 2024 (kT = 2.8 keV, N$_{\rm H}$ =
2.3 $\times$ 10$^{22}$ cm$^{-2}$), the corresponding mean unabsorbed
flux is $\langle$F$_{\rm x}$ (0.5 - 7 keV)$\rangle$ = 1.2 $\times$ 
10$^{-13}$ ergs cm$^{-2}$ s$^{-1}$ which equates to an unabsorbed
luminosity $\langle$L$_{\rm x}$ (0.5 - 7 keV)$\rangle$ = 2.5 $\times$
10$^{30}$ ergs s$^{-1}$. This average is about 1300 times larger
than the X-ray luminosity of the quiet Sun, underscoring the
prodigious X-ray output of pre-main-sequence stars.   
Summing over 283
sources gives a total luminosity L$_{\rm x,tot}$ (0.5 - 7 keV) $\sim$
7 $\times$ 10$^{32}$ ergs s$^{-1}$. 

The source of the ionizing radiation which excites NGC 2024
has been difficult to identify but previous studies based on energy
considerations have concluded that the exciting star is most
likely of spectral type  B0 - O9 (Hjellming 1968; Gordon 1969;
Thronson et al. 1984). Although the  O9.5Ib star $\zeta$ Ori may 
contribute to the ionization, it lies  $\sim$15$'$ away and is not 
believed to be the primary excitation source (Gordon 1969; B89).
The lack of an obvious ionizing source led Barnes et al. (1989)
to suggest that the combined contribution from the embedded IR population,
rather than a putative B0-O9 star, might be the source of the ionization.
It is thus relevant to ask if the total X-ray emission associated with the cloud
population might be an important contribution to the ionization. 
Our {\it Chandra} observation shows that the  known X-ray population
could contribute L$_{\rm x,tot}$ $\sim$ 10$^{33}$ ergs s$^{-1}$ toward the  
ionizing luminosity, assuming a distance of $\sim$415 pc. The actual X-ray 
contribution could be  higher if many
deeply embedded sources within the cloud have escaped detection
because of very high extinction. But, the above estimate of L$_{\rm x,tot}$
is much less than the Lyman continuum luminosity of a
a B0 - O9 star (Thompson 1984) and we conclude that the presently known
X-ray population cannot by itself account for the required 
ionizing radiation. 

Recently, it has been suggested by Bik et al. (2003) that 
the infrared source IRS 2b
is the long-sought ionizing source. As shown in Figure 2, this source lies
$\approx$5$''$ west of B89-IRS 2 at P.A. $\approx$ $-$76$^{\circ}$
(Jiang, Perrier, \& L\'{e}na 1984; Nisini et al. 1994).
Bik et al. derive a spectral type of O8V - B2V for IRS 2b.
IRS 2b was detected by Chandra (source 182) and its X-ray emission
may be variable (KS = 1.41, P$_{var}$ = 0.96). Due
to possible contamination from the nearby source 183
(see below), spectral fit results are somewhat uncertain.
The absorption column density determined from 1T {\scshape vapec} 
models gives A$_{\rm V}$ = 14.9 [12.6 - 18.9; 90\% conf.]
mag using the Gorenstein (1975) conversion, but 2T models
converge to an intrinsically softer spectrum with 
higher absorption (Table 1).

Interestingly, the Chandra image reveals two nearby
X-ray sources located $\approx$1.7$''$ - 2.9$''$ north
of IRS 2b (Fig. 2). These are the faint X-ray source
180 and the brighter source  183, the latter being hard and
variable. Interestingly, neither of
these two X-ray sources have known IR counterparts, but 
source 183  has been detected at 3.6 cm 
in recent {\em VLA}
observations (Rodriguez et al. 2003). The variability in
source 183 suggests that it is an embedded young star
and higher sensitive high-resolution searches for an IR counterpart 
are warranted.

\subsection{Interesting X-ray Sources}

Three {\it Chandra} sources showed exceptional
variability  that deserves further comment, namely
sources 94, 192 and 207. 

\subsubsection{Persistent X-ray Variability in the cTTS Haro 5-59 }

Source 94 is unusual in several respects. It was
categorized as a class II source (cTTS) by HLP and
is one of the few X-ray detections that has an optical
counterpart, being associated with the emission-line
star Haro 5-59 (V = 14.5). Its broad-band X-ray light curve
shows a slowly declining count rate for most of the
observation, with an abrupt increase near the end (Fig. 7).
Closer inspection reveals that the variability is much
more complex than suggested by Figure 7, as demonstrated
in the four-panel light
curves shown in Figure 8.

Remarkably,  the X-ray source is  in a 
state of nearly continuous variability throughout the
observation with significant variations  ($\geq$3$\sigma$)
present in all four parameters: observed flux,
$\langle$E$\rangle$,  L$_{\rm x}$, and kT. 
The observed flux and L$_{\rm x}$
generally undergo a slow decline
which is punctuated by abrupt increases near UT $\approx$
13 h and again about one hour prior to the end of
the observation. The increases in flux and L$_{x}$
are accompanied by increases in mean photon energy 
$\langle$E$\rangle$ and kT. 
These fluctuations suggest reheating of the flaring plasma
during the decay, but we cannot rule out the possibility
that the fluctuations are due to flaring of different plasma 
structures - perhaps within the same active region.

Figure 8 shows that the plasma temperature fluctuates
from a low of kT $\sim$ 2 keV up to kT $\sim$ 5 keV,
with the maximum value reached  near the
end of the observation. Because of these temperature
variations,  isothermal fits have difficulty in 
reproducing the time-averaged spectrum. More
sophisticated fits of the  time-averaged spectrum
using a DEM model ({\scshape c6pvmkl}) show 
a double-peaked structure with maxima near
$\sim$1 keV and $\sim$5 keV, confirming the 
multi-temperature nature of the plasma.

The persistent X-ray variability in  Haro 5-59 reflects 
strong surface magnetic activity and may be 
a consequence of nearly continuous low-level flaring
that produces the energy input needed to maintain 
the plasma at temperatures above kT $\sim$2 keV.
This behavior is, at least qualitatively, similar to that
recently detected in sensitive  {\it XMM-Newton} observations of 
the nearby magnetically-active star Proxima Centauri 
(G\"{u}del et al. 2003). The corona of Prox Cen displayed
a nearly uninterrupted sequence of low-level X-ray flares,
many of which were correlated with optical variability.
Evidence for continuued reheating during flares has
also been obtained in other active late-type stars
such as Algol (Schmitt \& Favata 1999).
The variability detected here in Haro 5-59 may reflect similar
processes occurring in  a much younger magnetically active
classical T Tauri star.

\subsubsection{Slow Variability }

Prominent variability that lasted several hours was detected
in sources 192 and 207. The variability is 
characterized by slow (non-impulsive) rise phases, as 
shown in Figure 7.  These two sources exhibited the
strongest X-ray variability of all detections as
measured by their respective values  KS = 27.4
and KS = 17.9 (Table 1). 
Source 192 is associated with the heavily reddened 2MASS
source 05414611-0154147 (J = 17.30, H = 14.48, K = 11.42).
Source 207 has a faint K = 14.8 IR counterpart (Meyer 1996).

We have analyzed the time behavior of these two sources using
light curves generated from time-binned event lists  
and by extracting time-partitioned spectra during the outbursts.
In order to mitigate pileup, the central 
pixel in the PSF core was excluded in the spectra extracted 
for source 192 near the end of the observation when its
count rate exceeded 0.1 c s$^{-1}$. The two analysis 
approaches give consistent results, as summarized below.

The distinguishing feature of  source 192 is its
high temperature. Both analysis techniques indicate that
the X-ray temperature reached very high values shortly after
the beginning of the outburst,  certainly in excess of  kT $\sim$ 6 keV and
quite possibly as high as  kT $\sim$ 10 keV. The temperature
remained at values above kT $\sim$ 4 - 5 keV during the
four hour rise phase monitored by {\it Chandra}, and the
observed X-ray flux continued to increase until the end of the
observation. 

A similar situation exists for source 207, as shown in
Figure 9. Remarkably, its temperature remained at 
levels above  kT $\sim$ 5  keV for most of the 
observation and may have risen to values in excess
of kT $\sim$ 10 keV during  events such as
that near UT $\approx$ 23 h. Spectral analysis 
confirms maximum temperatures of at least 
kT $\sim$ 6  keV. 

In summary,  sources 192 and 207 exhibit 
strong absorption (log N$_{\rm H}$ $\sim$ 22.6
cm$^{-2}$, or A$_V$ $\sim$ 18 mag) that is 
accompanied by spectacular variability and 
very high  temperatures (kT $\geq$ 5 - 6 keV). 
Increases in $\langle$E$\rangle$ (hardening
of the spectrum) and  temperature increases
appear to occur many hours after the beginning
of the outburst. We have noticed similar behavior
in {\em Chandra} observations of other young
stars, such as the wTTS  DoAr 21 in
$\rho$ Ophiuchus (Gagn\'{e} et al. 2003).
These results suggest that continued energy
input  may prolong the duration of
some large outbursts. 

\section{Comparison of X-ray and IR Properties}

We have used JHK magnitudes from  2MASS  and other IR
surveys  to make comparisons between
the X-ray and IR properties of sources in NGC 2024.
Figure 17 shows the distribution of K magnitudes
for those {\it Chandra} detections with known IR counterparts.
The X-ray detections show a strong peak near
K $\approx$ 11 - 12.
There is a strong tendency to detect the brightest
K-band sources in X-rays and there is a rapid falloff
in the number of X-ray
detections for K $>$ 12. Only 20\% of the catalogued
IR sources in the ACIS-I FoV with K $>$ 12 were 
detected in X-rays, while 74\% of the IR sources with
K $<$ 12 were detected. 

The J-K versus K color magnitude diagram in Figure 18
shows that {\it Chandra} detections span a broad range 
of J-K colors, but no sources fainter than K = 16.7 were 
detected. The faintest IR source detected 
was source 108 in Meyer (1996) with K = 16.74, which
corresponds to {\it Chandra} source 129.
Figure 18 also reveals some notable 
non-detections. The class I source HLP-29 (K = 9.1)
was not detected by {\it Chandra}, nor was the 
flat-spectrum source  HLP-6 (K = 7.6). But,
the latter object lies in a gap between CCDs where
sensitivity was reduced.

The J-H versus H-K color-color diagram in Figure 19
allows us to discriminate between normally reddened
stars and stars with true infrared excesses (as 
due, for example, to circumstellar disks). The
left and right dashed lines in Figure 19 show the
normal reddening loci of M0V and A0V stars assuming
intrinsic J-H and H-K colors from Bessell \& Brett (1988)
and the reddening law from Rieke \& Lebofsky (1985). 
A significant number of the {\it Chandra} detections lie
between the two dashed lines and are thus likely to be
normally reddened stars. More interesting are the 
sources lying to the right of the normal reddening 
band. These objects
have true IR excesses and several lack IR classifications
but were detected by {\it Chandra}. Thus, they are
good candidates for new embedded cluster members.

\section{Candidate Protostars}

The NGC 2024 region contains the class I infrared source HLP-29
and seven compact millimeter sources FIR 1-7 which may be 
protostellar in nature (M88; M92). These objects represent early phases of
stellar evolution about which little is known at X-ray wavelengths. We have
thus carefully inspected the broad-band 
astrometric-corrected {\it Chandra} images 
for any evidence of X-ray emission from these candidate protostars. 

\subsection{The Class I Source HLP-29}

No significant X-ray emission was detected from the class I source
HLP-29 within a circular region of radius 3$''$ centered on the
J2000 position given by HLP. Only one photon was
detected within the 95\% encircled energy region centered at the
IR position, which is consistent with the expected number of
background events for a 76 ks exposure. Since the extinction
toward HLP-29 is not well-known, any conversion of the 
upper limit on its {\it Chandra} count rate to an equivalent
upper limit on intrinsic L$_{\rm x}$ would be highly uncertain.

\subsection{The Millimeter Sources FIR 1-7}

We searched for X-ray emission within 3$''$ of the millimeter
positions of FIR 1-7, including the (precessed) positions
of FIR 1-6 from M88, FIR-7 from M92,
and FIR 2-7 from the OVRO observations of
Chandler \& Carlstrom (1996). Because of the high absorption
toward these dust condensations (M92), any X-ray photons
that escape are likely to be hard. Thus, we used images
over a slightly  broader energy range of   0.5 - 8 keV
(rather than 0.5 - 7 keV) to search for emission in the
vicinity of FIR 1-7. Even so, no significant X-ray detections
were found. However, we do find  weak
emission near FIR-4, as shown in Figure 4. Four photons
were detected within 1.2$''$ of the FIR-4 position given by 
M88, and three of these were localized in two adjacent pixels. 
The emission is hard and the mean energy of the four
photons is $\langle$E$\rangle$ = 5.9 keV. Such emission 
would most likely be due to a flaring source deeply embedded
within the cloud or an extragalactic  source seen through  
the cloud. However, contamination from extragalactic sources
along this line-of-sight is expected to be low due to the very
high column density toward the  FIR-4 dust condensation (M92).
No more than one of the four photons is attributable to quiescent
background in a 76 ks exposure (Table 6.6 of {\em POG}).  Furthermore,
the probability of detecting two background photons in the
same pixel is no larger than  $p$ $\approx$ 2.6 $\times$ 10$^{-4}$,
assuming a maximum quiescent background rate in the 0.5 - 8 keV range 
of $\approx$0.22 counts s$^{-1}$ per ACIS-I CCD. The
photon arrivals are well separated in time so the weak 
emission is not due to a cosmic ray afterglow event.
The proximity of the hard X-ray emission to the FIR-4 millimeter
peak and the unusual nature of this dust condensation 
warrant further comment.

FIR-4 is exceptional in that it is the only one of the seven
condensations that has a near-IR source at the position of
the millimeter peak. A faint K-band source was discovered by
Moore \& Chandler (1989) and our inspection of 2MASS images
confirms this K-band source at an offset of only 0.5$''$ from the 
FIR-4 mm peak position of M88 (Fig. 4). 
The IR source was detected at H, K, and L bands by
Moore \& Yamashita (1995 = MY95). It is heavily reddened 
(H - K = 3.7) and there is a bright reflection nebula 
extending $\approx$10$''$ to its northwest.
MY95 suggested that the reflection nebula is due
to scattering in a region channeled out by a molecular outflow,
and noted that the IR source might even be a localized peak
in the surface brightness of scattered light rather than
the actual illuminating source. 
The redshifted lobe of an outflow has been detected with the
Owens Valley millimeter array by Chandler \& Carlstrom (1996).

Even though the IR source is nearly coincident with the FIR-4
millimeter peak, the physical relationship between the two is
not yet clear. It was argued by M92 that the IR source does
not lie at the center of the FIR-4 dust condensation, since it
would then be obscured by an extinction A$_{V}$ $\sim$
1000 mag and be undetectable in the near-IR. The extinction 
of the IR source was estimated to be A$_V$ $\leq$  50 mag
by MY95.  Similarly, any physical connection between the
weak X-ray emission and either FIR-4 or the reddened IR
source is not yet established since the X-ray peak is 
displaced $\approx$1$''$ to the north (Fig. 4).

\section{Summary}

The embedded infrared cluster in NGC 2024 provides an interesting
contrast to other well-studied star-forming regions in Orion
such as the Orion Nebula Cluster (ONC). NGC 2024 lies at approximately
the same distance as the ONC but has a lower
stellar density  and is more heavily obscured, with
a dearth of optically visible stars. Although star-formation appears
to have already ceased in the ONC (Feigelson et al. 2002), active 
star-formation may still be underway in NGC 2024 and 
many of the cluster members are still in the class II accretion
phase.

The catalog of 283 {\it Chandra}
sources is large enough for statistical studies and provides
new information on the distribution of young stars within the 
cloud, absorption, X-ray spectral and variability properties, 
and energetics of the
X-ray emitting population. 
The most important results to emerge from this study are the following:

\begin{enumerate}

\item {\it Chandra}  detected 283 sources in NGC2024 within a $\approx$17$'$ 
      $\times$ 17$'$ ACIS-I region centered near the millimeter
      source FIR-5, a five-fold increase over previous X-ray 
      surveys. A total of 248 X-ray sources (88\%)
      were identified with  counterparts,  mostly  
      infrared sources.

\item Roughly half of the 35 sources without counterparts 
      are expected to be extragalactic based on source counts
      from {\em Chandra} Deep Field surveys. About a dozen
      of the unidentified sources have positions within the 
      projected boundary of the known IR cluster and some of these 
      could be embedded young stars.
      We have provided accurate
      X-ray positions based on registration of {\it Chandra} images
      with the 2MASS data base (Table 1) that will facilitate 
      future searches for counterparts to these X-ray sources in
      the infrared and other spectral regions.

\item The X-ray sources in NGC 2024 are in general characterized by
      hard heavily absorbed emission with strong variability. By fitting
      the spectra of more than 100 of the brightest  detections,
      we have derived a mean X-ray energy 
      $\langle$kT$\rangle$ = 2.8 keV and mean extinction 
      $\langle$A$_V$$\rangle$ = 10.5 mag. This latter value is in
      excellent agreement with extinction estimates from IR studies.
      However, considerable variation about these mean values was
      observed, with kT  $\sim$6 - 10 keV  derived
      for some strongly-variable sources and extinctions as large as 
      A$_V$ $\sim$ 68 mag for deeply embedded sources.

\item Variability was detected at high confidence levels in 
      25\% of the X-ray
      sources, but we have argued that the true variability fraction
      is likely to be larger because of the difficulty in detecting 
      variability in weak sources. We find no compelling evidence
      that variability and spectral hardness are correlated. Of
      particular interest are  sources
      which undergo poorly-understood slow variability lasting 
      many hours during which plasma temperatures are maintained
      at high values  kT $\geq$ 5 - 6 keV 
      such as sources 192 and 207 (Fig. 7).
       Such behavior is decidedly different from solar-like
      impulsive bursts (e.g. sources 27, 81 and 174) that signal magnetic reconnection
      and rapid energy release.  Our time-resolved spectroscopic analysis
      suggests that the plasma in such slowly varying sources is being reheated. However,
      we cannot spatially resolve individual flaring regions so it
      is possible that multiple unresolved flares are taking place,
      perhaps in the same active region.
      We also find clear evidence for persistent 
      low-level variability in the star Haro 5-59, which is 
      probably the consequence of nearly continuous low-level flaring
      in this magnetically active cTTS.

\item  A comparison with IR data shows that the brightest
       K-band sources are preferentially detected in X-rays.
       {\it Chandra} detected 74\% of the known IR sources 
       with K $<$ 12 but only 20\% of those with K $>$ 12.
       Near-infrared color-color diagrams reveal that many {\it Chandra}
       detections are heavily reddened, including extreme 
       objects such as the hard variable X-ray source 192
       (= 2MASS J05414611-0154147; J-H = 2.82, H-K = 3.06; Fig. 19).
       These heavily reddened sources could be young cluster members
       and follow-up observations in the mid-IR would be useful
       to determine if disks are still present.

\item  The detection rate of 27 known class II sources (cTTS) in 
       NGC 2024 is at least 96\%, and may be as high as 100\%. The
       X-ray and bolometric luminosities of these class II
       sources are correlated and we also find a correlation
       between the observed  luminosities in the 
       equally absorbed J band and
       1.1 - 1.4 keV X-ray band. A notable exception is 
       the IR-bright cTTS HLP-2, whose X-ray emission is at
       least an order of magnitude lower than expected from
       regression fits.

\item       The ionizing source of NGC 2024 has been elusive, but
       previous studies have determined that
       the required ionization flux is equivalent to that of a 
       B0-O9 star. We estimate the total X-ray luminosity from
       all 283 X-ray sources in the {\it Chandra} FoV to be 
       L$_{\rm x,tot}$ (0.5 - 7 keV) $\sim$ 10$^{33}$ ergs s$^{-1}$ at
       an assumed distance of $\sim$415 pc. This is
       significantly less than the Lyman continuum luminosity of a
       B0-O9 star.  Thus,  the X-ray contribution
       from the {\it known} population cannot account for all of the missing 
       ionizing flux. Recent work by Bik et al. (2003) suggests
       that the luminous IR source IRS 2b is a young O8V - B2V star 
       and may be the long-sought
       ionizing source. IRS 2b is detected by {\em Chandra} and
       its X-ray emission is likely variable, possibly a sign of
       magnetic activity in this embedded high-mass star.

\item  Close examination of the {\it Chandra} images shows no 
       significant X-ray detections ($\geq$6 counts) of any of the seven 
       millimeter-bright condensations FIR 1-7 in NGC 2024.
       These condensations, which may be protostellar, are thus not
       yet sufficiently evolved to produce X-rays or any intrinsic
       X-ray emission has been completely absorbed by intervening
       material. A potentially interesting exception is FIR-4, for
       which we detect  faint hard localized emission (4 counts)
       within 1.2$''$ of the mm position. Previous observations
       have detected a heavily-reddened near-IR source within 0.5$''$ of 
       FIR-4 as well as a CO outflow, indicating that  stars have
       recently formed and may still be forming in this region. 
       However, the physical relationship between the FIR-4 dust
       condensation, near-IR source, and faint X-ray emission is not yet clear 
       because of the small positional offsets between the millimeter, IR, and 
       X-ray peaks (Fig. 4).

\end{enumerate}


\acknowledgments

This work was supported by SAO grant GO1-2009A.
This research has made use of the SIMBAD
astronomical database, the Astrophysics Data
System (ADS), and the ASURV statistical analysis
package maintained by Penn State. We have utilized
data products from the Two Micron All Sky Survey (2MASS), which is
a joint project of the Univ. of Massachusetts and 
IPAC/CalTech, funded by NASA and NSF. We have also
utilized data and software from the 
High Energy Astrophysics Science Archive 
Research Center (HEASARC), provided by
NASA's Goddard Space Flight Center. We thank 
M. Meyer for providing JHK photometry in electronic
format from his Ph.D. thesis and L. Rodriguez for
providing a table of VLA radio sources in NGC 2024
prior to publication.

\clearpage

\figcaption{Smoothed broad-band (0.5 - 7 keV) {\it Chandra} ACIS-I 
image of the central 
$\approx$4$'$ $\times$ 4$'$ region of NGC 2024 where X-ray source
density is highest. Smoothing was done with the $wavelet$ smoothing
algorithm in the HEASARC XIMAGE software package (Rosati, Burg, \&
Giacconi 1994).
Source numbers corresponding to the first column
of Table 1 are shown for a few of the brightest sources.
Crosses mark the  peak positions of the compact millimeter cores FIR 1-6
from M88 (FIR 1-6) and M92 (FIR 7).
Coordinates are J2000.
\label{fig1}
}
\figcaption{Unsmoothed broad-band {\it Chandra} ACIS-I 
image of the region near IRS 2b (source 182), which has been
proposed as the ionizing source of NGC 2024 (Bik et al. 2003).
Source 183 lies 1.7$''$ north of IRS 2b and has no known IR
counterpart but has been detected with the {\em VLA}
(Rodriguez et al. 2003). Source 180 has no known counterpart.
Source 187 (B89-IRS 2) is a bright near-IR source with K = 4.6 - 5.8.
The pixel size is 0.492$''$ and coordinates are J2000.
\label{fig2}
}
\figcaption{Smoothed broad-band  {\it Chandra} ACIS-I image of
the region in the vicinity of the millimeter cores FIR 1-7.
The smoothing algorithm is the same as in Figure 1.
Crosses mark the positions of cores FIR 1-6 (M88) and 
FIR 7 (M92). Open circles mark the positions of 
FIR 2-7 from the OVRO observations of Chandler \&
Carlstrom (1996).  None of
the cores were detected in X-rays with the possible 
exception of weak emission naear FIR-4 (see Fig. 4). 
Two {\it Chandra} sources (nos. 122
and 139) in the vicinity of FIR 1 and FIR 2 are identified.
Coordinates are J2000.
\label{fig3}
}

\figcaption{Unsmoothed  {\it Chandra} ACIS-I image (0.5 - 8 keV) 
showing 4 X-ray photons within 1.2$''$ of the mm source FIR 4.
Pixel size is 0.492$''$. Dark pixels contain two photons and
light pixels contain one photon. The cross ($+$) marks the position of
the mm/sub-mm source FIR-4 from the IRTF observations of M88, and
a dashed circle of radius 1$''$ centered on FIR-4 is shown for scale.   
The small open circle marks the position of the 3 mm continuum
peak of FIR-4 determined with the Owens Valley millimeter 
array (Chandler \& Carlstrom 1996). An asterisk is at the  
position of the K-band source 2MASS 05414407-0154448 first 
identified by Moore \& Chandler (1989). The three X-ray photons
at the top of the dashed circle have energies
2.49, 7.39, and 7.77 keV while the photon just below 
FIR 4 has energy 5.98 keV. Coordinates are J2000.
\label{fig4}
}

\figcaption{Positions of all 283 {\it Chandra} detections
in the ACIS-I field of view (filled circles), overlaid
with positions of known IR sources (open circles).
The IR source positions were taken from the all-sky
release of the 2MASS Point Source Catalog and 
other surveys (B89, B03, HLL, HLP, M96). 
Coordinates are J2000.
\label{fig5}
}

\figcaption{Positions of the 35 {\it Chandra} detections
in the ACIS-I field of view  without known counterparts
at IR, optical, or radio wavelengths. Six circled sources have faint hard
non-variable X-ray emission and may be extragalactic (AGNs). The dotted line
marks the approximate boundary of the embedded IR cluster.
Coordinates are J2000.
\label{fig6}
}

\figcaption{Broad-band {\it Chandra} light curves of
ten sources showing interesting variability. Source
numbers correspond to the first column of Table 1.
The count rate of  source 192 may be slightly underestimated
toward the end of the observation due to
moderate pileup at count rates above 0.1 c s$^{-1}$.
\label{fig7}
}

\figcaption{{\it Chandra} light curves of variable source
no. 94 (= J054138.9-015936), whose optical counterpart is
the emission-line star Haro 5-59. The panels show, from top
to bottom, the absorbed flux f$_{\rm x}$
(0.5 - 7 keV), mean photon energy $\langle$E$\rangle$, 
unabsorbed luminosity L$_{\rm x}$ (0.5 - 7 keV), and
plasma temperature kT. Error bars are 1$\sigma$.
Note that the source is in a state
of continuous variability during the $\sim$21 hour observation,
and the plasma may have been reheated at  $\approx$13 h 
and again at $\approx$27 h. 
\label{fig8}
}

\figcaption{{\it Chandra} light curves of the hard variable
source no. 207 (= J054147.4-015526) showing the same
quantities as in Fig. 8. Error bars are 1$\sigma$.
\label{fig9}
}

\figcaption{{\it Chandra} ACIS-I spectra of source 85.
{\it Top panel}: Data points with 1$\sigma$ error bars show the 
background-subtracted flare spectrum, binned to a minimum of
15 counts per bin. Dotted line is the background-subtracted 
pre-flare spectrum binned
to a minimum of 10 counts per bin (error bars omitted for clarity).
Solid line is a best-fit 1T {\scshape vapec} model of the flare spectrum
using solar abundances except for iron, which was held fixed
at Fe = 0.3 solar. The best-fit model of the flare spectrum
converges to an absorption N$_{\rm H}$ = 1.8 [1.7 - 2.0] $\times$
10$^{22}$ cm$^{-2}$ and  kT = 3.8 [3.3 - 4.3] keV with 
$\chi^2$/dof = 85.8/104,  where
brackets enclose 90\% confidence intervals. The pre-flare
temperature is  kT $\approx$ 3.0 keV and the observed flux
increased by about a factor of two during the flare.
{\it Bottom panel}: Residuals of the best-fit 1T {\scshape vapec} flare model,
in the sense of data $-$ model.
\label{fig10}
}

\figcaption{Histogram showing the fraction of variable 
sources (KS $>$ 1.7) as a function of the number of 
source counts. The apparent monotonic increase in
variability fraction with source counts is very likely a
selection effect caused by the inherent difficulties in detecting
variability in weak sources.
\label{fig11}
}

\figcaption{Histogram showing number of X-ray sources as
a function of mean photon energy $\langle$E$\rangle$.
The distribution is truncated at low energies below
$\approx$1.4 keV due to the absorption of soft photons
by intervening cloud material. The mean for all detections
is $\langle$E$\rangle$ = 2.7 keV.
\label{fig12}
}

\figcaption{Distribution of mean photon energy $\langle$E$\rangle$
versus the KS variability statistic for all 283 X-ray detections in
NGC 2024. Variable sources have KS $>$ 1.7 and lie above the solid 
line. The vertical dashed line is the average value 
$\langle$E$\rangle$ = 2.7 keV for the entire sample.
Most {\it Chandra} detections without counterparts
(open squares) are non-variable and have 
above-average $\langle$E$\rangle$.
All four quadrants are well-populated and there is no
obvious correlation between KS and $\langle$E$\rangle$.
\label{fig13}
}

\figcaption{Absorption column density N$_{\rm H}$ versus time-averaged 
plasma temperature kT as determined from 1T {\scshape vapec} fits of the
spectra of the brightest {\it Chandra} detections. For clarity,  
error bars (90\% confidence) are shown only for non-variable sources.
Circled sources showed strong variability in their light curves.
Sources with the highest plasma temperatures and largest absorptions
are all variable. But, the converse is not true since
many sources with low temperature and absorption also varied.
Thus, variability alone is not a reliable  indicator of temperature
or absorption.
\label{fig14}
}

\figcaption{Kaplan-Meier estimator showing the cumulative
X-ray luminosity distribution for known class II sources (classical
T Tauri stars) in NGC 2024. The hard tail above log L$_{\rm x}$ 
$\approx$ 30.7 ergs s$^{-1}$ is due to the contribution of
four class II sources that experienced rapid-onset flares or other 
strong variability (sources 41, 94, 133, 174).
\label{fig15}
}

\figcaption{Absorbed  X-ray luminosity (1.1 - 1.4 keV) versus observed 
J magnitude for known class II sources in NGC 2024. No extinction
corrections have been applied. In this energy range the X-ray
and J-band extinctions are approximately equal. The source
shown as an X-ray upper limit is HLP-100, which was
possibly detected (see text). Four class II sources that
experienced  flares or other strong variability are omitted in 
this figure (sources 41, 94, 133, 174). 
\label{fig16}
}

\figcaption{Histogram showing the number of {\it Chandra}
detections versus K magnitude for those X-ray sources
with known IR counterparts. The dotted line shows the
corresponding histogram for all 912 known IR sources in
the  {\it Chandra} field-of-view.
\label{fig17}
}

\figcaption{Color-magnitude diagram for IR sources within
the  {\it Chandra} ACIS-I field-of-view having J and K photometry.
IR sources detected by {\it Chandra} are shown with filled 
symbols and undetected IR sources with open symbols.
Undetected IR sources in the field which lack classifications
based on  IR spectral index are not shown.
The class I source HLP-29 (circled cross) was not detected.
The undetected
flat-spectrum source HLP-6 (K = 7.6) lies in a gap between
CCDs where sensitivity is reduced. The faintest K-band source
detected by Chandra is source  129, whose IR counterpart is 
source 108 of Meyer 1996 (M-108; K = 16.74). {\it Chandra} source
no. 160 (= M-165) is not visible in this plot and has
K = 12.31, J-K = 8.59.
\label{fig18}
}

\figcaption{Color-color diagram for IR sources
within the ACIS-I field-of-view detected by
{\it Chandra}. Also shown are the colors of
the undetected class I source HLP-29 (circled 
cross) and the  undetected class II/flat-spectrum
sources listed in HLP.
{\it Chandra} detections are solid symbols and 
undetected IR sources are open symbols. 
Undetected IR sources in the field which lack
classifications based on spectral index are not shown.
The dotted line at lower left is the main sequence.
The sloping dashed lines mark the approximate
reddening band for normally reddened M0V
(left) and A0V (right) stars, based on data
from Bessel \& Brett (1988) and Rieke \& 
Lebofsky (1985). The slopes of the dashed lines
match that of the A$_{V}$ = 10.5 mag reddening
vector, which corresponds to the mean extinction 
toward NGC 2024. Sources to the right of the
band have excess reddening. The most heavily reddened
{\it Chandra} detections are identified.
\label{fig19}
}


\clearpage

\begin{figure}
\figurenum{1}
\epsscale{0.9}
\plotone{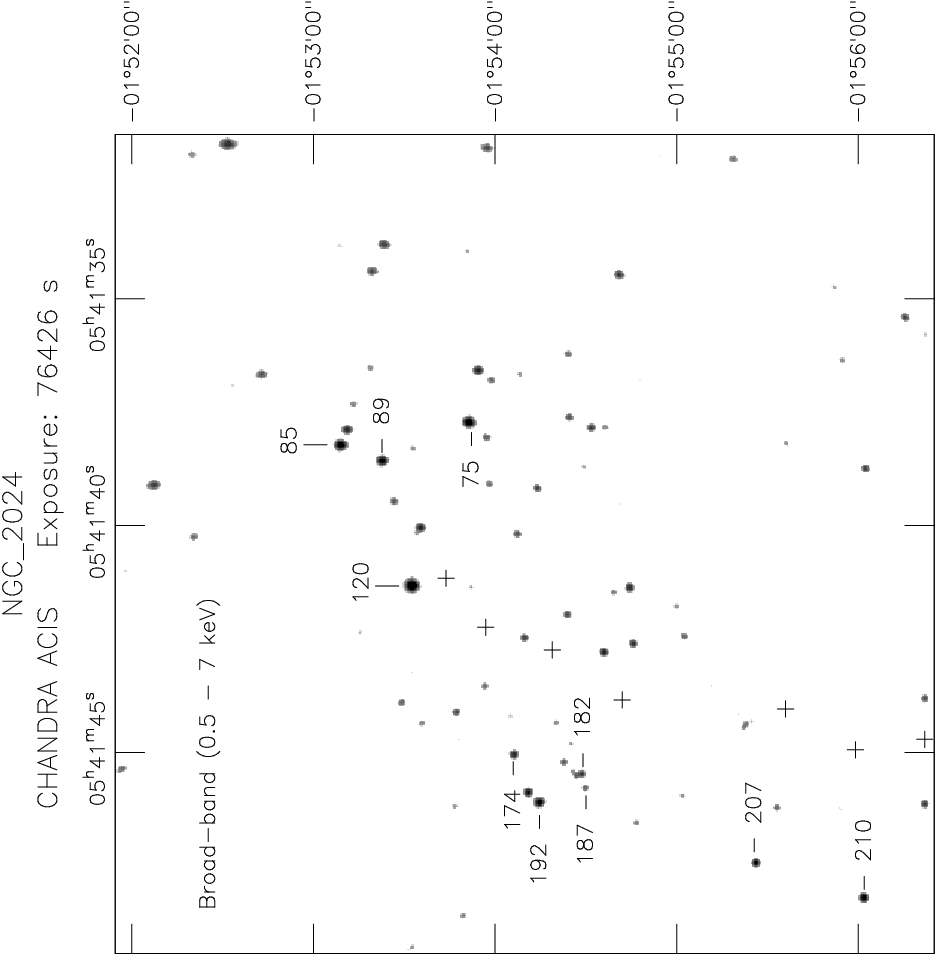}
\caption{}

\end{figure}
 
\clearpage

\begin{figure}
\figurenum{2}
\epsscale{0.9}
\plotone{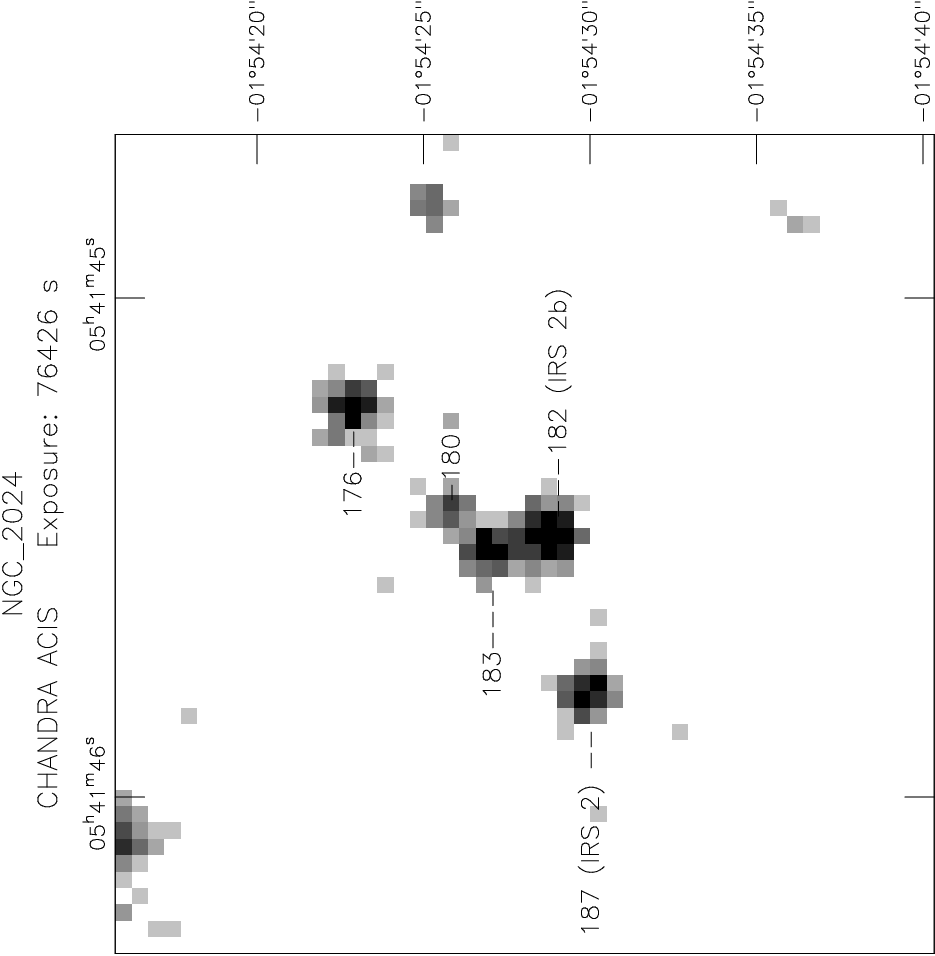}
\caption{}
\end{figure}

\clearpage

\begin{figure}
\figurenum{3}
\epsscale{0.5}
\plotone{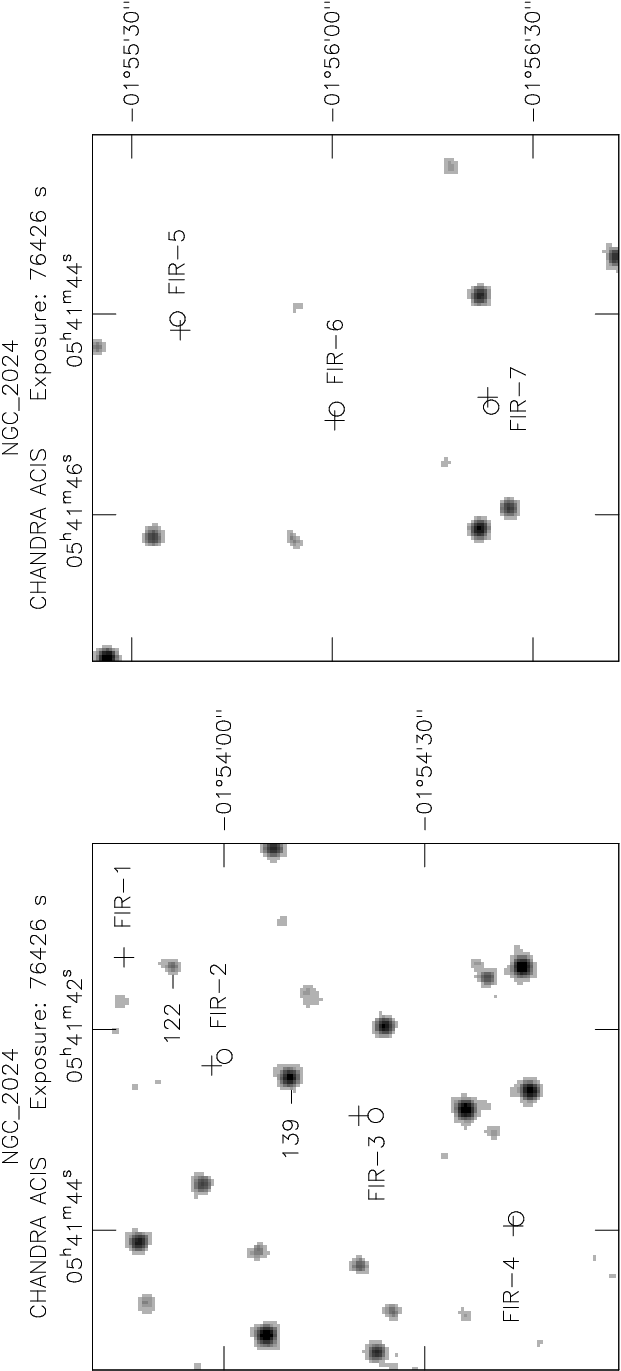}
\caption{}
\end{figure}

\clearpage

\begin{figure}
\figurenum{4}
\epsscale{0.7}
\plotone{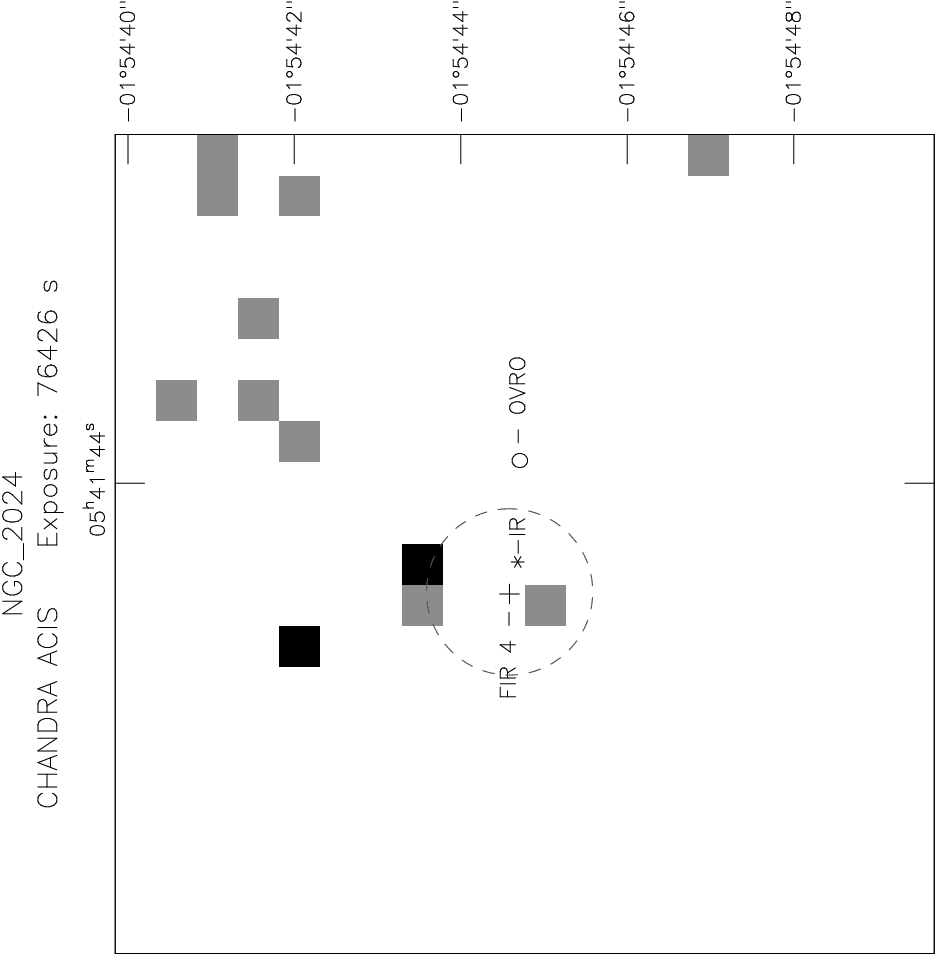}
\caption{}
\end{figure}

\clearpage

\begin{figure}
\figurenum{5}
\epsscale{0.9}
\plotone{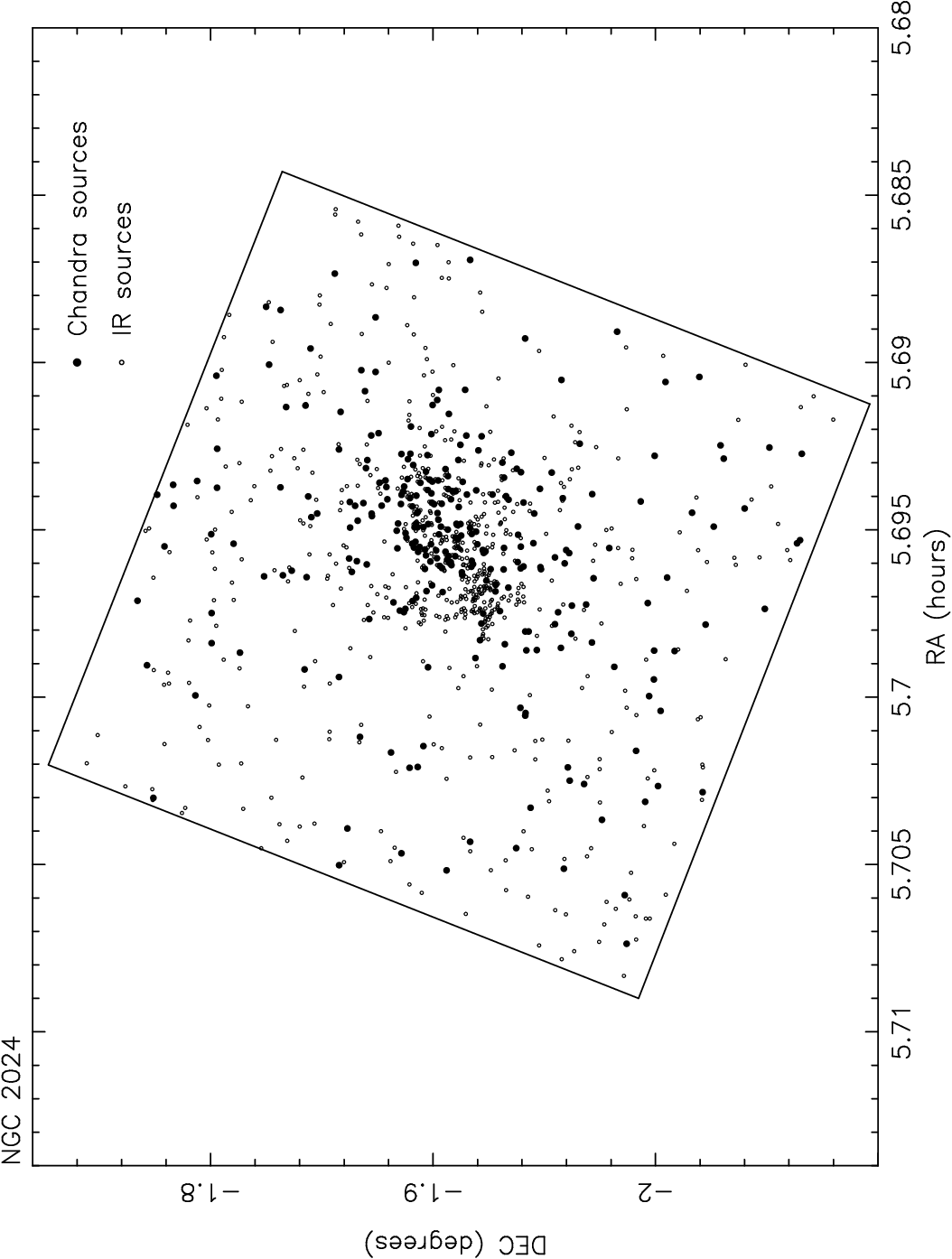}
\caption{}
\end{figure}

\clearpage

\begin{figure}
\figurenum{6}
\epsscale{0.9}
\plotone{f6.eps}
\caption{}
\end{figure}

\clearpage

\begin{figure}
\figurenum{7}
\epsscale{0.9}
\plotone{f7.eps}
\caption{}
\end{figure}

\clearpage

\begin{figure}
\figurenum{8}
\epsscale{0.7}
\plotone{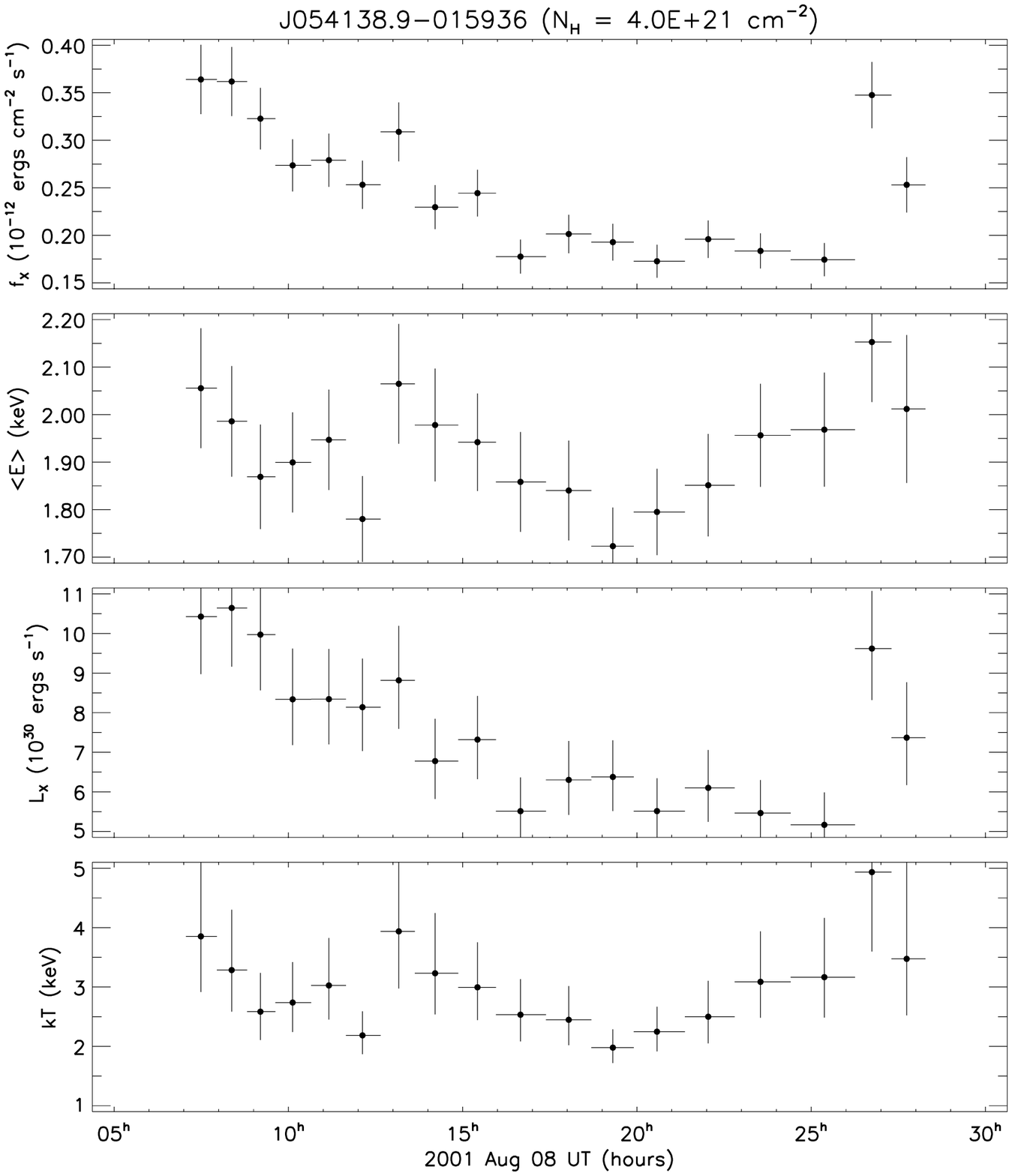}
\caption{}
\end{figure}

\clearpage

\begin{figure}
\figurenum{9}
\epsscale{0.9}
\plotone{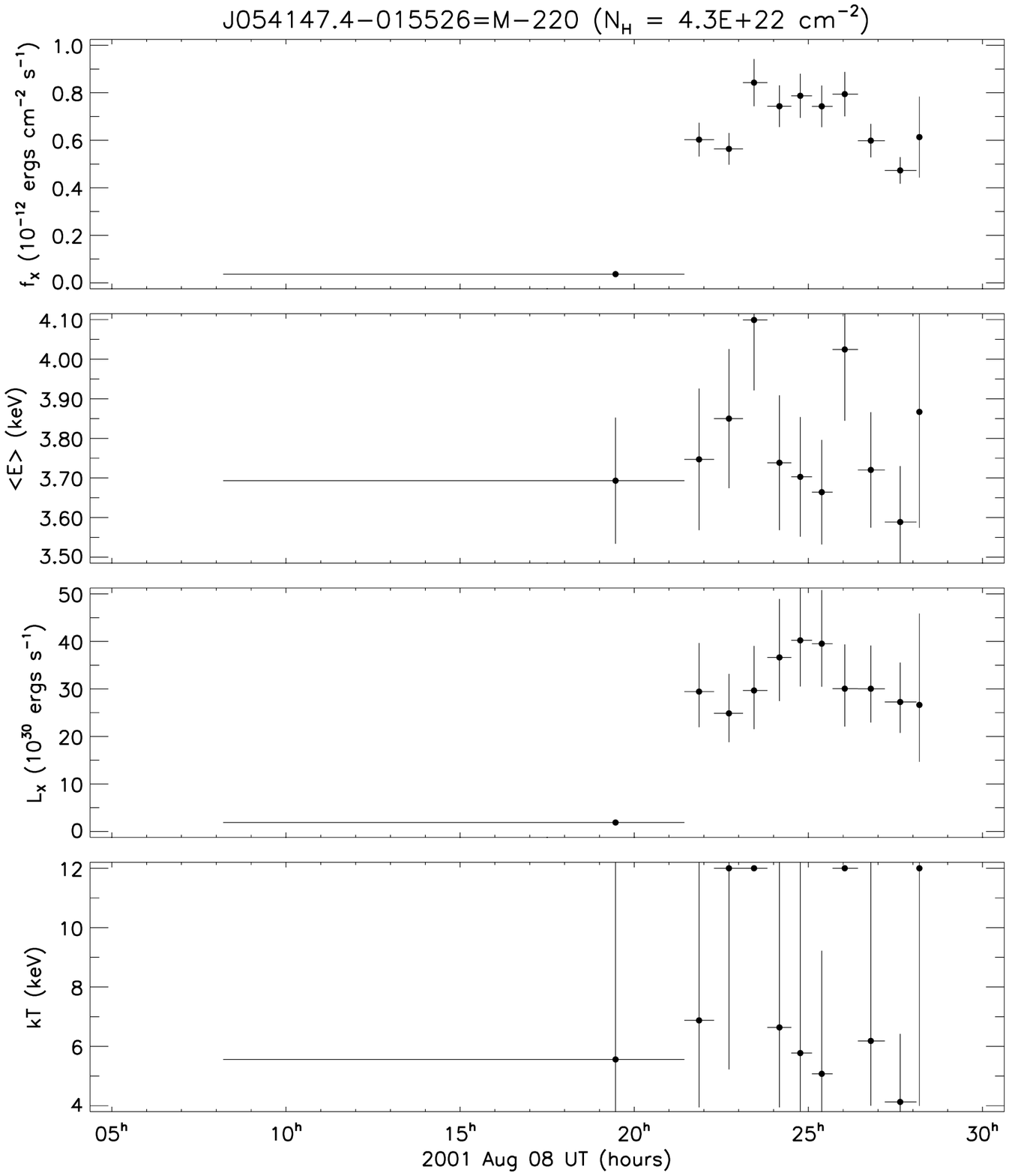}
\caption{}
\end{figure}

\clearpage

\begin{figure}
\figurenum{10}
\epsscale{0.9}
\plotone{f10.eps}
\caption{}
\end{figure}

\clearpage

\begin{figure}
\figurenum{11}
\epsscale{0.9}
\plotone{f11.eps}
\caption{}
\end{figure}

\clearpage

\begin{figure}
\figurenum{12}
\epsscale{0.9}
\plotone{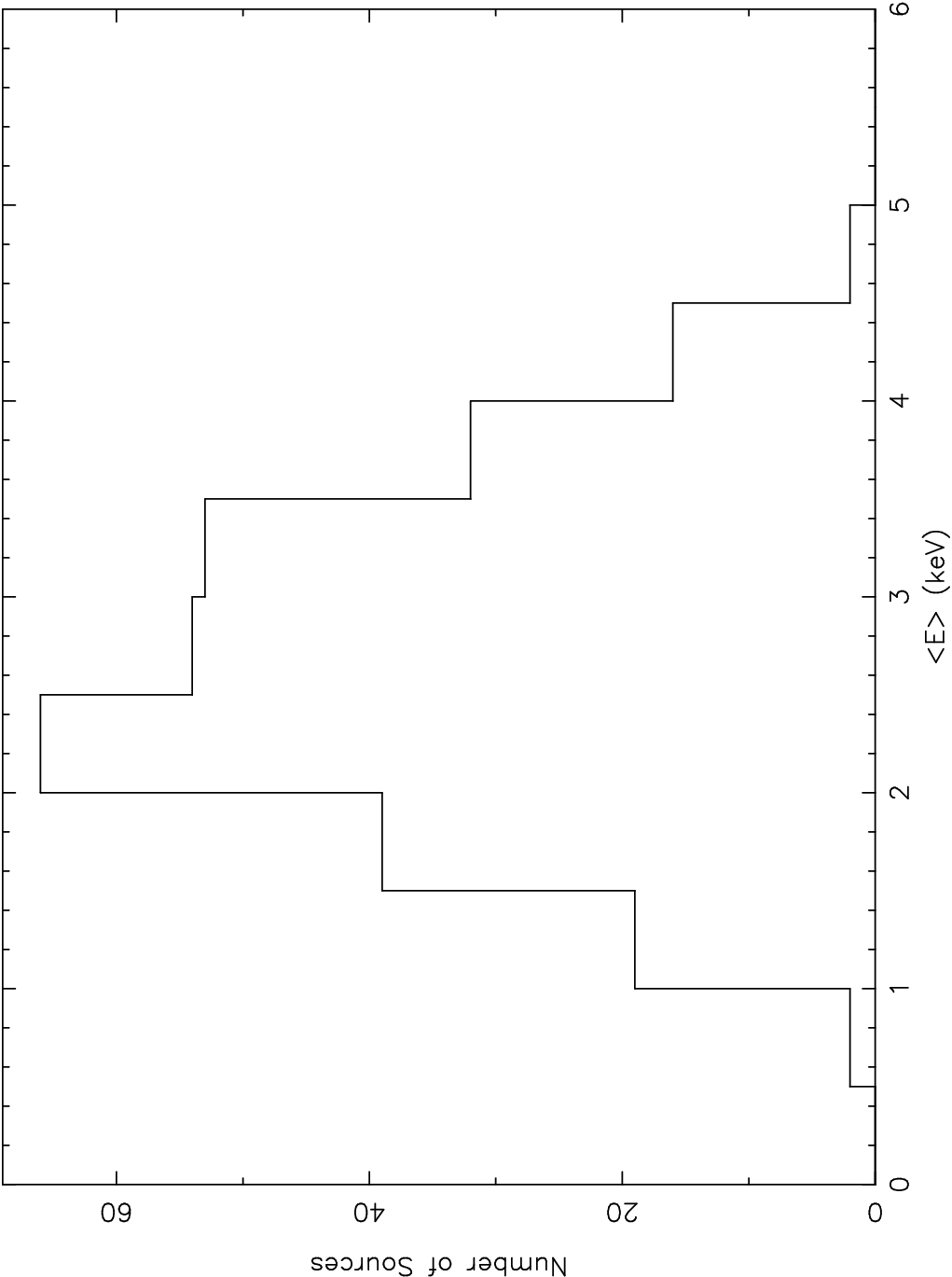}
\caption{}
\end{figure}

\clearpage

\begin{figure}
\figurenum{13}
\epsscale{0.9}
\plotone{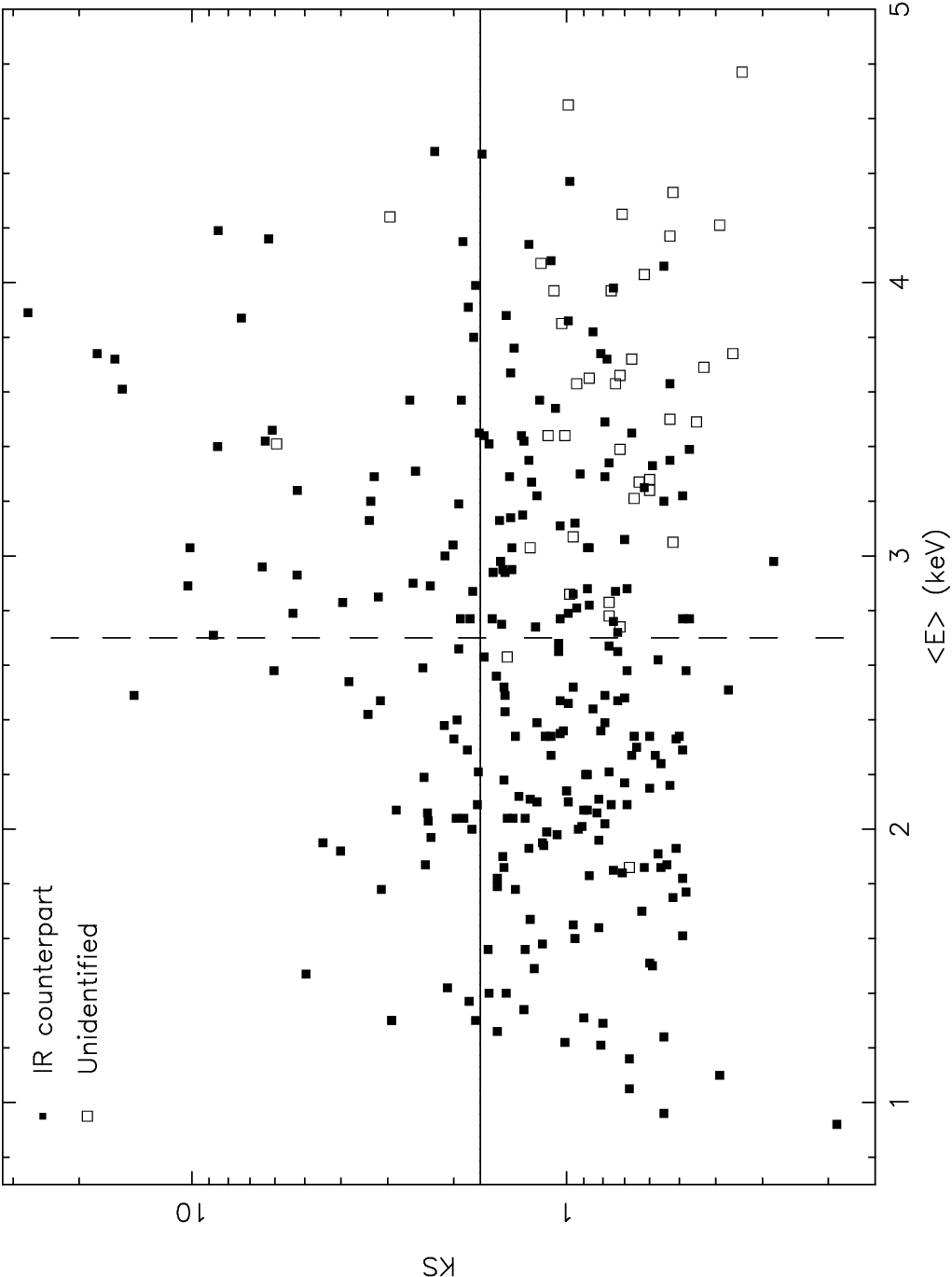}
\caption{}
\end{figure}

\clearpage

\begin{figure}
\figurenum{14}
\epsscale{0.9}
\plotone{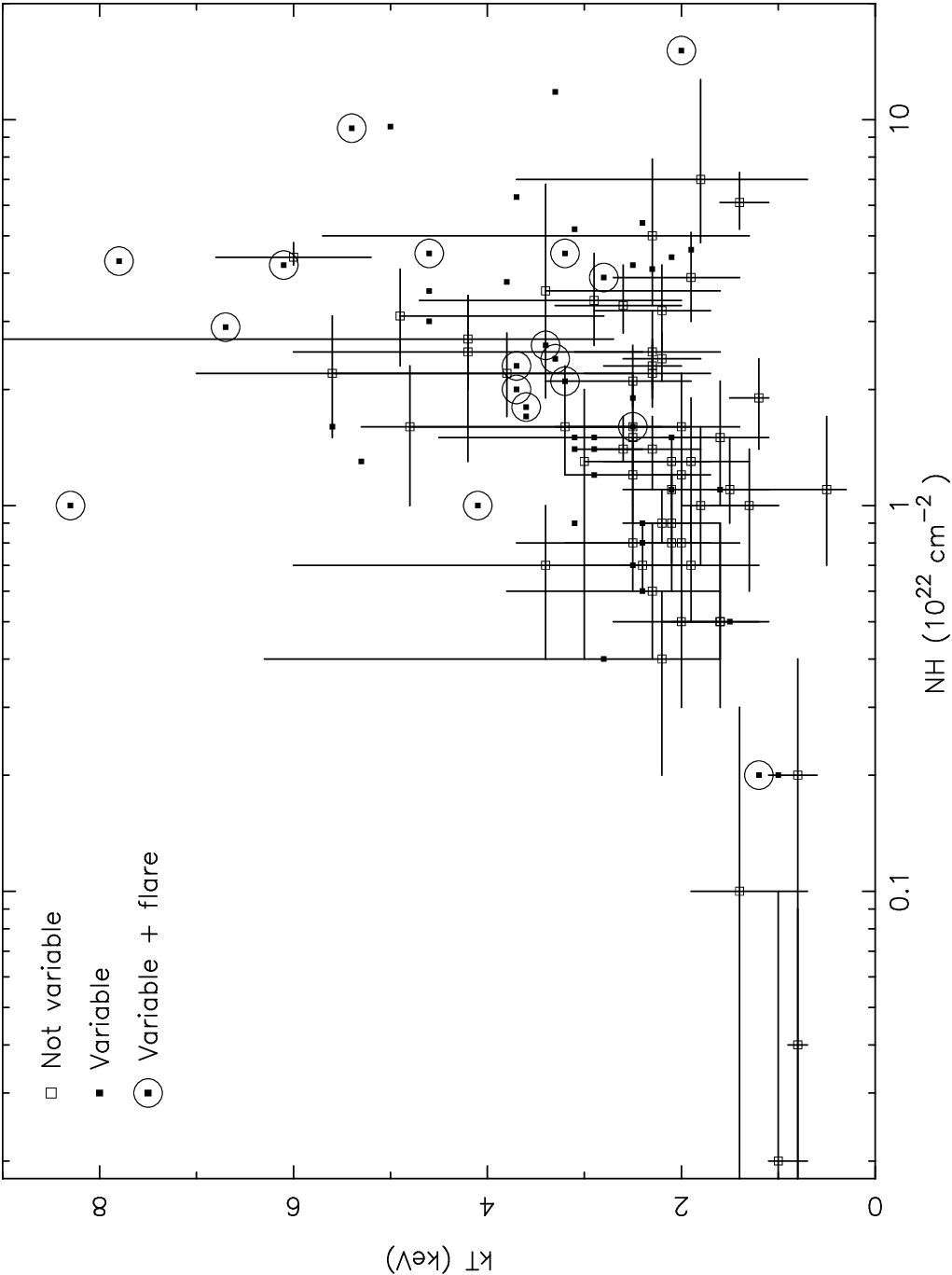}
\caption{}
\end{figure}

\clearpage

\begin{figure}
\figurenum{15}
\epsscale{0.9}
\plotone{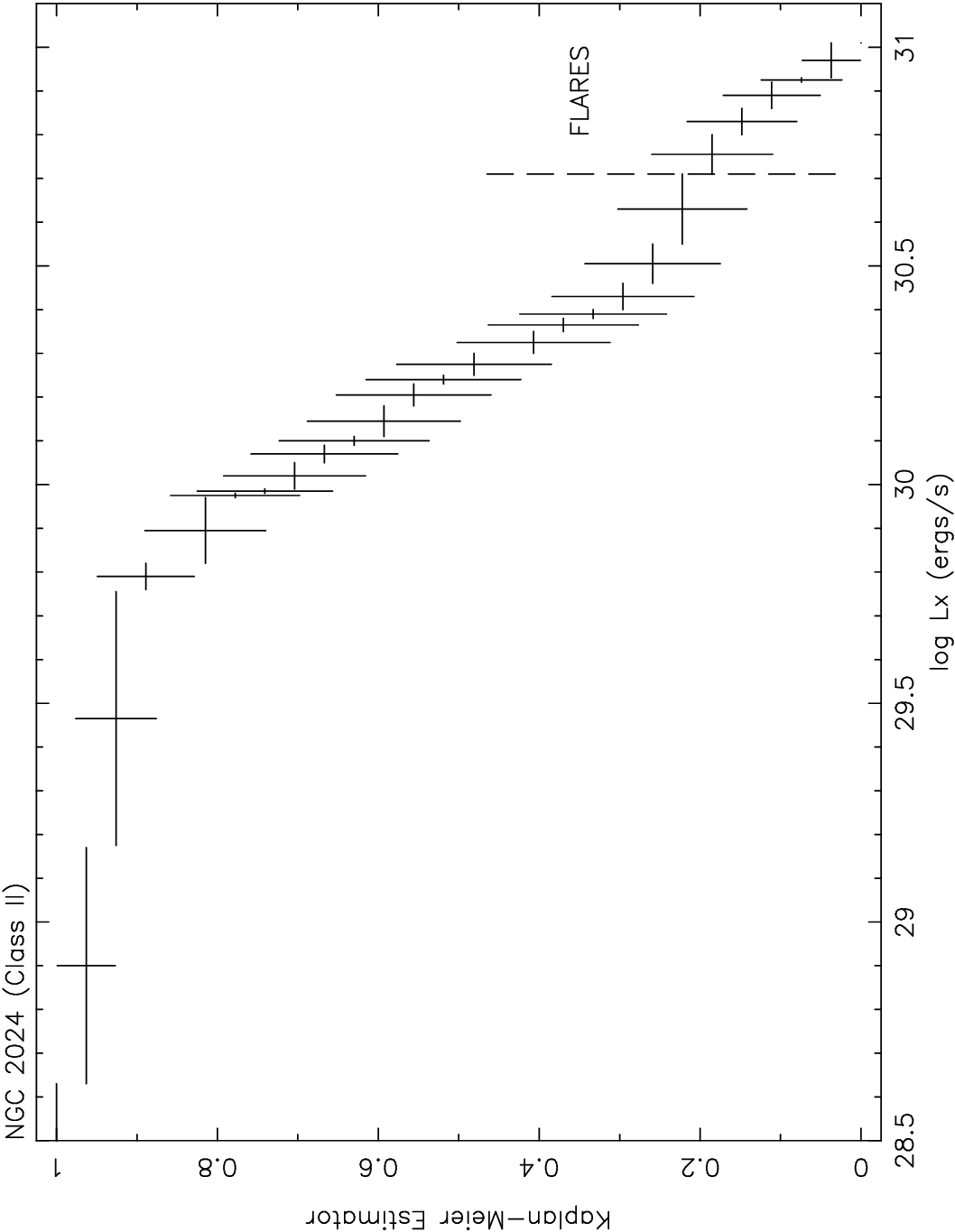}
\caption{}
\end{figure}

\clearpage

\begin{figure}
\figurenum{16}
\epsscale{0.9}
\plotone{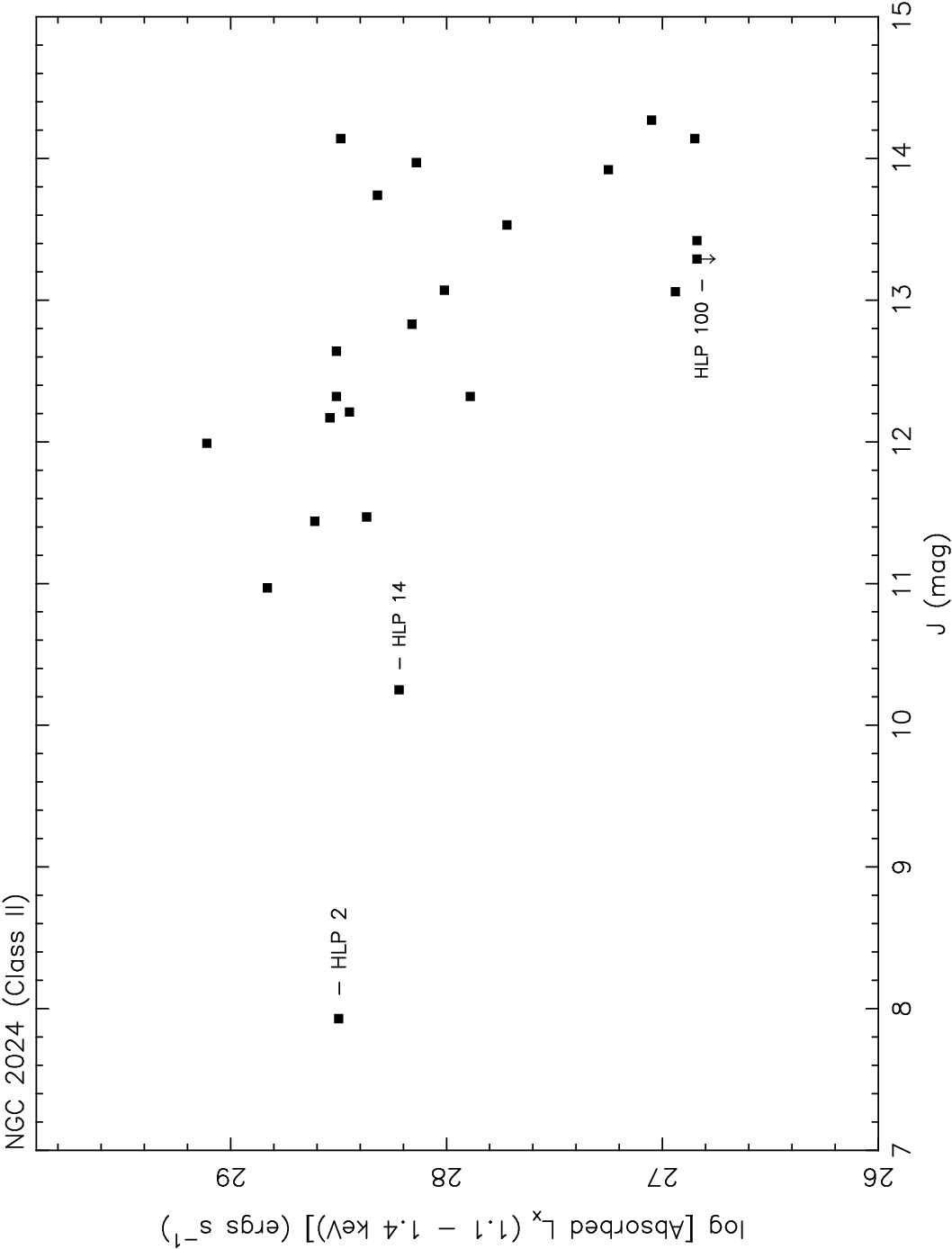}
\caption{}
\end{figure}

\clearpage

\begin{figure}
\figurenum{17}
\epsscale{0.9}
\plotone{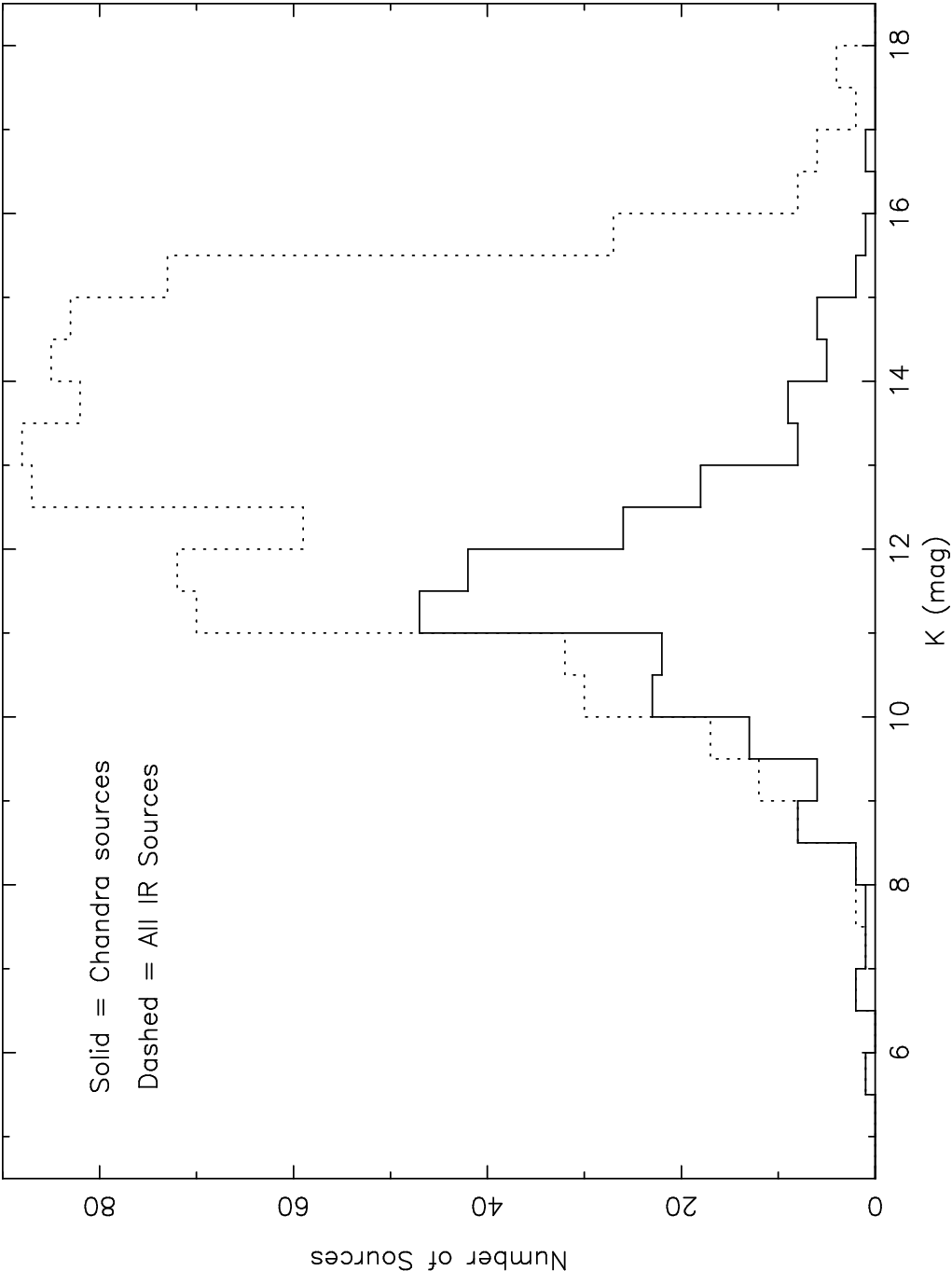}
\caption{}
\end{figure}

\clearpage

\begin{figure}
\figurenum{18}
\epsscale{0.9}
\plotone{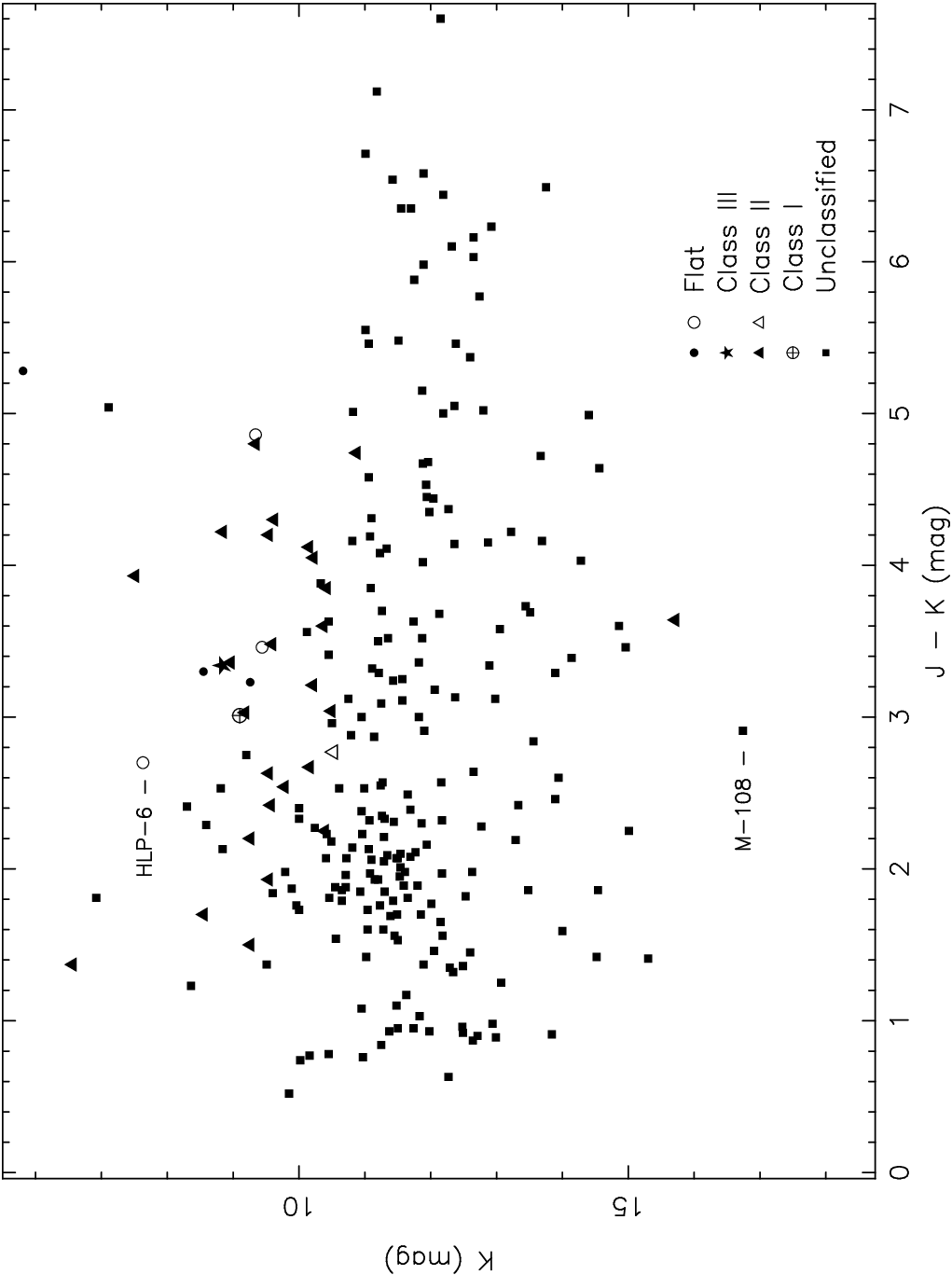}
\caption{}
\end{figure}

\clearpage

\begin{figure}
\figurenum{19}
\epsscale{0.9}
\plotone{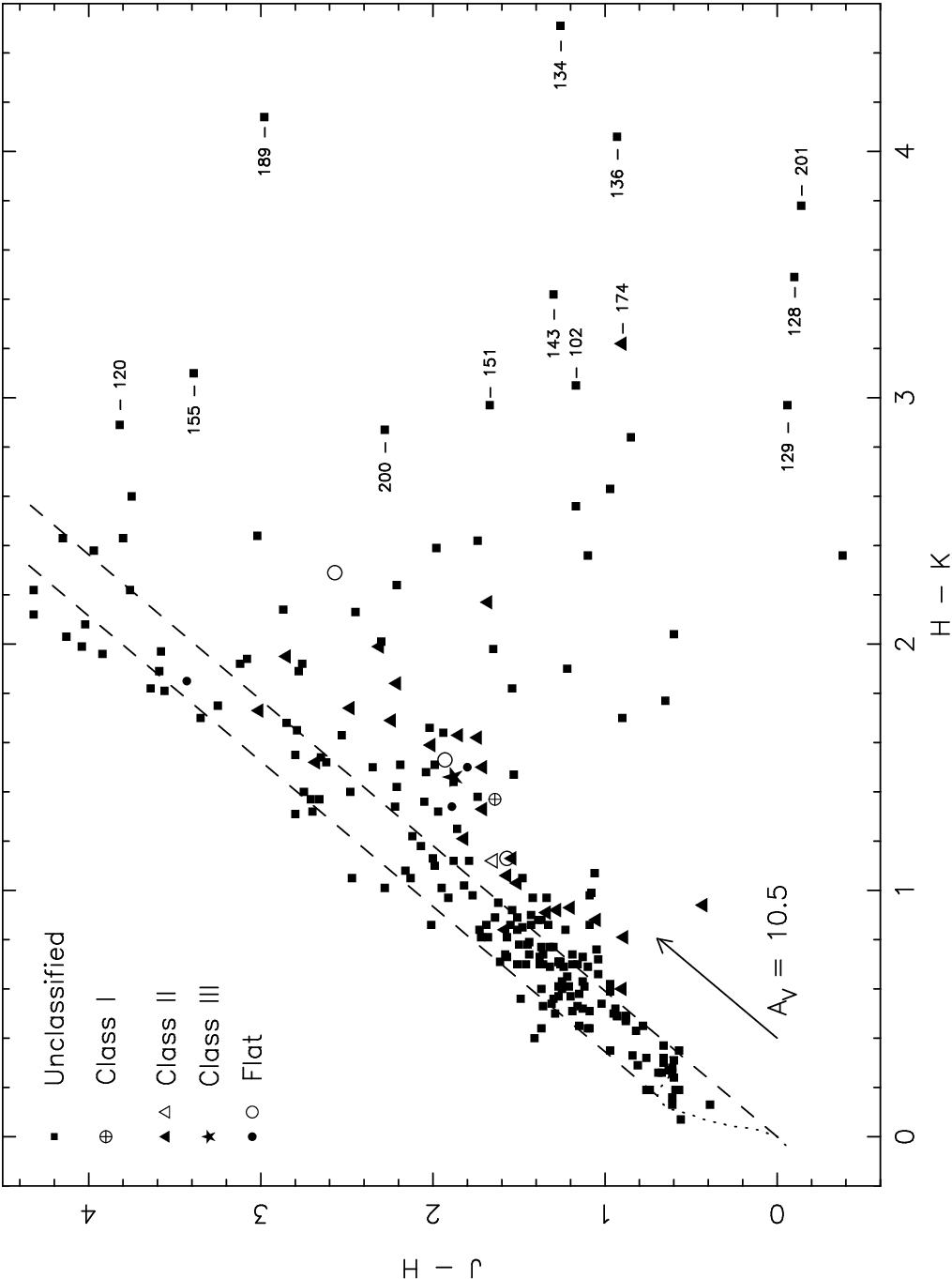}

\caption{}
\end{figure}

\clearpage

\begin{deluxetable}{lllllllllcll}
\rotate
\tablewidth{0pt}
\tablecaption{Chandra X-ray Sources in NGC 2024\tablenotemark{a} }
\tabletypesize{\scriptsize}
\tablehead{
\colhead{No.}      &
\colhead{Name}     &
\colhead{R.A.}       &
\colhead{Decl.}    &
\colhead{Net}   &
\colhead{KS}       &
\colhead{E}        & 
\colhead{N$_{\rm H}$}       &
\colhead{kT}       &
\colhead{F$_{\rm x,abs}$}       &
\colhead{log L$_{\rm x}$}   &
\colhead{Identification}    \\
\colhead{     }    &
\colhead{     }    &
\colhead{(J2000)     }    &
\colhead{(J2000)     }    &
\colhead{Counts }    &
\colhead{     }    &
\colhead{ (keV)}   & 
\colhead{(10$^{22}$ cm$^{-2}$)}       &
\colhead{ (keV)}       &
\colhead{ (erg/cm$^{2}$/s) }       &
\colhead{ (erg/s)}   &
\colhead{      } \\
\colhead{ (1)     }    &
\colhead{ (2)     }    &
\colhead{ (3)     }    &
\colhead{ (4)     }    &
\colhead{ (5)     }    &
\colhead{ (6)     }    &
\colhead{ (7)     }    & 
\colhead{ (8)     }       &
\colhead{ (9)     }       &
\colhead{ (10) }       &
\colhead{ (11)}   &
\colhead{ (12)     }
}
\startdata
   &                 &           &            &              &    &    &                    &                    &         &    &     \nl
  1&J054112.9-015459 & 5 41 12.97&  -1 55 00.0&  29 $\pm$ 7  &1.25&3.03& ...                & ...                & 5.33e-15& ...& ...     \nl 
  2&J054113.2-015332 & 5 41 13.29&  -1 53 32.0&  57 $\pm$ 9  &0.80&1.29& ...                & ...                & 5.33e-15& ... &
   05411324-0153306     \nl
  3&J054114.4-015121 & 5 41 14.43&  -1 51 21.0&  25 $\pm$ 7  &1.00&2.14& ...                & ...                & 2.97e-15&...  &
   05411449-0151204     \nl 
  4&J054118.0-014929 & 5 41 18.00&  -1 49 29.7&  41 $\pm$ 8  &1.21&2.74& ...                & ...                & 7.79e-15&...  &
    05411799-0149292     \nl 
  5*&J054118.3-014953 & 5 41 18.36&  -1 49 53.2& 120 $\pm$ 12 &0.88&2.88&1.3$^{+0.8}_{-0.5}$& ...                & 2.45e-14&30.21 &
    05411838-0149531     \nl 
  6&J054119.1-015226 & 5 41 19.14&  -1 52 26.9&  40 $\pm$ 8  &0.93&2.00& ...                & ...                & 5.09e-15&...   &
    05411918-0152268     \nl 
  7&J054120.6-015857 & 5 41 20.68&  -1 58 58.0& 545 $\pm$ 25 &2.08&1.42&0.2$^{+0.1}_{-0.1}$ &1.2$^{+0.1}_{-0.3}$ & 4.92e-14&30.16  & Haro 5-46,
    05412068-0158582   \nl 
  8&J054121.4-015629 & 5 41 21.42&  -1 56 29.0&  17 $\pm$ 6  &1.08&3.97& ... & ... & 7.18e-15&...  & ...    \nl 
  9&J054122.5-015041 & 5 41 22.50&  -1 50 41.8&  41 $\pm$ 8  &0.53&3.50& ... & ... & 1.20e-14&...    & ...    \nl
 10&J054124.2-014934 & 5 41 24.23&  -1 49 34.7& 666 $\pm$ 27 &0.68&1.16&0.04$^{+0.05}_{-0.04}$  &0.8$^{+0.1}_{-0.1}$ &5.47e-14&30.09 &
    HLL-2,05412424-0149348 \nl
 11&J054124.8-015203 & 5 41 24.83&  -1 52 03.9&   8 $\pm$ 5  &0.81&3.74& ... & ... & 3.76e-15&...    & 05412475-0152063     \nl     
 12&J054125.0-015226 & 5 41 25.01&  -1 52 26.9&  76 $\pm$ 10 &0.70&2.17& ...    & ...    & 1.08e-14&...    & HLL-31,05412501-0152267    \nl
 13&J054125.4-014809 & 5 41 25.43&  -1 48 09.3&  46 $\pm$ 8  &0.60&2.34& ...    & ...    & 5.68e-15&...    &05412553-0148092 \nl
 14&J054125.5-020111 & 5 41 25.55&  -2 01 11.1&  59 $\pm$ 9  &0.90&1.31& ...    & ...    & 3.92e-15&...    &05412562-0201116 \nl
 15*&J054125.8-015727 & 5 41 25.87&  -1 57 27.7&  25 $\pm$ 6  &0.69&2.09& ...    & ...    & 4.00e-15&...    &HLL-131,05412588-0157286   \nl 
 16*&J054126.1-020016 & 5 41 26.10&  -2 00 16.2& 194 $\pm$ 15 &1.04&2.35&1.6$^{+0.6}_{-0.6}$ &2.0$^{+1.2}_{-0.6}$& 2.98e-14& 30.35 &
    HLP-54,05412611-0200161    \nl 
 17&J054126.9-015451 & 5 41 26.94&  -1 54 51.8&  87 $\pm$ 10 &1.30&1.34& ...    & ...    & 6.24e-15&...    & HLL-93,05412694-0154517         \nl
 18*f&J054126.9-015409 & 5 41 26.96&  -1 54 09.3& 425 $\pm$ 22 &8.51&4.19&9.5$^{+2.4}_{-2.6}$ &5.4$^{+...}_{-2.1}$ & 2.16e-13&31.18  & 
    HLL-72,05412695-0154093         \nl
 19&J054127.0-015209 & 5 41 27.08&  -1 52 09.8&  65 $\pm$ 9  &2.30&1.97& ...    & ...    & 7.22e-15&...   & HLL-25,05412711-0152098   \nl
 20&J054128.0-015406 & 5 41 28.04&  -1 54 06.8&  16 $\pm$ 5  &0.43&3.69& ...    & ...    & 4.84e-15&...    & ...                \nl
 21&J054128.5-015359 & 5 41 28.57&  -1 53 59.3&   9 $\pm$ 4  &0.64&3.27& ...    & ...    & 1.81e-15&...    & ...               \nl 
 22&J054128.6-015033 & 5 41 28.63&  -1 50 33.6&  56 $\pm$ 9  &1.88&2.04& ...    & ...    & 7.11e-15&...    & HLL-9,05412868-0150331 \nl
 23&J054128.7-015002 & 5 41 28.79&  -1 50 02.3& 279 $\pm$ 18 &1.44&2.04&1.1$^{+0.4}_{-0.2}$ &1.5$^{+0.4}_{-0.4}$& 3.52e-14&30.29    & 
    HLL-6,05412880-0150024          \nl
 24&J054129.3-015130 & 5 41 29.31&  -1 51 30.5&  60 $\pm$ 9  &1.81&2.77& ...    & ...    & 1.25e-14&...    & 
    HLL-16,05412932-0151305         \nl
 25&J054129.5-015425 & 5 41 29.53&  -1 54 25.6& 244 $\pm$ 17 &1.53&1.82&0.5$^{+0.4}_{-0.1}$&1.6$^{+0.5}_{-0.5}$ & 2.63e-14&30.00 & 
     HLL-82,05412953-0154255         \nl
 26&J054130.8-015324 & 5 41 30.89&  -1 53 24.3&  26 $\pm$ 6  &2.12&2.38& ...    & ...    & 3.55e-15&...    & 05413089-0153246                \nl 
 27*f&J054131.6-015231 & 5 41 31.61&  -1 52 31.7&1358 $\pm$ 38 &14.3&2.49&1.0$^{+0.1}_{-0.2}$&4.1$^{+1.1}_{-0.4}$ & 2.22e-13&30.92  & 05413162-0152317 \nl
 28&J054131.6-015357 & 5 41 31.70&  -1 53 57.3& 488 $\pm$ 23 &1.72&2.21&1.1$^{+0.2}_{-0.3}$&2.1$^{+0.6}_{-0.4}$& 6.63e-14&30.50  & 
  HLL-66,05413170-0153573         \nl
 29&J054131.8-015219 & 5 41 31.85&  -1 52 19.9& 174 $\pm$ 14 &1.15&1.94&0.7$^{+0.3}_{-0.2}$ &1.9$^{+0.7}_{-0.7}$& 1.92e-14&29.90    & 
    HLL-29,05413186-0152200         \nl
 30&J054131.8-015453 & 5 41 31.87&  -1 54 53.9&  46 $\pm$ 8  &1.73&2.09& ...    & ...    & 5.55e-15&...    & HLL-95,05413186-0154537         \nl
 31&J054131.9-015518 & 5 41 31.94&  -1 55 18.6& 187 $\pm$ 15 &0.88&2.20&0.7$^{+0.3}_{-0.3}$ &3.4$^{+2.6}_{-1.4}$ & 2.52e-14&29.98 & 
     HLL-106,HLP-64, 05413194-0155186      \nl
 32&J054132.7-015757 & 5 41 32.74&  -1 57 57.3& 327 $\pm$ 19 &0.81&1.21&0.02$^{+0.08}_{-0.02}$ &1.0$^{+0.1}_{-0.3}$ & 2.64e-14&29.71 & 05413275-0157574 \nl
 33&J054132.8-015444 & 5 41 32.85&  -1 54 44.3&  27 $\pm$ 6  &1.39&2.04& ...    & ...    & 3.51e-15&...    & 05413284-0154442 \nl
 34&J054132.9-020145 & 5 41 32.92&  -2 01 45.3&  20 $\pm$ 6  &1.01&3.44& ...    & ...    & 6.15e-15&...    & ...               \nl
 35&J054133.1-020304 & 5 41 33.15&  -2 03 04.4&  32 $\pm$ 7  &0.55&1.24& ...    & ...    & 2.60e-15&...    & 05413319-0203051 \nl
 36*&J054133.2-014810 & 5 41 33.28&  -1 48 10.7&2434 $\pm$ 51 &1.96&2.40&1.2$^{+0.4}_{-0.1}$ & 2.9$^{+0.5}_{-0.3}$& 4.28e-13&31.26  & 05413329-0148108 \nl
 37&J054133.3-015127 & 5 41 33.35&  -1 51 27.7&  36 $\pm$ 7  &0.49&1.82& ...    & ...    & 3.28e-15&...    & HLL-15,05413337-0151271 \nl
 38&J054133.4-015513 & 5 41 33.46&  -1 55 13.3&      8   $\pm$ 4&    1.04&  2.47& ... & ... &1.73e-15&... & HLL-102,05413347-0155134 \nl
 39*& J054133.6-015606 & 5 41 33.70& -1 56 06.6&    13   $\pm$ 5&    1.47&  2.52& ... & ... &3.00e-14&... & HLL-115,05413371-0156068            \nl
 40& J054133.8-020356 & 5 41 33.81& -2 03 56.8&    17   $\pm$ 6&    0.45&  3.49& ... & ... &5.65e-15&... & ...                   \nl
 41*f& J054133.8-015323 & 5 41 33.82&  -1 53 23.4&  643    $\pm$ 26&   3.81&  2.54&1.6$^{+0.2}_{-0.2}$ &2.5$^{+0.6}_{-0.5}$ &1.02e-13&30.71 &
  HLL-47,HLP-39,05413382-0153234 \nl
 42& J054133.8-015308 & 5 41 33.85&  -1 53 08.8&    61   $\pm$ 9&    0.71&  1.84& ... & ... &5.92e-15&... &HLL-38,05413386-0153087   \nl
 43& J054133.9-015350 & 5 41 33.97&  -1 53 50.9&    71   $\pm$ 10&   0.60&  2.15& ... & ... &8.49e-15&... &HLL-62,05413397-0153510   \nl
 44& J054134.0-015958 & 5 41 34.04&  -1 59 58.7&    24   $\pm$ 6 &   1.03&  3.85& ... & ... &9.03e-15&... & ...        \nl
 45& J054134.3-020150 & 5 41 34.35&  -2 01 50.4&  1809   $\pm$ 44&   1.75&  1.30&0.2$^{+0.1}_{-0.1}$ &1.0$^{+0.3}_{-0.3}$ &1.41e-13&30.65 &
     05413436-0201507 \nl
 46& J054134.4-015319 & 5 41 34.40&  -1 53 19.4&   442   $\pm$ 22&   1.10&  2.27&1.3$^{+0.2}_{-0.2}$ &2.1$^{+0.6}_{-0.4}$&5.83e-14&30.48 &
     HLL-44,05413441-0153194  \nl
 47& J054134.4-015441 & 5 41 34.49&  -1 54 41.0&   552   $\pm$ 25&   0.90&  2.07&0.7$^{+0.2}_{-0.1}$ &2.4$^{+0.6}_{-0.6}$&6.36e-14&30.43 &
      HLL-88,05413448-0154410  \nl
 48& J054134.5-015213 & 5 41 34.51&  -1 52 13.5&    58   $\pm$ 9 &   0.82&  1.96& ... & ... &6.12e-15&... &05413451-0152135 \nl
 49& J054134.7-015552 & 5 41 34.76&  -1 55 52.1&    61   $\pm$ 9 &   1.77&  3.80& ... & ... &3.43e-14&... &05413476-0155522        \nl
 50& J054135.0-015327 & 5 41 35.03&  -1 53 28.0&    21   $\pm$ 6 &   0.53&  2.16& ... & ... &3.31e-15&... &
     05413502-0153286,B-1,M-1 \nl
 51& J054135.3-015211 & 5 41 35.37&  -1 52 11.4&     6   $\pm$ 4 &   0.73&  2.65& ... & ... &9.67e-16&... &HLL-24,05413537-0152116   \nl
 52*f& J054135.4-015615 & 5 41 35.43&  -1 56 15.5&   315   $\pm$ 19&   4.96&  1.47&0.2$^{+0.3}_{-0.1}$&1.2$^{+0.3}_{-0.3}$ &2.77e-14&29.88 
    &M-6,05413543-0156154          \nl
 53& J054135.4-015420 & 5 41 35.47&  -1 54 20.1&    13   $\pm$ 5 &   1.75&  3.99& ... & ... &3.70e-15&... &
      05413547-0154195,B-2,M-8 \nl
 54& J054135.7-015348 & 5 41 35.74&  -1 53 48.6&    25   $\pm$ 6 &   1.71&  3.45& ... & ... &6.35e-15&... &
      05413575-0153483,B-3,M-11 \nl
 55& J054135.8-015622 & 5 41 35.80&  -1 56 22.2&    49   $\pm$ 8 &   1.31&  3.15& ... & ... &1.09e-14&... &HLL-118,M-12,05413581-0156222    \nl
 56& J054135.8-015712 & 5 41 35.85&  -1 57 12.1&   238   $\pm$ 16&   0.74&  2.87&3.2$^{+1.0}_{-0.7}$ &2.2$^{+0.7}_{-0.5}$&4.48e-14&30.55 &
      HLL-126,05413585-0157122   \nl
 57& J054136.2-015424 & 5 41 36.23&  -1 54 24.1&   184  $\pm$ 15&   1.51&  3.13&2.2$^{+0.9}_{-0.7}$ &5.6$^{+...}_{-2.5}$ &3.81e-14&30.30 &
     HLL-81,HLP-26,05413623-0154241,M-17,B-6    \nl
 58& J054136.3-015554 & 5 41 36.38&  -1 55 54.8&    86   $\pm$ 10&   0.62&  1.86& ... & ... &1.74e-14&... &HLL-113,M-22,05413638-0155548     \nl
 59& J054136.5-015318 & 5 41 36.54&  -1 53 18.8&   114   $\pm$ 12&   3.39&  2.42&1.5$^{+0.5}_{-0.5}$ &2.1$^{+1.2}_{-0.7}$ &1.67e-14&29.95 &
      HLL-43,05413654-0153188,B-12,M-25      \nl
 60& J054136.5-015354 & 5 41 36.60&  -1 53 54.4&   985   $\pm$ 32&   0.89&  2.20&1.2$^{+0.2}_{-0.1}$ &2.0$^{+0.3}_{-0.3}$ &1.24e-13&30.80 &
     HLL-63,HLP-56,05413659-0153544,B-13,M-27    \nl
 61& J054136.6-015242 & 5 41 36.69&  -1 52 42.9&   413   $\pm$ 21&   2.35&  2.06&0.8$^{+0.2}_{-0.2}$ &2.4$^{+0.7}_{-0.5}$ &4.86e-14&30.29 &
     HLL-33,05413669-0152432,B-16      \nl
 62& J054136.6-015408 & 5 41 36.69&  -1 54 08.2&    68   $\pm$ 9 &   1.79&  2.00& ... & ...  &7.32e-15&... &
      HLL-71,05413668-0154082,B-15,M-30      \nl
 63& J054136.7-014738 & 5 41 36.75&  -1 47 38.3&    42   $\pm$ 8 &   0.87&  1.83& ... & ...  &1.13e-14&... &05413671-0147381 \nl
 64& J054136.8-015358 & 5 41 36.81&  -1 53 58.8&   202   $\pm$ 15&   1.34&  2.12&0.9$^{+0.3}_{-0.2}$ &2.1$^{+0.3}_{-0.5}$ &2.40e-14&30.05 &
      HLL-67,HLP-53,05413681-0153589,B-18,M-32    \nl
 65& J054136.8-015447 & 5 41 36.84&  -1 54 47.9&    31   $\pm$ 7 &   1.14&  2.34& ... & ... &4.04e-15&... &
     HLL-92,05413682-0154479,B-19,M-33 \nl
 66& J054136.9-015233 & 5 41 36.93&  -1 52 33.3&    52   $\pm$ 8 &   1.48&  1.90& ... & ... &5.24e-15&... &
     HLL-32,05413693-0152333,B-24       \nl
 67& J054136.9-015339 & 5 41 36.93&  -1 53 39.5&    21   $\pm$ 6 &   0.66&  2.34& ... & ... &2.86e-15&... &
     HLL-55,05413693-0153393,B-21,M-34       \nl
 68& J054137.1-014659 & 5 41 37.16&  -1 46 59.4&   286   $\pm$ 18&   1.06&  1.98&1.0$^{+0.6}_{-0.3}$ &1.8$^{+0.2}_{-0.8}$ &4.35e-14&30.54 &
      05413714-0146597 \nl
 69& J054137.3-015313 & 5 41 37.34&  -1 53 13.3&   115   $\pm$ 12&   0.51&  2.33&1.3$^{+0.6}_{-0.5}$ &1.9$^{+0.9}_{-0.6}$ &1.75e-14&30.09 &
      HLL-42,HLP-14,05413733-0153132,B-25,M-35      \nl
 70& J054137.3-015244 & 5 41 37.39&  -1 52 44.6&    19   $\pm$ 6 &   1.54&  2.56& ... & ... &3.86e-15&... &
     HLL-34,05413739-0152447,B-26        \nl
 71& J054137.4-014953 & 5 41 37.43&  -1 49 53.1&   181   $\pm$ 15&   0.73&  2.47&1.5$^{+0.7}_{-0.6}$ &2.5$^{+2.0}_{-0.8}$ &2.94e-14&30.14 &
     HLL-3,HLP-12,05413744-0149532      \nl
 72& J054137.4-014810 & 5 41 37.47&  -1 48 10.5&   204   $\pm$ 15&   1.20&  2.10&0.8$^{+0.4}_{-0.3}$ &2.0$^{+1.7}_{-0.6}$ &6.68e-14&30.44 &
     05413748-0148108\nl
 73& J054137.5-015652 & 5 41 37.59&  -1 56 52.9&    11   $\pm$ 4 &   0.52&  4.33& ... & ... &3.61e-15&... & ...             \nl
 74& J054137.6-015424 & 5 41 37.62&  -1 54 24.5&   297   $\pm$ 18&   1.10&  2.34&1.2$^{+0.3}_{-0.2}$ &2.5$^{+0.7}_{-0.6}$ &4.58e-14&30.33 &
     HLL-80,05413762-0154246,B-28,M-38 \nl
 75*& J054137.7-015351 & 5 41 37.73&  -1 53 51.4&  3084   $\pm$ 57&   0.99&  2.46&1.4$^{+0.3}_{-0.1}$ &2.6$^{+0.3}_{-0.2}$ &4.97e-13&31.38 &      HLL-61,05413774-0153514,B-30,M-43        \nl
 76& J054137.7-015509 & 5 41 37.75&  -1 55 09.9&    10     $\pm$ 4 &   1.05&  2.68& ... & ... &1.51e-15&... &B-31,M-44      \nl
 77& J054137.8-015436 & 5 41 37.85&  -1 54 36.5&   97   $\pm$ 11&   0.60&  1.51&1.1$^{+0.6}_{-0.4}$ &0.5$^{+1.2}_{-0.2}$ &7.54e-15&30.30 &
      HLL-86,HLP-2,05413785-0154364,B89-1,B-33 \nl
 78& J054137.8-015431 & 5 41 37.86&  -1 54 31.8&   426   $\pm$ 22&   0.79&  2.39&1.4$^{+0.3}_{-0.3}$ &2.3$^{+0.8}_{-0.5}$ &6.44e-14&30.52 &
     05413786-0154323,V-2 \nl
 79& J054137.9-015311 & 5 41 37.90&  -1 53 11.2&   825   $\pm$ 30&   1.47&  2.18&1.1$^{+0.1}_{-0.1}$ &2.1$^{+0.5}_{-0.3}$ &1.03e-13&30.68 &
     B-35,M-48,05413789-0153112 \nl
 80& J054138.0-015357 & 5 41 38.08&  -1 53 57.2&   238   $\pm$ 16&   2.00&  2.33&1.4$^{+0.4}_{-0.3}$ &2.9$^{+2.0}_{-0.9}$ &3.30e-14&30.15 &
     HLL-64,05413807-0153572,B-37,M-51  \nl
 81*f& J054138.1-015817 & 5 41 38.16&  -1 58 17.6&   433   $\pm$ 22&   8.76&  2.71&1.0$^{+0.3}_{-0.2}$ &8.3$^{+...}_{-3.6}$ &7.86e-14&30.44
      &HLL-137,05413816-0158176 \nl
 82& J054138.1-015454 & 5 41 38.20&  -1 54 54.8&    17   $\pm$ 5 &   0.70&  3.06& ... & ...  &3.39e-15&... &
     HLL-94,05413818-0154549,B-40,M-54 \nl
 83& J054138.2-015536 & 5 41 38.21&  -1 55 36.2&    67   $\pm$ 9 &   0.79&  2.02& ... & ...  &2.49e-14&... &
     HLL-109,05413821-0155361,M-56   \nl
 84& J054138.2-014633 & 5 41 38.22&  -1 46 33.6&    42   $\pm$ 8 &   0.56&  1.86& ... & ...  &7.53e-15&... &05413819-0146300  \nl
 85*f& J054138.2-015309 & 5 41 38.24&  -1 53 09.1&  2496   $\pm$ 51&
 5.37&  2.79&1.8$^{+0.2}_{-0.2}$ &3.6$^{+0.5}_{-0.4}$ &4.47e-13&31.34 & 
       HLL-37,05413824-0153090,B-43,B89-11   \nl
 86& J054138.3-015332 & 5 41 38.32&  -1 53 32.9&    79   $\pm$ 10&   0.68&  1.05& ... & ...  &4.79e-15&... &
     05413832-0153331,B-45,M-60 \nl
 87& J054138.3-015558 & 5 41 38.38&  -1 55 58.2&    17   $\pm$ 5 &   0.70&  2.48& ... & ...  &2.69e-15&... & M-61,05413840-0155582 \nl
 88& J054138.4-015037 & 5 41 38.41&  -1 50 38.0&   309   $\pm$ 19&   0.73&  2.72&1.6$^{+0.7}_{-0.4}$ &3.2$^{+2.1}_{-1.0}$ &5.75e-14&30.46 &
     HLL-10,HLP-33,05413843-0150383 \nl
 89& J054138.5-015322 & 5 41 38.59&  -1 53 22.7&  1869   $\pm$ 44&   1.66&  2.63&1.9$^{+0.2}_{-0.1}$ &2.5$^{+0.3}_{-0.2}$ &3.01e-13&31.22 &
     HLL-45,05413858-0153227,B89-10,B-47 \nl
 90& J054138.5-015730 & 5 41 38.60&  -1 57 30.0&    61   $\pm$ 9 &   0.83&  2.06& ... & ... &7.78e-15&... &HLL-130,05413861-0157301 \nl
 91& J054138.7-015246 & 5 41 38.72&  -1 52 46.0&    25   $\pm$ 6 &   0.69&  2.58& ... & ... &5.87e-15&... &05413871-0152463 \nl
 92& J054138.7-015429 & 5 41 38.72&  -1 54 29.1&    57   $\pm$ 9 &   1.46&  2.94& ... & ... &1.05e-14&... &B-48,M-67,05413871-0154289 \nl
 93*f& J054138.7-015602 & 5 41 38.76&  -1 56 02.3&  385   $\pm$ 21&  10.2&  2.89&2.0$^{+0.4}_{-0.4}$ &3.7$^{+1.8}_{-1.0}$ &7.49e-14&30.59 &
       HLL-114,M-68,05413876-0156024 \nl
 94*& J054138.9-015936 & 5 41 38.95&  -1 59 36.1&  2108   $\pm$ 47&   4.01&  1.92&0.4$^{+0.1}_{-0.1}$ & 2.8$^{+0.3}_{-0.3}$ &2.36e-13&30.86     &
     HLP-40,Haro 5-59,05413895-0159362 \nl
 95& J054139.0-015145 & 5 41 39.00&  -1 51 45.3&    26   $\pm$ 6 &   0.67&  2.27& ... & ... &4.25e-15&... &HLL-21,05413901-0151453 \nl
 96f& J054139.0-015626 & 5 41 39.06&  -1 56 26.4&   214   $\pm$ 16&   5.23&  3.24&4.4$^{+1.8}_{-0.6}$ &2.1$^{+1.1}_{-0.6}$ &4.92e-14&30.72 &      HLL-122,M-71,05413906-0156264 \nl
 97& J054139.0-015358 & 5 41 39.09&  -1 53 58.1&   153   $\pm$ 13&   1.20&  2.39&1.9$^{+0.5}_{-0.5}$ &1.2$^{+0.3}_{-0.1}$ &2.17e-14&30.18 &
     HLL-65,HLP-24,05413908-0153582,B-50,M-74 \nl
 98& J054139.1-015207 & 5 41 39.12&  -1 52 07.3&   672   $\pm$ 27&   1.58&  2.77&2.4$^{+0.4}_{-0.3}$ &2.2$^{+0.4}_{-0.4}$ &1.40e-13&30.92 &
      HLL-23,HLP-4,05413913-0152073  \nl
 99*& J054139.1-015414 & 5 41 39.19&  -1 54 14.0&   334   $\pm$ 19&   0.57&  2.62&1.6$^{+0.5}_{-0.4}$ &2.5$^{+0.8}_{-0.6}$ &5.28e-14&30.38 &      HLL-75,HLP-45,05413919-0154140,B-51,M-75 \nl
100& J054139.2-015307 & 5 41 39.23&  -1 53 08.0&    12   $\pm$ 5 &   0.37&  2.51& ... & ... &1.95e-15&... &05413920-0153075,B-53 \nl
101& J054139.2-015401 & 5 41 39.24&  -1 54 01.7&    19   $\pm$ 5 &   1.40&  2.95& ... & ... &3.64e-15&... &B-52,M-76  \nl
102& J054139.4-015236 & 5 41 39.40&  -1 52 36.8&    27   $\pm$ 6 &   0.59&  3.33& ... & ... &6.59e-15&... & 05413939-0152368 \nl
103& J054139.4-015153 & 5 41 39.41&  -1 51 53.9&    13   $\pm$ 5 &   0.39&  4.21& ... & ... &4.94e-15& ... & ...  \nl
104*f& J054139.4-014700 & 5 41 39.42&  -1 47 00.0&   296   $\pm$ 19&   8.54&  3.40&3.6$^{+1.1}_{-0.8}$ &4.6$^{+5.7}_{-1.6}$&7.95e-14&30.70 &
     05413956-0147020 \nl 
105*& J054139.4-015326 & 5 41 39.49&  -1 53 26.7&   276  $\pm$ 18&   2.31&  2.89&1.6$^{+0.3}_{-0.4}$ &5.6$^{+...}_{-2.0}$ &5.62e-14&30.40 &
      HLL-49;HLP-49,05413948-0153268,B-55,M-78 \nl
106& J054139.5-015441 & 5 41 39.55&  -1 54 41.6&    32   $\pm$ 7 &   0.49&  2.77& ... & ... &5.69e-15&... &
      05413956-0154416,B-57,M-80 \nl
107& J054139.7-020224 & 5 41 39.70&  -2 02 24.2&   380   $\pm$ 21&   2.40&  2.19&0.9$^{+0.3}_{-0.2}$ &2.4$^{+0.8}_{-0.5}$ &5.20e-14&30.40 &
     05413972-0202242 \nl
108*f& J054140.0-015335 & 5 41 40.05&  -1 53 35.6&   855   $\pm$ 30&   6.49&  2.96&2.4$^{+0.3}_{-0.3}$ &3.3$^{+1.0}_{-0.6}$ &1.71e-13&30.96      & B-58,M-84 \nl
109& J054140.1-020059 & 5 41 40.16&  -2 00 59.2&    15   $\pm$ 5 &   0.81&  2.36& ... & ... &3.53e-15&... &05414012-0201000 \nl
110& J054140.1-015407 & 5 41 40.20&  -1 54 07.2&   258   $\pm$ 17&   1.66&  3.44&5.4$^{+2.5}_{-1.2}$ &2.4$^{+1.2}_{-0.9}$ &5.85e-14&30.75 &
     05414019-0154073,B-60,M-87 \nl
111& J054140.2-015334 & 5 41 40.20&  -1 53 34.0&    111   $\pm$ 12 &  1.78&  2.87&2.7$^{+...}_{-1.6}$  &2.6$^{+...}_{-2.0}$  &2.21e-14&29.76 &
      HLL-53,HLP-61,05414016-0153344,M-86 \nl
112& J054140.2-015052 & 5 41 40.24&  -1 50 52.5&    61   $\pm$ 9 &   0.65&  2.30& ... & ... &1.00e-14&... &HLL-12,05414026-0150522   \nl
113& J054140.2-015643 & 5 41 40.25&  -1 56 43.8&    85   $\pm$ 10&   3.26&  3.29& 3.8$^{+2.9}_{-1.9}$ & 3.8$^{+...}_{-2.2}$ &1.95e-14&30.14
      & 05414025-0156439  \nl
114*& J054140.2-015220 & 5 41 40.26&  -1 52 20.5&   205   $\pm$ 15&   3.18&  2.85&1.3$^{+0.5}_{-0.3}$ &5.3$^{+...}_{-2.2}$&3.95e-14&30.27 &
      HLL-28,05414026-0152205 \nl
115& J054140.5-015221 & 5 41 40.52&  -1 52 21.1&    46   $\pm$ 8 &   1.02&  2.36& ... & ... &6.65e-15&... &05414051-0152212 \nl
116& J054140.6-015043 & 5 41 40.64&  -1 50 43.1&    31   $\pm$ 7 &   1.46&  2.43& ... & ... &4.88e-15&... &05414067-0150431 \nl
117& J054140.9-015408 & 5 41 40.91&  -1 54 08.8&    12   $\pm$ 5 &   0.78&  3.72& ... & ... &3.12e-15&... &05414091-0154087,M-92 \nl
118& J054141.0-015158 & 5 41 41.02&  -1 51 58.1&    45   $\pm$ 8 &   0.96&  2.86& ... & ... &8.46e-15&... &05414102-0151577 \nl
119& J054141.3-015303 & 5 41 41.32&  -1 53 03.3&    17   $\pm$ 5 &   1.26&  3.35& ... & ... &3.61e-15&... &05414134-0153033  \nl
120*& J054141.3-015332 & 5 41 41.34&  -1 53 32.6&  7379   $\pm$ 87&   1.41&  3.67&4.4$^{+0.4}_{-0.2}$  &6.0$^{+0.8}_{-0.8}$  &2.11e-12&32.13 &HLL-52,05414134-0153326,M-95,V-3 \nl
121& J054141.3-015437 & 5 41 41.35&  -1 54 37.8&    16   $\pm$ 5 &   1.50&  2.98& ... & ... &2.97e-15&... &M-96   \nl
122& J054141.3-015352 & 5 41 41.38&  -1 53 52.0&     47   $\pm$ 8 &   0.47&  2.77& ... & ... &9.45e-15&... & 
     HLL-60,05414138-0153522,B-66,M-98 \nl
123*& J054141.3-015444 & 5 41 41.38&  -1 54 44.5&   692   $\pm$ 27&   3.33&  3.20&4.1$^{+0.6}_{-0.5}$ &2.3$^{+0.5}_{-0.4}$&1.57e-13&31.13 &
     05414138-0154445,B-65,M-97,V-4 \nl
124& J054141.4-015439 & 5 41 41.49&  -1 54 39.2&    90   $\pm$ 11&   1.94&  3.19&4.6$^{+2.8}_{-1.3}$  &1.9$^{+2.0}_{-0.8}$  &2.06e-14&30.33 & HLL-87,05414148-0154390,B-67,M-102,V-5 \nl
125& J054141.6-015754 & 5 41 41.64&  -1 57 54.5&    94   $\pm$ 11&   0.92&  3.30&3.6$^{+3.2}_{-1.7}$  &3.4$^{+...}_{-1.8}$ &2.13e-14&30.21 &
     HLL-136,HLP-38,05414164-0157545 \nl
126& J054141.6-020134 & 5 41 41.66&  -2 01 34.2&    30   $\pm$ 7 &   1.12&  3.44& ... & ...  &1.10e-14&... & ...  \nl
127& J054141.6-015412 & 5 41 41.66&  -1 54 12.7&    31   $\pm$ 7 &   0.28&  2.98& ... & ... &8.50e-15&... &
      HLL-73,05414167-0154128,M-107 \nl
128& J054141.7-015344 & 5 41 41.71&  -1 53 44.3&    18   $\pm$ 5 &   0.53&  3.63& ... & ... &5.54e-15&... &M-109,05414170-0153444  \nl
129& J054141.7-015334 & 5 41 41.73&  -1 53 34.8&    24   $\pm$ 6 &   0.99&  3.86& ... & ... &7.55e-15&... &M-108  \nl
130& J054141.7-015330 & 5 41 41.73&  -1 53 30.1&    23   $\pm$ 6 &   0.71&  4.25& ... & ... &1.13e-14&... & ... \nl
131& J054141.7-015145 & 5 41 41.76&  -1 51 45.7&   109   $\pm$ 11&   1.97&  2.04&0.8$^{+0.4}_{-0.3}$ &2.4$^{+1.5}_{-0.8}$ &1.29e-14&29.67 &
     HLL-20,05414177-0151456 \nl
132& J054141.7-015459 & 5 41 41.79&  -1 55 00.0&    66   $\pm$ 9 &   0.69&  2.88& ... & ... &1.27e-14&... &05414179-0154599,M-110 \nl
133*f& J054141.9-015423 & 5 41 41.98&  -1 54 23.9&   314   $\pm$ 19&   7.38&  3.87&6.3$^{+1.3}_{-1.5}$ &3.7$^{+...}_{-1.2}$&1.17e-13&30.93 &
     HLL-78,HLP-80,05414198-0154240,B-73,M-113 \nl
134& J054142.0-015301 & 5 41 42.09&  -1 53 01.2&    18   $\pm$ 5 &   0.55&  3.20& ... & ... &3.69e-15&... &05414211-0153014 \nl
135& J054142.1-015509 & 5 41 42.20&  -1 55 09.6&     7   $\pm$ 4 &   0.87&  2.82& ... & ... &2.94e-15&... &
      HLL-100,05414220-0155099,M-118  \nl
136& J054142.3-015315 & 5 41 42.36&  -1 53 15.5&    56   $\pm$ 9 &   1.10&  4.08& ... & ... &1.71e-14&... &
     05414236-0153152,M-120 \nl
137& J054142.4-015502 & 5 41 42.45&  -1 55 02.6&   164   $\pm$ 14&   1.57&  2.94&2.2$^{+0.6}_{-0.5}$ &3.8$^{+3.2}_{-1.3}$ &5.07e-14&30.47 &
      HLL-97,05414245-0155025,M-121 \nl
138& J054142.4-014800 & 5 41 42.48&  -1 48 00.9&   133   $\pm$ 13&   1.16&  1.95&0.5$^{+0.4}_{-0.2}$ &1.6$^{+0.6}_{-0.4}$&1.52e-14&29.86 &
      05414248-0148013 \nl
139*f& J054142.4-015409 & 5 41 42.49&  -1 54 09.6&   403   $\pm$ 21&   5.24&2.93&2.1$^{+0.5}_{-0.4}$ &3.2$^{+1.8}_{-0.9}$&9.02e-14&30.72 &
       HLL-70,05414249-0154097,B-74,M-122 \nl
140& J054142.5-015617 & 5 41 42.54&  -1 56 17.6&    26   $\pm$ 6 &   1.25&  1.67& ... & ... &1.97e-15&... &HLL-117,05414254-0156175,M-124 \nl
141*f& J054142.6-015445 & 5 41 42.62&  -1 54 45.6&   513   $\pm$ 24&   2.57&  2.90&2.6$^{+0.5}_{-0.5}$&3.4$^{+2.8}_{-0.8}$&1.16e-13&30.80 &
      HLL-90,05414261-0154456,B-76,M-128 \nl
142& J054142.8-015436 & 5 41 42.81&  -1 54 36.0&   686   $\pm$ 27&   1.49&  2.75&2.3$^{+0.3}_{-0.4}$ &2.3$^{+0.5}_{-0.3}$ &1.35e-13&30.90 
   & HLL-85,05414280-0154360,B-78,M-133 \nl
143& J054142.8-015317 & 5 41 42.85&  -1 53 17.0&    13   $\pm$ 5 &   0.77&  3.34& ... & ... &2.76e-15&... &05414288-0153163,M-135 \nl
144& J054143.0-015440 & 5 41 43.03&  -1 54 40.2&    23   $\pm$ 6 &   1.18&  3.57& ... & ... &6.48e-15&... &M-136   \nl
145*& J054143.1-020354 & 5 41 43.12&  -2 03 54.0&  462   $\pm$ 23&   2.93&  1.30& 0.2$^{+0.2}_{-0.1}$ & ... &4.28e-14&... &
    05414316-0203541 \nl
146& J054143.2-015331 & 5 41 43.29&  -1 53 31.5&    30   $\pm$ 7 &   3.96&  2.83& ... & ... &5.40e-15&... &
     HLL-51 \nl
147& J054143.4-015642 & 5 41 43.44&  -1 56 42.4&   285   $\pm$ 18&   1.05&  2.65&2.2$^{+0.5}_{-0.4}$ &2.3$^{+0.7}_{-0.6}$ &4.78e-14&30.48 &
     HLL-124,B89-24,05414344-0156425,V-6 \nl
148*& J054143.4-020349 & 5 41 43.45&  -2 03 49.0&   464   $\pm$ 23&   2.93&  1.30&0.2$^{+0.2}_{-0.2}$  & ... &4.37e-14&... &
       05414344-0203482 \nl
149& J054143.4-015326 & 5 41 43.49&  -1 53 26.3&    20   $\pm$ 6 &   1.68&  4.47& ... & ... &6.87e-15&... &
      M96-146 \nl
150& J054143.4-014837 & 5 41 43.50&  -1 48 37.1&    16   $\pm$ 6 &   0.94&  3.63& ... & ... &4.35e-15&... & ... \nl
151& J054143.5-015511 & 5 41 43.55&  -1 55 11.7&    24   $\pm$ 6 &   2.25&  4.48& ... & ... &2.02e-14&... &05414362-0155114,M-147  \nl
152*& J054143.5-015356 & 5 41 43.55&  -1 53 56.6&   140   $\pm$ 13&  0.87&  3.03&6.1$^{+1.2}_{-0.9}$& 1.4$^{+0.2}_{-0.3}$&2.62e-14&30.83 &
      05414356-0153567,M-149,V-7 \nl
153*& J054143.7-014645 & 5 41 43.79&  -1 46 45.4&   743   $\pm$ 29&   1.62&  1.56&0.3$^{+0.1}_{-0.1}$ & ... &7.13e-14&30.41 &
      05414380-0146457 \nl
154& J054143.8-015621 & 5 41 43.82&  -1 56 22.0&   258   $\pm$ 17&   2.62&  3.57&5.2$^{+1.7}_{-1.3}$ &3.1$^{+2.2}_{-1.0}$ &6.36e-14&30.69 &
      05414383-0156220,M-153,B89-31   \nl
155& J054143.8-015338 & 5 41 43.85&  -1 53 38.3&    19   $\pm$ 5 &   0.88&  3.03& ... & ... &3.43e-15&... &05414384-0153383,M-154 \nl
156& J054143.9-015329 & 5 41 43.91&  -1 53 29.1&   187   $\pm$ 15&   1.42&  3.29&2.7$^{+0.8}_{-0.7}$&4.2$^{+6.0}_{-1.5}$&4.16e-14&30.39 &
     05414391-0153291,M-156  \nl
157& J054143.9-015554 & 5 41 43.94&  -1 55 54.8&    10   $\pm$ 4 &   0.87&  3.65& ... & ... &2.17e-15&... & ... \nl
158& J054143.9-015845 & 5 41 43.96&  -1 58 45.1&    22   $\pm$ 6 &   2.01&  3.04& ... & ... &4.12e-15&... &
      HLL-139,05414398-0158454 \nl
159& J054143.9-015302 & 5 41 43.99&  -1 53 02.2&     9   $\pm$ 4 &   0.62&  3.25& ... & ... &1.95e-15&... & 05414397-0153019 \nl
160*f& J054144.1-015347 & 5 41 44.12&  -1 53 47.1&   301   $\pm$ 18& 6.24&  4.16&15.1$^{+5.1}_{-4.8}$ &2.0$^{+1.3}_{-0.6}$ &8.54e-14&31.38 &
      05414412-0153472,M-165   \nl
161& J054144.1-015520 & 5 41 44.17&  -1 55 20.4&    34   $\pm$ 7 &   0.85&  3.82& ... & ... &9.45e-15&... &M-167 \nl
162& J054144.1-015735 & 5 41 44.18&  -1 57 35.5&     8   $\pm$ 4 &   0.54&  1.87& ... & ... &6.53e-16&... &05414420-0157353 \nl
163& J054144.2-015404 & 5 41 44.22&  -1 54 04.9&    48   $\pm$ 8 &   1.07&  3.54& ... & ... &1.14e-14&... &05414422-0154051,M-170 \nl
164& J054144.3-015524 & 5 41 44.33&  -1 55 24.8&    34   $\pm$ 7 &   0.58&  2.27& ... & ... &4.21e-15&... &M-173  \nl
165& J054144.3-015420 & 5 41 44.36&  -1 54 20.2&    74   $\pm$ 10&   1.48&  2.95& ... & ... &1.34e-14&... &
     05414436-0154203,B-84,M-174 \nl
166& J054144.3-015335 & 5 41 44.37&  -1 53 35.8&    97   $\pm$ 11&   1.32&  3.44&5.0$^{+2.9}_{-1.7}$ &2.3$^{+3.4}_{-1.0}$ &2.19e-14&30.30 &
      05414436-0153358,M-176  \nl
167& J054144.4-015522A& 5 41 44.40&  -1 55 22.8&   138   $\pm$ 13&   0.85&  2.44&2.1$^{+0.5}_{-0.3}$ &2.5$^{+0.9}_{-0.6}$ &2.07e-14&30.11 &
      HLL-107,HLP-58,05414440-0155229,B89-3 \nl
168& J054144.4-015522B& 5 41 44.49&  -1 55 22.0&    56   $\pm$ 9 &   0.60&  3.28& ... & ... &1.28e-14&... & ... \nl
169& J054144.5-015740 & 5 41 44.50&  -1 57 40.2&   266   $\pm$ 17&   1.13&  1.99&0.8$^{+0.2}_{-0.2}$ &2.1$^{+0.4}_{-0.4}$ &3.09e-14&30.09 &
     HLL-133,05414451-0157402,B89-29 \nl
170& J054144.7-015348 & 5 41 44.74&  -1 53 48.5&    39   $\pm$ 7 &   1.83&  3.91& ... & ... &1.06e-14&... &M-182,05414474-0153485  \nl
171& J054144.8-015425 & 5 41 44.82&  -1 54 25.1&    47   $\pm$ 8 &   1.04&  3.11& ... & ... &8.77e-15&... &
     HLL-77,HLP-22,05414482-0154251,B-85,V-11 \nl
172& J054144.8-015435 & 5 41 44.84&  -1 54 35.9&    15   $\pm$ 5 &   0.79&  3.29& ... & ... &3.08e-15&... &
     05414483-0154357,B-86,M-184,HLP-73 \nl
173& J054144.9-015717 & 5 41 44.96&  -1 57 17.3&    41   $\pm$ 7 &   0.55&  0.96& ... & ... &2.59e-15&... &05414496-0157174  \nl
174*f& J054145.0-015406 & 5 41 45.06&  -1 54 06.2&   651   $\pm$ 27&   16.0&  3.72&4.2$^{+0.7}_{-0.5}$&6.1$^{+3.1}_{-2.1}$&1.70e-13&31.01 &
      HLP-74,05414506-0154063,M-188 \nl
175*f& J054145.0-015144 & 5 41 45.08&  -1 51 44.3&  1810   $\pm$ 44&   10.1&  3.03&2.3$^{+0.2}_{-0.2}$ &3.7$^{+0.7}_{-0.5}$ &3.70e-13&31.26 &
     HLL-19,05414508-0151443 \nl
176& J054145.2-015422 & 5 41 45.23&  -1 54 22.7&   196   $\pm$ 15&   1.40&  3.03&3.9$^{+1.2}_{-0.9}$ &1.9$^{+0.8}_{-0.5}$&3.75e-14&30.59 &
     B-89,M-190 \nl
177& J054145.3-015502 & 5 41 45.33&  -1 55 02.6&    16   $\pm$ 5 &   2.11&  3.00& ... & ... &5.90e-15&... &05414532-0155025,M-193 \nl
178& J054145.3-015156 & 5 41 45.38&  -1 51 56.6&   224   $\pm$ 16&   0.49&  3.22&3.4$^{+1.1}_{-0.8}$ &2.9$^{+1.8}_{-0.9}$&4.98e-14&30.52 &
      05414538-0151566 \nl
179& J054145.3-015355 & 5 41 45.40&  -1 53 55.7&    13   $\pm$ 5 &   0.48&  2.58& ... & ... &2.17e-15&... &M-195 \nl
180& J054145.4-015425 & 5 41 45.42&  -1 54 25.7&    64   $\pm$ 9 &   0.76&  3.97& ... & ... &1.74e-14&... & ... \nl
181& J054145.4-015616 & 5 41 45.48&  -1 56 17.0&     8   $\pm$ 4 &   0.19&  0.92& ... & ... &3.96e-16&... &
     HLL-116,05414548-0156169,M-196,B89-22 \nl
182*& J054145.4-015428 & 5 41 45.48&  -1 54 28.6&   371   $\pm$ 20&   1.41&  3.14&3.3$^{+0.9}_{-0.5}$ &2.6$^{+0.7}_{-0.6}$&7.42e-14&30.65 &
     05414550-0154286,B-90,M-197,IRS2b,V-15    \nl
183*f& J054145.5-015426 & 5 41 45.52&  -1 54 26.9&   220   $\pm$ 16&   5.94&  3.41&2.9$^{+1.0}_{-1.0}$ &6.7$^{+...}_{-3.1}$ &5.28e-14&30.39 
     &V-16  \nl
184& J054145.6-015732 & 5 41 45.61&  -1 57 33.0&    34   $\pm$ 7 &   0.95&  1.60& ... & ... &3.47e-15&... &05414562-0157330 \nl
185& J054145.6-015504 & 5 41 45.69&  -1 55 04.6&    14   $\pm$ 5 &   0.75&  3.98& ... & ... &6.56e-15&... &M-200  \nl
186& J054145.7-015212 & 5 41 45.72&  -1 52 12.8&     10   $\pm$ 4 &   0.57&  1.91& ... & ... &8.49e-16&... &05414576-0152130 \nl
187*& J054145.7-015429 & 5 41 45.80&  -1 54 29.8&   162   $\pm$ 14&   1.30&  3.42&3.1$^{+1.0}_{-0.8}$ &4.9$^{+...}_{-2.1}$ &3.74e-14&30.37 &
    HLP-1,B89-2,0541458-0154297,B-92,M-205,V-19 \nl
188*f& J054145.8-015410 & 5 41 45.89&  -1 54 10.9&  1065   $\pm$ 34&   2.53&  3.31&3.9$^{+0.4}_{-0.3}$ &2.8$^{+0.5}_{-0.4}$&2.47e-13&31.25 &
      M-206,V-20 \nl
189*& J054145.9-015626 & 5 41 45.94&  -1 56 26.4&   176   $\pm$ 14&   1.45&  3.88&5.5$^{+2.6}_{-1.9}$ & ... &5.03e-14&30.54 &
     05414594-0156263,M-207,B89-23  \nl
190& J054145.9-015501 & 5 41 45.97&  -1 55 01.9&    65   $\pm$ 9 &   1.61&  3.41& ... & ... &2.42e-14&... & 
     HLL-96,HLP-92,M-208 \nl
191& J054146.0-015653 & 5 41 46.02&  -1 56 53.8&    33   $\pm$ 7 &   1.38&  3.76& ... & ... &8.40e-15&... & 05414602-0156538 \nl
192*f& J054146.1-015414 & 5 41 46.11&  -1 54 14.7&  1753   $\pm$ 43&   27.4&3.89&3.7$^{+0.5}_{-0.5}$ & ... &5.20e-13&var. &
     05414611-0154147,B-93,M-210 \nl
193*& J054146.1-015621 & 5 41 46.15&  -1 56 21.9&   397   $\pm$ 21&   2.96&  4.24&9.6$^{+2.4}_{-1.8}$ &5.0$^{+5.4}_{-1.9}$ &1.27e-13&31.07 &
      V-21  \nl
194& J054146.2-015346 & 5 41 46.20&  -1 53 46.7&    66   $\pm$ 9&   0.79&  2.49& ... & ... &2.00e-14&... &
     HLL-58,05414621-0153466,M-212,B89-7 \nl
195& J054146.2-015533 & 5 41 46.23&  -1 55 33.1&   165   $\pm$ 14&   1.89&  4.15&11.8$^{+10.0}_{-5.3}$ &3.3$^{+...}_{-1.9}$ &4.84e-14&30.86 &      M-211 \nl
196*& J054146.2-015654 & 5 41 46.23&  -1 56 54.6&   150   $\pm$ 13&   1.17&  4.07&5.2$^{+2.4}_{-1.4}$ & ... &4.50e-14&30.41 & V-22  \nl
197& J054146.2-015554 & 5 41 46.26&  -1 55 54.1&    32   $\pm$ 7 &   0.34&  4.77& ... & ... &1.18e-14&... & ... \nl
198& J054146.4-015011 & 5 41 46.40&  -1 50 11.3&    47   $\pm$ 8 &   0.91&  2.01& ... & ... &5.64e-15&... &
      HLL-5,05414641-0150111 \nl
199f& J054146.5-015148 & 5 41 46.52&  -1 51 48.6&    91   $\pm$ 11&   6.37&  3.42&3.0$^{+1.3}_{-0.5}$ &4.6$^{+...}_{-2.1}$ &2.33e-14&30.17 &
       05414652-0151487 \nl
200& J054146.5-015446 & 5 41 46.57&  -1 54 46.6&     94   $\pm$ 11&   1.24&  3.27&7.0$^{+5.7}_{-2.2}$ &1.8$^{+1.9}_{-1.1}$ &6.68e-14&31.06 
   & 05414655-0154469,B-94,M-214,V-23 \nl
201& J054146.8-015447 & 5 41 46.86&  -1 54 47.4&    16   $\pm$ 5 &   1.26&  4.14& ...                & ...                &2.70e-14&... &
     M-217 \nl
202& J054146.8-014957 & 5 41 46.89&  -1 49 57.1&   370   $\pm$ 20&   0.88&  2.07&0.8$^{+0.2}_{-0.2}$ &2.5$^{+0.7}_{-0.6}$ &4.48e-14&30.25 &
     HLL-4,HLP-23,05414690-0149573 \nl
203& J054147.0-014926 & 5 41 47.01&  -1 49 26.6&   449   $\pm$ 22&   3.14&  2.47&1.5$^{+0.3}_{-0.4}$ &2.9$^{+1.5}_{-0.7}$ &7.13e-14&30.55 &
     HLL-1,05414702-0149265 \nl
204& J054147.0-015035 & 5 41 47.10&  -1 50 35.4&   158   $\pm$ 14&   0.51&  1.93&1.0$^{+0.4}_{-0.4}$ &1.3$^{+0.4}_{-0.3}$ &1.67e-14&29.99 &
      HLL-8,HLP-90,05414712-0150353 \nl
205& J054147.1-020018 & 5 41 47.14&  -2 00 18.8&    29   $\pm$ 7 &   0.96&  1.65& ... & ... &2.27e-15&... &05414714-0200190 \nl
206& J054147.2-015819 & 5 41 47.21&  -1 58 19.7&    12   $\pm$ 5 &   0.94&  2.81& ... & ... &1.98e-15&... &HLL-138,05414723-0158202 \nl
207*f& J054147.4-015526 & 5 41 47.45&  -1 55 26.3& 899   $\pm$ 31&   17.9&  3.74&4.3$^{+0.7}_{-0.6}$ &7.8$^{+...}_{-2.8}$ &2.40e-13&31.11 &       M-220 \nl
208& J054147.9-015541 & 5 41 47.93&  -1 55 41.3&    24   $\pm$ 6 &   0.55&  4.06& ... & ... &6.80e-15&... & 05414793-0155416,M-225 \nl
209& J054147.9-015358 & 5 41 47.97&  -1 53 58.4&    17   $\pm$ 5 &   0.67&  3.45& ... & ... &4.13e-15&... &05414796-0153585,M-226 \nl
210*f& J054148.2-015601& 5 41 48.21&  -1 56 01.8&  1616   $\pm$ 41&   15.4&  3.61&4.5$^{+0.5}_{-0.3}$ &4.6$^{+1.3}_{-0.9}$ &4.17e-13&31.45 
     & 05414821-0156020,M-230,V-24  \nl
211& J054148.6-015349 & 5 41 48.60&  -1 53 49.5&    95   $\pm$ 11&   0.96&  2.52&1.3$^{+0.7}_{-0.9}$ &3.0$^{+...}_{-1.3}$ &1.50e-14&29.80 &
     HLL-59,05414861-0153495 \nl
212& J054148.6-015540 & 5 41 48.63&  -1 55 41.0&     9   $\pm$ 4 &   0.99&  4.65& ... & ... &3.49e-15&... & ...  \nl
213& J054148.6-015415 & 5 41 48.69&  -1 54 15.6&    32   $\pm$ 7 &   2.42&  2.59& ... & ... &5.46e-15&... &HLL-74,05414868-0154159 \nl
214& J054149.3-015332 & 5 41 49.31&  -1 53 32.7&    76   $\pm$ 10&   1.94&  2.66&1.4$^{+0.7}_{-1.0}$  &3.1$^{+...}_{-1.6}$ &1.25e-14&29.79 &
      HLL-50,05414934-0153318 \nl
215& J054149.6-014601 & 5 41 49.63&  -1 46 01.7&   146   $\pm$ 13&   2.34&  2.03&1.1$^{+0.5}_{-0.4}$ &1.6$^{+0.7}_{-0.5}$ &1.70e-14&30.05 &
     05414961-0146015 \nl
216& J054149.6-015326 & 5 41 49.66&  -1 53 26.9&    89   $\pm$ 10&    0.99&  2.79& 1.6$^{+0.7}_{-0.6}$ &4.8$^{+...}_{-2.3}$&1.53e-14&29.78 &
      HLL-48,05414966-0153271 \nl
217& J054149.8-015255 & 5 41 49.81&  -1 52 55.8&    21   $\pm$ 6 &   0.98&  4.37& ... & ... &6.63e-15&... &HLL-36,05414983-0152560 \nl
218& J054149.8-015947 & 5 41 49.88&  -1 59 47.4&    11   $\pm$ 5 &   0.47&  3.39& ... & ... &2.76e-15&... &05414984-0159475 \nl
219& J054150.0-015807 & 5 41 50.04&  -1 58 07.8&    11   $\pm$ 4 &   1.29&  1.56& ... & ... &9.11e-16&... &05415006-0158083 \nl
220& J054150.1-015744 & 5 41 50.15&  -1 57 44.7&   154   $\pm$ 13&   0.95&  3.12&2.5$^{+1.0}_{-1.2}$ &4.2$^{+1.8}_{-1.8}$ &3.26e-14&30.18 &
     HLL-135,HLP-98,05415014-0157449 \nl
221& J054150.5-015315 & 5 41 50.50&  -1 53 15.2&   558   $\pm$ 25&   1.25&  2.11&0.9$^{+0.2}_{-0.1}$ &2.2$^{+0.4}_{-0.3}$ &6.78e-14&30.48 &
      HLL-41,05415050-0153153,B89-6 \nl
222*f& J054150.5-020256 & 5 41 50.50&  -2 02 56.9&   242   $\pm$ 17&   6.10&  3.46&4.5$^{+1.3}_{-0.9}$&3.2$^{+2.5}_{-1.1}$&6.34e-14&30.67 &
      05415053-0202570 \nl
223*& J054150.6-015306 & 5 41 50.69&  -1 53 06.2&    22   $\pm$ 6 &   0.68&  1.86& ... & ... &2.05e-15&... & see notes  \nl
224& J054150.7-015548 & 5 41 50.74&  -1 55 48.3&    13   $\pm$ 5 &   0.75&  2.76& ... & ... &2.32e-15&... & 05415074-0155486  \nl
225& J054150.8-015312 & 5 41 50.86&  -1 53 12.7&    22   $\pm$ 6 &   0.77&  2.67& ... & ... &3.40e-15&... &HLL-40,05415087-0153128 \nl
226& J054150.8-015722 & 5 41 50.87&  -1 57 22.2&     7   $\pm$ 4 &   0.62&  4.03& ... & ... &2.63e-15&... & ... \nl
227& J054150.9-014801 & 5 41 50.95&  -1 48 01.6&     6   $\pm$ 4 &   0.74&  3.63& ... & ... &2.59e-15&... & ... \nl
228& J054150.9-015520 & 5 41 51.00&  -1 55 21.0&    13   $\pm$ 5 &   0.39&  1.10& ... & ... &6.24e-16&... &05415100-0155213 \nl
229& J054151.6-015216 & 5 41 51.60&  -1 52 16.6&    66   $\pm$ 9 &   0.50&  2.34& ... & ... &1.00e-14&... &05415161-0152167 \nl
230& J054152.0-015518 & 5 41 52.08&  -1 55 18.1&    18   $\pm$ 5 &   1.16&  1.58& ... & ... &1.58e-15&... &05415209-0155183 \nl
231& J054152.1-015717 & 5 41 52.15&  -1 57 17.3&    30   $\pm$ 7 &   1.37&  2.34& ... & ... &1.35e-14&... &
      HLL-127,HLP-75,B89-26,05415215-0157174 \nl
232*& J054152.1-020121 & 5 41 52.20&  -2 01 21.1&   171   $\pm$ 14&   0.77&  2.21&1.5$^{+0.6}_{-0.5}$ &1.6$^{+0.6}_{-0.5}$&2.41e-14&30.17 &
      05415220-0201211 \nl
233& J054152.9-015629 & 5 41 52.93&  -1 56 29.3&    55   $\pm$ 8 &   0.53&  3.35& ... & ... &1.42e-14&... &HLL-121,B89-25,05415292-0156297 \nl
234*f& J054152.9-015634 & 5 41 52.94&  -1 56 34.8&   783   $\pm$ 29&   6.04&  2.58&1.5$^{+0.3}_{-0.2}$ &3.1$^{+1.1}_{-0.7}$ &1.31e-13&30.80 &
       HLL-123,05415294-0156351 \nl
235& J054153.1-015743 & 5 41 53.19&  -1 57 43.8&    59   $\pm$ 9 &   1.91&  3.57& ... & ... &2.96e-14&... &
     HLL-134,HLP-78,05415319-0157441,B89-27  \nl
236& J054153.8-015516 & 5 41 53.87&  -1 55 16.1&   304   $\pm$ 18&   1.04&  2.77&2.5$^{+0.7}_{-0.5}$ &2.3$^{+0.8}_{-0.7}$ &5.19e-14&30.55 &
      HLL-101,HLP-88,05415387-0155163,B89-21 \nl
237& J054154.1-015817 & 5 41 54.12&  -1 58 17.1&    12   $\pm$ 5 &   0.59&  1.50& ... & ... &1.57e-15&... &05415406-0158170 \nl
238& J054154.1-014801 & 5 41 54.19&  -1 48 01.7&    77   $\pm$ 10&   1.26&  1.93& ... & ... &1.02e-14&... &05415419-0148017 \nl
239& J054154.3-015556 & 5 41 54.31&  -1 55 56.2&    25   $\pm$ 6 &   1.01&  1.22& ... & ... &1.67e-15&... &HLL-112,05415430-0155566 \nl
240& J054154.7-015726 & 5 41 54.70&  -1 57 27.0&    33   $\pm$ 7 &   0.49&  1.61& ... & ... &2.74e-15&... &HLL-128,05415472-0157273 \nl
241& J054154.9-015647 & 5 41 54.93&  -1 56 47.9&    22   $\pm$ 6 &   0.56&  2.24& ... & ... &3.16e-15&... &HLL-125,05415492-0156483 \nl
242& J054154.9-015630 & 5 41 54.96&  -1 56 30.9&    18   $\pm$ 5 &   1.20&  3.22& ... & ... &3.82e-15&... &05415497-0156312  \nl
243& J054154.9-015957 & 5 41 54.99&  -1 59 57.8&    19   $\pm$ 6 &   0.67&  3.72& ... & ... &5.00e-15&... & ... \nl
244& J054155.0-020030 & 5 41 55.04&  -2 00 30.9&   111   $\pm$ 12&   3.36&  3.13&4.2$^{+1.6}_{-0.5}$ &2.5$^{+1.4}_{-0.9}$ &2.34e-14&30.30 
   &05415505-0200309 \nl
245& J054155.2-014847 & 5 41 55.21&  -1 48 47.4&    15   $\pm$ 6 &   0.75&  1.85& ... & ... &1.21e-15&... &05415518-0148485 \nl
246& J054155.7-015508 & 5 41 55.79&  -1 55 08.5&   135   $\pm$ 13&   1.61&  1.40&0.1$^{+0.2}_{-0.1}$ &1.4$^{+0.5}_{-0.7}$ &9.95e-15&29.41 &
     HLL-99,05415579-0155092 \nl
247& J054156.5-014616 & 5 41 56.57&  -1 46 17.0&   217  $\pm$ 16&   0.82&  2.11&0.6$^{+0.3}_{-0.2}$ &2.3$^{+1.5}_{-0.7}$ &2.78e-14&30.17 &
     05415656-0146168 \nl
248& J054156.6-015552 & 5 41 56.68&  -1 55 52.4&   170   $\pm$ 14&   3.12&  1.78&0.5$^{+0.3}_{-0.2}$ &1.5$^{+0.5}_{-0.5}$&1.56e-14&29.78 &
     HLL-110,05415668-0155526 \nl
249& J054156.7-015853 & 5 41 56.74&  -1 58 53.4&    28   $\pm$ 6 &   0.49&  2.29& ... & ... &4.33e-15&... &05415674-0158538 \nl
250& J054156.7-015351 & 5 41 56.79&  -1 53 51.8&   264   $\pm$ 17&   1.92&  2.77&1.7$^{+0.5}_{-0.4}$ &3.6$^{+2.2}_{-1.2}$ &1.18e-13&30.71 &
     05415680-0153521 \nl
251& J054157.0-015032 & 5 41 57.02&  -1 50 32.0&    16   $\pm$ 5 &   1.47&  1.86& ... & ... &1.61e-15&... &HLL-7,05415697-0150331   \nl
252& J054157.8-015127 & 5 41 57.83&  -1 51 27.7&    38   $\pm$ 7 &   1.22&  1.49& ... & ... &2.64e-15&... &HLL-14,05415786-0151278  \nl
253& J054158.0-015957 & 5 41 58.10&  -1 59 57.3&    39   $\pm$ 7 &   1.82&  1.37& ... & ... &3.09e-15&... &05415807-0159575 \nl
254& J054159.8-014735 & 5 41 59.82&  -1 47 35.3&    39   $\pm$ 8 &   1.29&  2.04& ... & ... &3.23e-15&... &05415976-0147363 \nl
255& J054159.8-015949 & 5 41 59.89&  -1 59 49.6&     7   $\pm$ 4 &   0.79&  3.49& ... & ... &2.20e-15&... &05415997-0159496  \nl
256& J054201.1-015621 & 5 42 01.14&  -1 56 21.3&    23   $\pm$ 6 &   0.63&  1.70& ... & ... &2.66e-15&... &05420114-0156214  \nl
257& J054201.4-020008 & 5 42 01.47&  -2 00 08.3&    15   $\pm$ 5 &   0.36&  3.74& ... & ... &3.77e-15&... & ...  \nl
258*& J054201.7-015629 & 5 42 01.71&  -1 56 29.6&   160   $\pm$ 14&   1.84&  2.29&0.9$^{+0.3}_{-0.3}$ & 3.1$^{+2.4}_{-1.1}$ &2.19e-14&29.94 &
       HLL-120,05420172-0156299 \nl
259& J054201.9-015629 & 5 42 01.98&  -1 56 29.2&    19   $\pm$ 5 &   0.48&  1.77& ... & ... &1.64e-15&... &05420201-0156289 \nl
260& J054204.2-015201 & 5 42 04.27&  -1 52 01.8&    16   $\pm$ 5 &   0.66&  3.21& ... & ... &3.40e-15&... & ... \nl
261& J054205.2-015344 & 5 42 05.27&  -1 53 44.4&    65   $\pm$ 9 &   2.38&  1.87& ... & ... &9.36e-15&... &05420528-0153447 \nl
262& J054205.7-015928 & 5 42 05.77&  -1 59 28.7&    42   $\pm$ 8 &   1.46&  2.49& ... & ... &7.67e-15&... &05420580-0159289 \nl
263& J054205.9-015252 & 5 42 05.95&  -1 52 52.0&   136   $\pm$ 13&   1.53&  1.79&0.4$^{+0.2}_{-0.2}$  & 2.2$^{+4.1}_{-0.6}$ &3.63e-14&30.08 &
     HLL-35,05420597-0152522 \nl
264& J054207.5-015335 & 5 42 07.51&  -1 53 35.1&    22   $\pm$ 6 &   0.52&  1.75& ... & ... &1.68e-15&... &05420744-0153368 \nl
265& J054207.5-015738 & 5 42 07.55&  -1 57 38.2&    18   $\pm$ 6 &   0.72&  3.66& ... & ... &6.05e-15&... & ... \nl
266& J054207.6-015322 & 5 42 07.60&  -1 53 22.1&     9   $\pm$ 4 &   0.52&  3.05& ... & ... &1.73e-15&... & ... \nl
267& J054208.9-015741 & 5 42 08.98&  -1 57 41.2&    25   $\pm$ 6 &   0.77&  2.83& ... & ... &4.68e-15&... & ...  \nl
268& J054209.3-015804 & 5 42 09.35&  -1 58 04.5&   528   $\pm$ 24&   1.37&  1.78&0.5$^{+0.1}_{-0.2}$ &2.0$^{+0.7}_{-0.4}$ &5.32e-14&30.27 &
     05420934-0158048 \nl
269*& J054209.5-020004 & 5 42 09.57&  -2 00 04.2&  1207   $\pm$ 36 &   4.47&  1.95&0.6$^{+0.1}_{-0.1}$&2.4$^{+0.3}_{-0.3}$&1.36e-13&30.69 &
      05420956-0200044 \nl
270& J054210.2-020116 & 5 42 10.22&  -2 01 16.3&     10   $\pm$ 5 &   0.72&  2.74& ... & ... &1.25e-15&... & ...  \nl
271& J054211.0-014626 & 5 42 11.07&  -1 46 26.2&    24   $\pm$ 6 &   0.76&  2.09& ... & ... &2.99e-15&... &05421113-0146255 \nl
272*& J054211.2-015943 & 5 42 11.25&  -1 59 43.3&  1352   $\pm$ 38&   1.53& 1.26&0.2$^{+0.2}_{-0.1}$  & ... &1.18e-13&... &05421125-0159437 \nl
273& J054211.8-015637 & 5 42 11.89&  -1 56 37.9&     9   $\pm$ 4 &   0.98&  2.86& ... & ... &1.78e-15&... & ...  \nl
274*& J054213.1-015833 & 5 42 13.19&  -1 58 33.3&  1251  $\pm$ 36&   2.85&  2.07&0.7$^{+0.2}_{-0.2}$& 2.5$^{+0.4}_{-0.3}$ &1.72e-13&30.81 &
      05421319-0158336 \nl
275& J054214.1-015141 & 5 42 14.12&  -1 51 41.3&   13    $\pm$ 6  &   0.53&  4.17& ... & ... &6.90e-15&... & ... \nl
276& J054215.5-015500 & 5 42 15.55&  -1 55 00.1&    15   $\pm$ 5 &    0.96&  3.07& ... & ... &3.03e-15&... & ... \nl
277& J054216.2-015614 & 5 42 16.24&  -1 56 14.6&   180   $\pm$ 15&   1.45&  1.40&0.2$^{+0.2}_{-0.2}$ &0.8$^{+0.3}_{-0.2}$ &1.44e-14&29.80 & 
      Haro 5-53,05421620-0156148  \nl
278& J054216.7-015308 & 5 42 16.79&  -1 53 08.9&    19   $\pm$ 6 &   0.99&  2.10& ... & ... &2.80e-15&... &05421677-0153091 \nl
279& J054218.0-015127 & 5 42 18.08&  -1 51 27.9&    20   $\pm$ 6 &   0.77&  2.78& ... & ... &4.74e-15&... & ... \nl
280& J054218.4-015731 & 5 42 18.45&  -1 57 31.7&    77   $\pm$ 10&   0.72&  3.39& ... & ...  &2.31e-14&... & ...  \nl
281& J054218.6-015421 & 5 42 18.63&  -1 54 21.8&    16   $\pm$ 6 &   0.60&  3.24& ... & ... &5.04e-15&... & ...  \nl
282& J054221.2-015910 & 5 42 21.29&  -1 59 10.2&    88   $\pm$ 11&   0.82&  1.64&0.0$^{+0.1}_{-...}$ &1.0$^{+0.3}_{-0.2}$ &1.01e-14&29.75 &
     05422123-0159104 \nl
283& J054226.5-015913 & 5 42 26.53&  -1 59 13.2&     8   $\pm$ 5 &   1.44&  2.63& ... & ... &2.19e-15&... &  ...
\enddata
\tablenotetext{a}{
Col. (1) running source number; an asterisk indicates that additional source notes follow
below; an $f$ inidicates that a flare or other obvious variability is seen in the
source light curve, (2) J2000 IAU source name, (3)-(4) J2000 {\it Chandra} position
with astrometric correction applied, (5) net source counts (= total counts - background counts)
$\pm$ net counts uncertainty (Gehrels 1986) in the 0.5 - 7.0 keV range; source counts
were measured in PSF-corrected 95\% encircled energy regions except for the brightest
sources where 99\% encircled energy regions were used; smaller regions were used 
for some closely-spaced sources, 
(6) KS variability statistic, sources
with KS $>$ 1.7 are variable (see text), 
(7) mean photon energy as determined from source events in
the 0.5 - 7 keV range, (8)-(9) absorption
column density N$_H$ and time-averaged characteristic temperature kT for sources with 
$\geq$90 counts as determined from spectral fits with 1T VAPEC optically thin plasma model
using a fixed iron abundance Fe =  0.3 solar (errors are 90\% confidence),
(10) absorbed flux (0.5 - 7 keV) as determined from source
event list (see text), (11) unabsorbed luminosity (0.5 - 7 keV) at d = 415 pc   
determined from spectral fits of brighter sources where an estimate of
N$_{\rm H}$ and kT is available,
(12) candidate IR or optical identification from 2MASS, SIMBAD, and  
B89 = Barnes et al. 1989; B = Beck et al. 2003; HLL = Haisch, Lada \& Lada
2000; HLP = Haisch et al. 2001, M = Meyer 1996. IR identifications found in
the all-sky release of the 2MASS Point Source Catalog are listed by their
coordinates (e.g. 05411324-0153306). Radio identifications (V-xx) are from
3.6 cm VLA observations of Rodriguez et al. 2003.  } \\ 
\end{deluxetable}
\clearpage

\footnotesize{
\noindent 5.  1T model gives a high but uncertain temperature kT $\geq$ 4 keV. \\
15. HLL-131 is a likely IR counterpart, but there is a $\approx$4$''$ positional offset even after
    correcting for systematic differences between CXO and HLL coordinates. \\
16. HLP-54 is a likely IR counterpart, but there is a $\approx$4$''$ positional offset between 
    CXO and HLP coordinates. \\
18. Flare spectrum gives kT = 5.4$^{+...}_{-2.2}$ keV and log Lx = 31.58 ergs s$^{-1}$ (unabsorbed). \\
27. Flare spectrum gives kT = 4.2$^{+0.6}_{-0.7}$ keV 
    and log Lx = 30.99 ergs s$^{-1}$ (unabsorbed). (Fig. 7) \\
36. Variability is present throughout the observation.
    Complex or rapidly varying temperature structure suspected. \\ 
39. Source lies in CCD gap. Flux is uncertain. \\
41. Weak flare whose  spectrum gives kT = 2.6$^{+1.0}_{-0.5}$ keV 
    and log Lx = 30.80 ergs s$^{-1}$ (unabsorbed). \\
52. Weak flare. Insufficient counts to accurately determine flare parameters.  \\ 
75. 1T models give $\chi^2_{red}$ = 1.09 but
    2T models give a slightly better fit with $\chi^2_{red}$ = 1.05 and converge to 
    N$_{\rm H}$ = 1.7$^{+0.2}_{-0.2}$ $\times$ 10$^{22}$ cm$^{-2}$, kT$_{1}$ = 0.6$^{+0.3}_{-0.2}$ keV, 
    kT$_{2}$ = 2.7$^{+0.4}_{-0.3}$  keV, and log Lx = 31.50 ergs s$^{-1}$ (unabsorbed). \\
81. Flare spectrum  gives kT = 9.8$^{+...}_{-4.5}$ keV 
    and log Lx = 31.08 ergs s$^{-1}$ (unabsorbed). (Fig. 7) \\
85. Flare spectrum gives kT = 3.8$^{+0.5}_{-0.5}$ keV and 
    log Lx = 31.34 ergs s$^{-1}$ (unabsorbed). (Figs. 7 \& 10) \\
93. Flare spectrum gives kT = 3.5$^{+2.1}_{-1.1}$ keV and log Lx = 30.98 ergs s$^{-1}$ (unabsorbed). \\
94.  Slow variability is present throughout the observation. DEM model shows a 
     double-peaked structure with cool and hot components.  \\
99. 1T models give $\chi^2_{red}$ = 0.94 but
    2T models give a slightly better fit with $\chi^2_{red}$ = 0.92 and converge to a larger absorption
    N$_{H}$ = 2.7$^{+0.4}_{-0.6}$ $\times$ 10$^{22}$ cm$^{-2}$. \\
104. Declining count rate may indicate decay phase of a flare. \\
105. Weak flare. Insufficient counts to accurately determine flare parameters.  \\
108. Flare spectrum gives same temperature kT = 3.3$^{+1.0}_{-0.6}$ keV and log Lx = 31.01 
     ergs s$^{-1}$ (unabsorbed). \\
114. Declining count rate may indicate decay phase of a flare. Possible 
     spectral contamination from nearby source. \\
120. Source may be affected by moderate pileup ($\approx$10 - 15\%) during short periods
     when the count rate fluctuated to levels of $\approx$0.15 c s$^{-1}$. \\
123. Count rate increased throughout the observation (Fig. 7). \\
133. 1T models are only marginally acceptable with $\chi^2_{red}$ = 1.35. DEM model peaks near 1.7 keV but
     shows substantial emission up to $\sim$4 keV. Declining count rate may indicate decay phase of a flare. \\
139. Slowly rising variability and outburst spectrum gives 
     kT = 4.2$^{+...}_{-1.6}$ keV and log Lx = 30.79 ergs s$^{-1}$ (unabsorbed). \\
141. The spectrum used to determine N$_{\rm H}$, kT excludes data from a weak flare
     that occurred during the last 5 ksec of the observation. \\
145. No reliable spectral fit because of contamination from nearby source no. 148. Possible
     optical counterpart is GSC 04771-00589. \\
148. No reliable spectral fit because of contamination from nearby source no. 145. \\
152. 1T models are only marginally acceptable with $\chi^2_{red}$ = 1.38. DEM model peaks near 1.4 keV. \\
153. 1T models are not acceptable with $\chi^2_{red}$ = 1.91 but 2T models are 
     acceptable with $\chi^2_{red}$ = 1.00. 
      N$_{\rm H}$ is from 2T model (kT$_{1}$ = 0.8 keV, kT$_{2}$ = 3.4 keV). \\
160. Weak flare. Insufficient counts to accurately determine flare parameters. \\
174. Flare spectrum gives kT = 4.5$^{+0.6}_{-0.7}$ keV and log Lx = 31.45 ergs s$^{-1}$ (unabsorbed). 
     Moderate pileup may have occurred near flare peak. (Fig. 7) \\
175. Long duration flare with a secondary peak during decay phase (Fig. 7). 
     Flare spectrum gives kT = 5.2$^{+2.6}_{-1.3}$ keV and log Lx = 31.47 ergs s$^{-1}$ (unabsorbed). \\
182. The counterpart name IRS 2b is from Nisini et al. (1994). 
     Spectral fit parameters are uncertain because of possible contamination from
     source 183 located 1.7$''$ to the north (Fig. 2). Some improvement is obtained with
     2T models which place most of the emission measure in a cool component
     at kT$_{1}$ $\approx$ 0.2 keV and require a larger absorption column
     N$_{H}$ = 5.5$^{+2.5}_{-2.0}$ $\times$ 10$^{22}$ cm$^{-2}$. \\
183. Weak flare. Insufficient counts to accurately determine flare parameters.  
     Spectral fit parameters are uncertain due to possible contamination from 
     source 182. \\ 
187. 1T models give $\chi^2_{red}$ = 1.10 but 
     2T models give a slightly better fit with $\chi^2_{red}$ = 0.88 and converge to a higher absorption 
     N$_{H}$ = 6.1$^{+4.9}_{-2.7}$ $\times$ 10$^{22}$ cm$^{-2}$,
     kT$_{1}$ = 0.6$^{+1.0}_{-0.3}$ keV, and a hotter component with uncertain
     temperature kT$_{2}$ $\geq$ 3 keV.  \\
188. Weak flare.
     Flare spectrum gives kT = 3.2$^{+0.8}_{-0.6}$ keV and same log Lx = 31.25 ergs s$^{-1}$ (unabsorbed). \\
189. 1T  models are not acceptable ($\chi^2_{red}$ = 1.52) nor are 2T models  ($\chi^2_{red}$ = 1.68).
     Hard heavily-absorbed spectrum with uncertain temperature.  \\
192. Slowly rising outburst (Fig. 7) with high absorption. The X-ray temperature shortly after
     onset reached values of at least kT $\sim$ 6 keV, and possibly as high as
     kT $\sim$ 10 keV. The unabsorbed luminosity varied from a pre-outburst value
     log Lx = 30.5 ergs s$^{-1}$ to at least log Lx = 31.8 ergs s$^{-1}$ during the outburst. 
     Source is affected by moderate pileup ($\approx$27\%) near end of observation. \\
193. Declining light curve may indicate decay phase of a flare. \\
196. 1T models give a high but uncertain temperature kT $\geq$ 3.1 keV. \\
207. Slowly rising outburst (Fig. 7). Very hard source. Time-averaged outburst 
     spectrum gives kT = 7.8$^{+...}_{-3.3}$ keV and  
     log Lx = 31.4 ergs s$^{-1}$ (unabsorbed). Temperature peaked about 4 hours
     after onset followed by a general decline, but there is evidence for at least one 
     reheating event.  
     DEM model of outburst spectrum suggests that most of the emission measure comes from
     plasma above $\sim$4 keV. Insufficient counts to analyze pre-outburst spectrum. \\
210. Declining light curve during first $\sim$20 ksec of observation may indicate decay
     phase of a flare. Decay spectrum gives kT = 4.9$^{+1.7}_{-1.1}$ keV 
     and log Lx = 31.59 ergs s$^{-1}$ (unabsorbed). 
     Source may be affected by moderate pileup ($\approx$16\%) at beginning of observation. \\
222. Weak flare. Insufficient counts to accurately determine flare parameters. \\
223. A faint IR source is visible at the X-ray position in 2MASS H and K-band images
     (2MASS J05415068-0153063), but is flagged as a filter glint artifact. Because
     of the close positional agreement with {\em Chandra}, the  IR source flagged as an 
     artifact may be real. \\
232. 1T models are only marginally acceptable with $\chi^2_{red}$ = 1.39.  \\
234. Declining light curve may indicate decay phase of a flare. \\
258. Possible double source. \\
269. Count rate declined throughout the observation (Fig. 7). \\
272. Soft source (low absorption). 1T models underestimate the flux below 1 keV and are not 
     acceptable ($\chi^2_{red}$ = 3.2). 2T models give improved fits 
     when the Fe abundance is allowed to vary ($\chi^2_{red}$ = 1.24), converging to Fe = 0.8 solar with 
     kT$_{1}$ = 0.77 keV and kT$_{2}$ = 1.9 keV. 
     DEM model peaks near 0.7 keV. \\
274. 1T  models are only marginally acceptable with $\chi^2_{red}$ = 1.42.  
     DEM model peaks near 2 keV. Variability is
     present throughout the observation. \\ 
}

\end{document}